\documentclass{article}
\usepackage[T1]{fontenc}
\usepackage[utf8]{inputenc}
\usepackage{lmodern,geometry}
\usepackage{amsmath,amssymb}
\usepackage{graphicx}
\usepackage{booktabs,longtable,array}
\usepackage{caption}
\usepackage{pdflscape,placeins}
\usepackage{microtype}
\usepackage{hyperref}
\usepackage{xurl}
\hypersetup{hidelinks,pdftitle={Information-theoretic Reconstruction of Geometric and Structural Aspects of Undeciphered Historical Messages}}
\title{Reconstruction of Geometric and Structural Aspects of Undeciphered Messages by Information-theoretic Methods}
\author{Hector Zenil$^{1,2}$, Abicumaran Uthamacumaran$^{1}$\\[2mm]
\small $^{1}$ Algocyte | Oxford Immune Algorithmics, Oxford University Innovation \\ \& 
\small London Institute for Healthcare Engineering, UK\\
\small $^{2}$ Algorithmic Dynamics Lab, Department of Biomedical Computing,\\
\small School of Biomedical Engineering and Imaging Sciences, King's Institute for AI,\\
\small King's College London, UK\\
}
\date{}

\begin{document}
\maketitle
\begin{abstract}
In previous work, we established sufficient conditions for recovering geometry and
topology from non-random information, demonstrating reconstruction with the
Arecibo message. Undeciphered historical records present a related problem:
one-way communication without access to their makers' intentions or encoding
conventions. Here we extend this framework to the Phaistos Disc, Andean khipu,
Rongorongo, Indus inscriptions and the Voynich manuscript, using Egyptian
layouts and Arecibo as controls. Algorithmic information dynamics (AID) guides
structural perturbations, combining classical information measures, predictive
coding, compression and algorithmic-probability-based estimates. Phaistos
retains 0.714 bits of excess adjacent-sign mutual information after positional
controls; its authentic segmentation lies within a broad information basin,
and cross-face boundary prediction achieves AUCs of 0.76 and 0.86. Repetition
explains its principal recurrence, while the tested spiral embeddings provide
no evidence of additional cross-winding structure. Across several corpora,
comparative transfer persists beyond exact shared local pairs. Khipu knot forms
distinguish authentic attachments after controlling for degree, depth, fibre
and twist (adjusted \(p=0.009\)); constrained assignment recovers 8.75\% of
concealed parents against 7.87\% expected by chance. Row-alignment tests yield
no comparable evidence after correction. Calibrated BDM recovers Arecibo's
23-column width (search-corrected \(p=0.005\)), while executable controls
detect spatial dependence and eliminate it through interventions on the
generator. These findings connect information-based structural reconstruction
with interventional reasoning, distinguishing recoverable organisation from
semantic decipherment.
\end{abstract}


\section{Introduction}\label{introduction}

Information can carry evidence of the space in which it was organised. A
picture transmitted as a stream of bits may retain its row width; a
sequence of signs may retain the boundaries of its compartments; and a
set of knot forms may retain information about their positions along a
cord. Physical embodiment alone does not establish recoverability. It concerns whether the surviving
data make a particular organisation inferentially recoverable after that
organisation has been concealed.

In previous work, we proved sufficient conditions for this proposition
in an algorithmic framework: under specified complexity and perturbation
assumptions, the correct object-space representation becomes
asymptotically favoured \cite{zenil2024geometry}. Our earlier message-reconstruction
study demonstrated the approach with Arecibo \cite{zenil2024arecibo}. The 1974
transmission contains 1,679 bits, intended as 73 rows of 23 columns.
Searching alternative arrangements exposed the intended width through
reductions in estimated complexity. A comparison of six further images
placed the intended dimension among the three leading candidates in
every case, with BDM selecting it first in four. Perturbation and
resolution experiments also demonstrated estimator-dependent noise
sensitivity. These are complementary achievements: a conditional
mathematical result and finite computational demonstrations, rather than
a theorem asserting recovery of every finite message.

Arecibo is particularly useful because it is a genuine one-way
transmission: it was broadcast without an interactive exchange that
could teach a receiver its code. Its intended layout is nevertheless
known to us, making reconstruction objectively testable. Historical
inscriptions create an analogous one-way problem across time. The
archaeologist cannot ask a deceased maker which direction to read,
whether a divider marks a word or an account, or whether neighbouring
rows should be interpreted together. Surviving objects may provide
neither a bilingual key nor an accepted mapping from marks to sounds.
This analogy concerns the receiver's access to the transmitter and
encoding, not the assertion that the original inscriptions were
intentionally written for unknown future readers.

The analogy also has limits that make the historical extension
scientifically interesting. Arecibo has a known binary alphabet,
complete ground truth and an intentionally pictorial organisation.
Ancient records can be damaged, repeatedly copied, inconsistently
transcribed and shaped by material constraints. Some components may
already be understood, as with numerical information in many khipu,
while others remain unresolved. In a known-script control such as
Egyptian, meaning is available to specialists but withheld from the
reconstruction algorithm. Across these cases, the relevant question is
which structural relations can be recovered without using a proposed
translation.

The Phaistos Disc supplies the most extensive benchmark. It is a clay
object from Bronze Age Crete, bearing stamped pictorial signs along a
spiral on each face \cite{heraklionPhaistos}. Faces A and B are modern labels, not a
demonstrated reading order. A sign type is a category in the inventory;
a sign occurrence is one impression of that type. Catalogue numbers such
as 02 and 12 are names, not numerical magnitudes. Incised dividers
separate the impressions into compartments, also called groups. Initial,
medial and final describe a sign's position within a compartment. These
compartments need not be words, and the short oblique strokes attached
to some groups need not be punctuation. Our transcription contains 123
positions in 31 groups on A and 119 positions in 30 groups on B,
including one damaged position assigned a provisional identity. There
are 45 sign types and 18 stroked groups.

The comparative objects supply distinct spatial configurations. Khipu
are Andean records made from a main cord and attached cords, sometimes
with further branches; knot forms and their relative positions can
participate in numerical organisation \cite{urton2005khipu,openKhipu}. Rongorongo consists of
signs on wooden tablets from Rapa Nui. Its conventional traversal
alternates successive lines and sign orientation, a configuration termed
reverse boustrophedon; the digital transcription already incorporates an
editorial reading convention \cite{rongopy}. Indus inscriptions are generally
short sequences on objects such as seals and tablets; a subset of the
digital corpus records more than one row \cite{indusCorpus}. The Voynich manuscript
provides pages, lines and word-like units in an unresolved script
\cite{reddy2011voynich,evaVoynich}. Egyptian hieroglyphs provide a known system in which signs
are arranged both horizontally and vertically within compact layout
groups, conventionally called quadrats \cite{rosmorducMdcDoc,rosmorducMdcTexts}. A quadrat is a
spatial grouping, not necessarily a linguistic word.

These objects distinguish order, topology and geometry. Order specifies
which occurrence follows which. Topology concerns relations such as
belonging to the same connected path, branch or compartment. Geometry
adds quantities such as distance, angle, row width or radial position.
Recovering a boundary supplies structural information, but does not by
itself reconstruct a two-dimensional metric. Similarly, locating a
statistical change between two sequences does not establish their
physical connectedness. We therefore treat the full Phaistos analysis
and the comparative experiments as related tasks within one framework,
with each result tied to the structure actually concealed or perturbed.

\FloatBarrier
{
\small
\setlength{\tabcolsep}{3pt}
\renewcommand{\arraystretch}{1.12}
\begin{longtable}{@{}>{\raggedright\arraybackslash}p{\dimexpr 0.200000\linewidth-4.8000pt\relax}>{\raggedright\arraybackslash}p{\dimexpr 0.200000\linewidth-4.8000pt\relax}>{\raggedright\arraybackslash}p{\dimexpr 0.200000\linewidth-4.8000pt\relax}>{\raggedright\arraybackslash}p{\dimexpr 0.200000\linewidth-4.8000pt\relax}>{\raggedright\arraybackslash}p{\dimexpr 0.200000\linewidth-4.8000pt\relax}@{}}
\caption{Corpus units, spatial structures and evaluation targets.}\label{tab:main-1}\\
\toprule
Record & Configuration & Analysed units & Structural target & Withholding \\
\midrule
\endfirsthead
\toprule
Record & Configuration & Analysed units & Structural target & Withholding \\
\midrule
\endhead
\bottomrule
\endlastfoot
Phaistos & Two spiral paths & 242 positions; 61 groups & Dividers, face break, recurrence & Faces; group families \\
Khipu & Cords and branches & 532 eligible khipu; 15,558 cords & Farthest eligible knot cluster & Whole khipu \\
Rongorongo & Alternating lines & 13 tablets; 11,666 tokens & Traversal; line breaks & H/P/Q together \\
Indus & Short multiline texts & 75 unique multiline sequences & Recorded row breaks & Mohenjo-daro site \\
Voynich & Page and line layout & 207 pages; 34,963 word units & Recorded line breaks & Whole pages \\
Egyptian & Horizontal/vertical joins & 186 files; 40,986 pairs & Vertical join operator & Whole files \\
Arecibo & Binary rectangular image & 1,679 bits; widths 10--100 & Intended width of 23 & Known-answer control \\
\end{longtable}
}

Counts refer to the stated selection and transcription, not all
surviving objects. Arecibo is an artificial transmission; Egyptian is a
known-script control.

\begin{figure}[h]
\centering
\includegraphics[width=\linewidth,height=0.52\textheight,keepaspectratio]{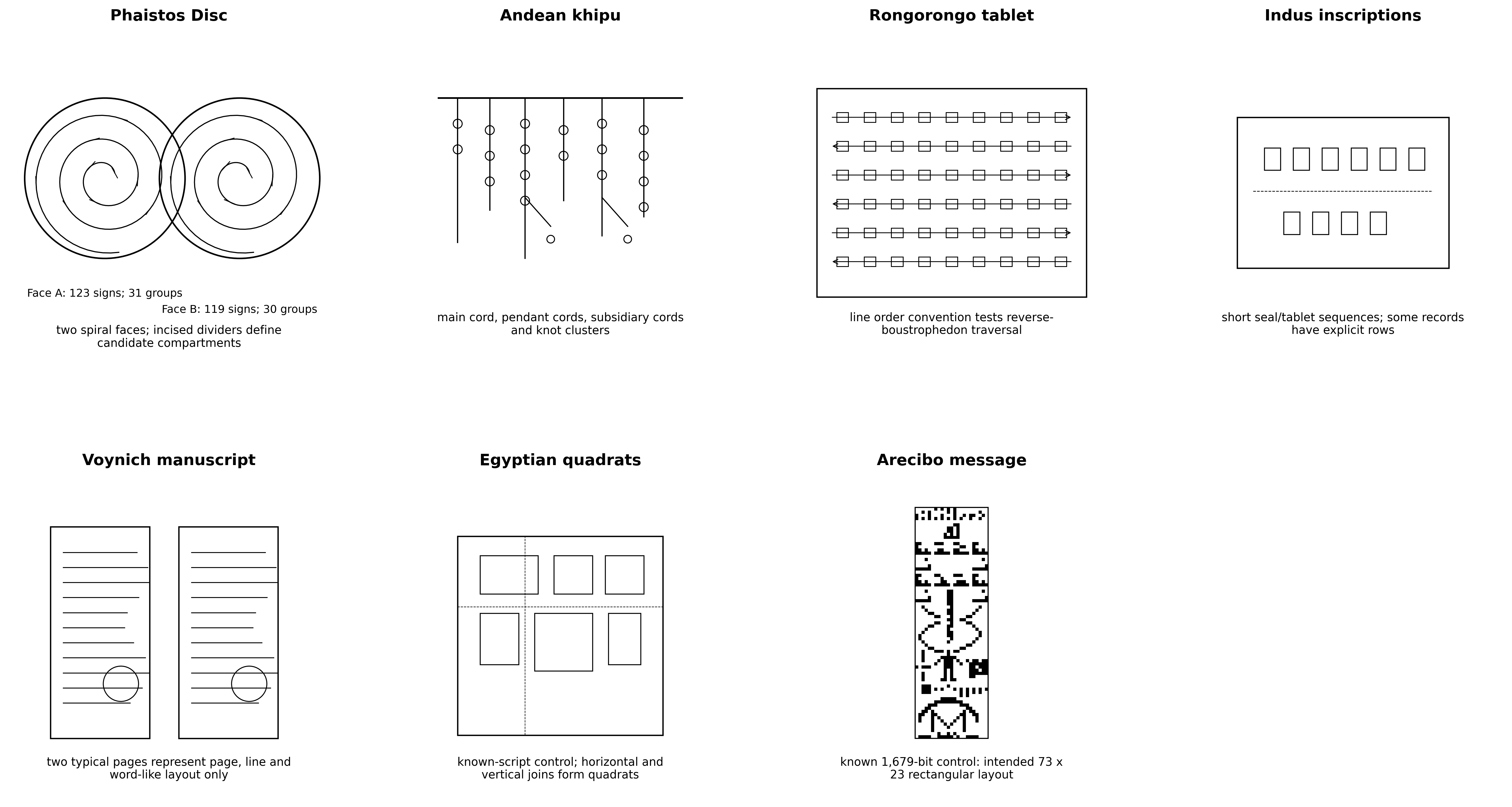}
\caption{Schematic object configurations used in the comparative reconstruction experiments. The panels identify the structural relations tested in the paper: the two spiral faces and divider compartments of Phaistos; main, pendant and subsidiary khipu cords; reverse-boustrophedon line traversal in Rongorongo; explicit row breaks in short Indus inscriptions; page and line layout in the Voynich manuscript; horizontal and vertical Egyptian quadrat joins; and the known 73 by 23 Arecibo layout. These diagrams represent the spatial relations analysed; they are not facsimile photographs.}
\label{fig:object-schematics}
\end{figure}

The contribution is an integrated reconstruction study. The Phaistos
benchmark combines conventional statistics and algorithmic perturbation
estimates; the larger corpora test transfer to new physical or editorial
units. Our primary extension of the earlier geometry work is
consequently empirical and operational: it asks which aspects of
historical records exhibit recoverability under explicit null models.
Prior statistical, syllabotactic and cryptanalytic studies remain
relevant \cite{jackson1999phaistos,davis2026phaistos,giorgi2026phaistos}. Unequal sign frequencies and repeated passages are established
observations; the present contribution concerns their role in structural
reconstruction under explicit controls.

\section{Mathematical framework}\label{mathematical-framework}

\subsection{Objects, representations and
recoverability}\label{objects-representations-and-recoverability}

Let \(X=(x_1,\ldots,x_n)\) denote a categorical sequence, with \(x_t\) a
sign identity rather than a number. A candidate structure can be
represented by

\begin{equation}
\label{eq:1}
M=(F,B,D,T,G).
\end{equation}

Here \(F\) allocates occurrences to faces or records; \(B\) marks
boundaries; \(D\) is a traversal convention; \(T\) specifies
connectivity or membership relations; and \(G\) describes a geometric
embedding. For a khipu, \(T\) could represent attachments and \(G\)
distances along cords. For a tablet, \(B\) could specify line breaks
while \(G\) supplies coordinates. The present experiments vary selected
components, not the full Cartesian product of all possible models.

The ideal two-part objective is

\begin{equation}
\label{eq:2}
L(M,X)=K(M)+K(X\mid M),\qquad M^*\in\operatorname*{arg\,min}_M L(M,X).
\end{equation}

\(K\) is prefix Kolmogorov complexity: the length of a shortest
self-delimiting programme that produces an object on a fixed universal
machine. The first term charges for specifying the model; the second
charges for the data conditional on it. Arbitrarily adding structural
detail is therefore not free. Exact \(K\) is uncomputable in general.
Our experiments instead use specified computable statistics or coding
costs \(C\), applied to representations \(R_M(X)\), and compare them
under matched perturbations:

\begin{equation}
\label{eq:3}
\Delta_C(M\to M';X)=C(R_{M'}(X))-C(R_M(X)).
\end{equation}

A positive value means that the altered representation has a greater
measured cost. This is the operational AID experiment. The
archaeological tests do not implement an exact universal minimisation of
equation \eqref{eq:2}, nor do classifier scores constitute direct
estimates of \(K\).

\subsection{Prior theorem and the finite-data
bridge}\label{prior-theorem-and-the-finite-data-bridge}

For completeness, the sufficient conditions in the earlier geometry
theorem can be written, using its notation for an object \(y\), space
\(\mathcal S\) and algorithmic perturbation \(\mathcal P\), as

\begin{equation}
\label{eq:4}
K(\mathcal S)=o(K(y\mid\mathcal S)),\quad K(\mathcal P)=O(\log|y|),\quad K(y\mid\mathcal S)=o(\log|y|).
\end{equation}

Under the theorem's uniform measure over admissible perturbation
outcomes, every fixed computable probability measure \(P\) obeys the
stated asymptotic comparison

\begin{equation}
\label{eq:5}
P[y',\mathcal S']=o\!\left(P_U[y\mid\mathcal S]\right)
\end{equation}

with probability approaching one as \(|y|\) grows \cite{zenil2024geometry}. \(P_U\)
denotes universal algorithmic probability; \(O\) denotes bounded growth
up to a constant factor, and \(o\) means the ratio tends to zero. These
conditions are not established for any finite inscription analysed here.
The empirical bridge is therefore measured recoverability, synthetic
controls and perturbation sensitivity.

Two elementary observations sharpen this bridge. First, for any fixed
computable bijection \(f\) with a computable inverse,

\begin{equation}
\label{eq:6}
K(f(X))\leq K(X)+K(f)+O(1),\qquad K(X)\leq K(f(X))+K(f^{-1})+O(1).
\end{equation}

The proof composes a programme for \(X\) with a programme for \(f\), and
then repeats the argument with the inverse. In particular, reversal
changes ideal complexity by at most a constant independent of sequence
length. Finite block estimators can nevertheless change because a
transformation changes their accessible local regularities. This is why
label invariance, block alignment and encoding overhead must be checked
explicitly, and why a single complexity difference cannot establish an
ancient reading direction.

Second, suppose a finite estimator satisfies
\(|\widehat C(M)-C(M)|\leq\epsilon\) on a specified set of models. Then

\begin{equation}
\label{eq:7}
C(M')-C(M)>2\epsilon\quad\Longrightarrow\quad \widehat C(M')>\widehat C(M).
\end{equation}

This follows by subtracting the two error bounds. In practice,
\(\epsilon\) is not known for an arbitrary archaeological object. The
statement explains what robust reconstruction requires: separation
larger than estimator uncertainty, rather than merely the existence of a
numerical minimum. Null distributions assess one part of this
uncertainty; alternative encodings and positive controls assess others.

\subsection{AID as a bridge between information and
intervention}\label{aid-as-a-bridge-between-information-and-intervention}

Algorithmic information dynamics (AID) supplies the experimental
organisation of this study \cite{zenil2019causal,aidScholarpedia}. It combines an information
measure with an explicitly defined perturbation and asks which
dependencies the intervention preserves, removes or creates. Its
operational procedure is agnostic to the particular underlying measure:
classical entropy, Bayesian predictive codes, dictionary compression and
algorithmic-probability-based estimates can each quantify the response.
These quantities retain their individual mathematical meanings. A
Shannon calculation inside an AID experiment remains a Shannon
calculation; using the same intervention makes its response comparable
with responses obtained from other information measures.

We describe an AID experiment by the tuple

\begin{equation}
\label{eq:34}
\mathcal A=(\mathcal R,\mathcal J,C,\mathcal N,\mathcal V).
\end{equation}

Here \(\mathcal R\) is a representation family, \(\mathcal J\) a family
of computable interventions, \(C\) an information statistic or coding
cost, \(\mathcal N\) a constrained randomisation model, and
\(\mathcal V\) an independent validation procedure. The representation
identifies the relation being investigated: a divider, line orientation,
attachment or spatial correspondence. The intervention changes that
relation. The randomisation model specifies what must remain fixed, such
as sign counts, the number of boundaries, parent degrees or object
identity. Validation asks whether the inferred constraint predicts
material excluded from its construction, or recovers a known relation in
a controlled generator. AID therefore links information-theoretic
descriptions to interventional questions through an explicit
experimental design.

For an intervention \(J\) on a representation \(R\), the elementary
information response and its null-centred value are

\begin{equation}
\label{eq:35}
\delta_C(J;R)=C(JR)-C(R),\qquad \widetilde\delta_C(J;R)=\delta_C(J;R)-\mathbb E_{N\sim\mathcal N}\delta_C(J;N).
\end{equation}

The second expression is appropriate when a perturbation itself changes
a nuisance quantity such as length or degree. In the historical
experiments, we usually preserve these quantities directly and compare
the authentic cost with the costs of constrained alternatives. A
collection of responses over increasing intervention sizes defines a
perturbation landscape. Agreement across different scoring functions
establishes that a structural feature is supported by several
descriptions; differences identify the scales or regularities to which
those descriptions respond.

The connection to statistical causal inference is precise at the level
of mechanisms. A structural causal model writes each variable as a
function of its parents and exogenous inputs,

\begin{equation}
\label{eq:36}
V_j=f_j(\operatorname{Pa}_j,U_j),\qquad \operatorname{do}(V_j=v):\ f_j\leftarrow v.
\end{equation}

An intervention replaces one of these mechanisms while retaining the
others \cite{pearl2009}. A more general mechanism intervention replaces \(f_j\)
by another specified function \(g_j\). AID asks how such modifications
alter the information needed to describe the resulting system. When the
generator is known, we can implement the mechanism replacement directly,
as in the synthetic row and tree controls below. When only an ancient
output survives, moving a boundary or rewiring an encoded attachment is
an intervention on the candidate representation. It establishes the
informational consequence of that proposed change. Identifying the
historical mechanism additionally requires assumptions connecting the
representation and intervention to the original production process. The
distinction specifies the causal question that each experiment can
answer.

\subsection{Algorithmic probability and interventions in software
space}\label{algorithmic-probability-and-interventions-in-software-space}

Algorithmic probability extends the hypothesis space from a chosen
family of probability distributions to computable generative
descriptions. Let \(U\) be a fixed universal prefix machine: a computing
system whose valid programmes are self-delimiting, so no complete
programme is the beginning of another valid programme. For an output
\(x\), its universal algorithmic probability is

\begin{equation}
\label{eq:37}
m_U(x)=\sum_{p:\ U(p)=x}2^{-|p|}.
\end{equation}

The length \(|p|\) is measured in bits. Shorter programmes receive
greater weight, and the prefix condition bounds the total weight by one.
This is the software-space interpretation of the inference problem:
different programmes can implement different mechanisms that produce the
same observed record. The coding theorem connects this distribution to
shortest-programme complexity. CTM makes the programme-output connection
concrete on finite collections of small machines; BDM carries those
local estimates into larger representations through block decomposition
and multiplicity. Their contribution to AID is to supply information
estimates informed by computable generative descriptions, complementing
statistics of observed frequencies and dictionary reuse.

A computable hypothesis intervention is an operator \(\mathcal J\) that
transforms one programme into another. The operator may replace a
copying rule, reverse a traversal, change a boundary rule or reassign a
parent relation. For a computable deterministic transformation,

\begin{equation}
\label{eq:38}
K(\mathcal J(p))\leq K(p)+K(\mathcal J)+O(1),\qquad x^{\mathcal J}=U(\mathcal J(p)).
\end{equation}

The inequality follows by describing \(p\) and the transformation that
is applied to it. It explains why the intervention itself belongs in the
accounting: a complicated transformation should not be treated as free
evidence for a simple mechanism. If randomness is required, its seed or
probability law must also be specified. For an explicitly enumerated
finite set of executable hypotheses \(\mathcal P\) and a specified
observation model, the corresponding Bayesian weights would be

\begin{equation}
\label{eq:39}
w(p\mid D)=\frac{2^{-|p|}P(D\mid U(p))}{\sum_{r\in\mathcal P}2^{-|r|}P(D\mid U(r))},\quad p\in\mathcal P.
\end{equation}

The observation model allows noise or missing signs; an exact-output
model is a special case. One could then average interventional
predictions across these weighted hypotheses. Equation \eqref{eq:39}
defines the connection between algorithmic hypothesis selection and
mechanistic intervention; the historical analyses below do not evaluate
this programme posterior. They implement explicit computable
transformations of the surviving representations, with CTM-BDM,
classical information and predictive codes used where appropriate. The
synthetic controls go one step further by intervening on a known
executable generator. This hierarchy connects statistical causal models,
information perturbations and algorithmic software-space hypotheses
without equating a digital rearrangement with an experiment on an
ancient maker.

\subsection{Phaistos-guided selection of comparative
experiments}\label{phaistos-guided-selection-of-comparative-experiments}

Phaistos identifies three productive experimental questions: whether
information survives transfer to a withheld record; whether shared
material explains that transfer; and whether changing a particular
structural relation alters the measured information. These questions
determine the comparative extensions. We apply a common structural-role
perturbation experiment to all five historical comparison corpora; add a
stricter feature-pair exclusion to the existing held-object tests;
compare three CTM block scales with dictionary and Bayesian coding in
the Rongorongo orientation experiment; and test spatial correspondences
specifically for row layouts and khipu attachments. Positive and
independent-generator controls accompany the new spatial tests.

The selection concerns a scientific target and available data. Spectral
analysis remains a targeted explanation of the Phaistos recurrence,
where a specific repeated passage supplied a testable alternative. The
comparative analyses preserve boundaries between unrelated objects
rather than imposing a global period. Likewise, recorded line order supports
traversal tests, while actual cord-parent identifiers support attachment
rewiring. The resulting design highlights AID through the interventions
it defines and the constraints those interventions test. It is an
exploratory extension selected after the Phaistos findings, rather than
a preregistered comparison of universally interchangeable estimators.

\section{Results}\label{results}

Numerical table entries are rounded to two decimal places. Small nonzero
values, including probabilities below 0.1, use scientific notation with
two decimal places in the mantissa; counts and identifiers remain exact.
Rounding is for presentation only and does not determine statistical
significance. Additional precision is retained in the text where needed
to distinguish small effects.

\subsection{Arecibo establishes an explicit geometric
control}\label{arecibo-establishes-an-explicit-geometric-control}

We first connect the historical analysis to the one-way-message
benchmark through a new PyBDM calculation on the 1,679-bit Arecibo
transcription \cite{areciboTranscription}. Candidate row widths range from 10 to 100. The
raw BDM minimum occurs at width 11, whereas shape-normalised BDM selects
the intended width 23. After independently calibrating every width with
99 bit-count-preserving shuffles, the largest low-complexity standard
score is 10.174 at width 23. Comparison with 199 additional shuffled
full-width scans gives a search-corrected \(p=0.005\).

This result makes an otherwise hidden methodological choice visible.
Unequal array shapes interact with finite block decomposition, padding
and residual-edge handling. A raw score across unequal shapes need not
select the intended geometry. Normalisation was examined after an
exploratory scan, so this is an estimator-sensitivity analysis, not a
preregistered replication of every setting in the earlier Arecibo paper.
The successful calibrated reconstruction nevertheless supplies a
concrete positive control with a known answer.

With 99 noisy replicates per nonzero level, the correct width remains
among the best three normalised-BDM candidates in 92.9\% of trials with
3\% of bits inverted and 84.8\% with 5\% inverted. At 10\% inversion
this falls to 44.4\%. Rank-one recovery is less stable. These
experiments invert selected bits exactly once; they should not be
equated numerically with experiments that replace selected bits by
random binary values, which change only about half of the selected
positions on average.

\FloatBarrier
{
\small
\setlength{\tabcolsep}{3pt}
\renewcommand{\arraystretch}{1.12}
\begin{longtable}{@{}>{\raggedright\arraybackslash}p{\dimexpr 0.200000\linewidth-4.8000pt\relax}>{\raggedright\arraybackslash}p{\dimexpr 0.200000\linewidth-4.8000pt\relax}>{\raggedright\arraybackslash}p{\dimexpr 0.200000\linewidth-4.8000pt\relax}>{\raggedright\arraybackslash}p{\dimexpr 0.200000\linewidth-4.8000pt\relax}>{\raggedright\arraybackslash}p{\dimexpr 0.200000\linewidth-4.8000pt\relax}@{}}
\caption{Arecibo width recovery under exact bit inversion.}\label{tab:main-2}\\
\toprule
Bits inverted & Replicates & Width 23 first & Width 23 in top 3 & Median rank \\
\midrule
\endfirsthead
\toprule
Bits inverted & Replicates & Width 23 first & Width 23 in top 3 & Median rank \\
\midrule
\endhead
\bottomrule
\endlastfoot
0\% & 1 & 1.00 & 1.00 & 1 \\
1\% & 99 & 0.47 & 1.00 & 2 \\
3\% & 99 & 0.54 & 0.93 & 1 \\
5\% & 99 & 0.42 & 0.85 & 2 \\
10\% & 99 & 0.18 & 0.44 & 5 \\
15\% & 99 & 0.09 & 0.19 & 15 \\
\end{longtable}
}

Noise probabilities are empirical fractions from the normalised-BDM
width scan. The zero-noise entry has one deterministic observation.

\begin{figure}[h]
\centering
\includegraphics[width=0.9\linewidth,height=0.58\textheight,keepaspectratio]{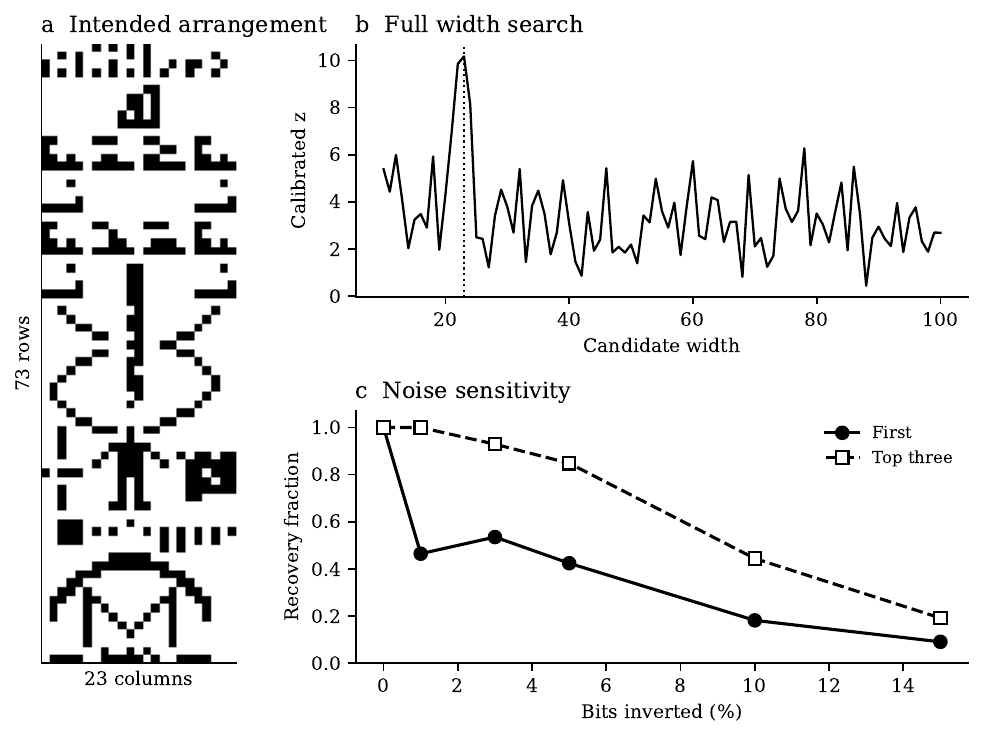}
\caption{Arecibo connects geometric reconstruction to a known one-way transmission. (a) The actual 1,679 bits at the intended 23-column width. (b) Width-specific calibration selects 23; the maximum-scan p value is 0.005 while peaks represent shadow configurations at multiples of 23. (c) Normalised-BDM recovery under exact bit inversion. The finite estimator and preprocessing differ from the earlier published experiments.}
\label{fig:1}
\end{figure}

\subsection{Phaistos exhibits dependencies beyond sign frequencies and
positional
roles}\label{phaistos-exhibits-dependencies-beyond-sign-frequencies-and-positional-roles}

The full Phaistos transcription has empirical single-sign entropy
\(H_0=4.995\) bits, below \(\log_2 45=5.492\). This describes the
sign-frequency distribution, not the entropy rate of its unknown
generator. Under randomisations preserving face allocation and the sign
inventory within initial, medial and final roles, LZW and Bayesian
first-order coding are substantially shorter for the inscription than
for surrogates (Table~\ref{tab:main-3}). CTM-BDM also detects pooled
structure; the block-size and repetition analyses identify the scale and
material contributing to its response.

\FloatBarrier
{
\small
\setlength{\tabcolsep}{3pt}
\renewcommand{\arraystretch}{1.12}
\begin{longtable}{@{}>{\raggedright\arraybackslash}p{\dimexpr 0.198473\linewidth-5.9542pt\relax}>{\raggedright\arraybackslash}p{\dimexpr 0.190840\linewidth-5.7252pt\relax}>{\raggedright\arraybackslash}p{\dimexpr 0.152672\linewidth-4.5802pt\relax}>{\raggedright\arraybackslash}p{\dimexpr 0.152672\linewidth-4.5802pt\relax}>{\raggedright\arraybackslash}p{\dimexpr 0.152672\linewidth-4.5802pt\relax}>{\raggedright\arraybackslash}p{\dimexpr 0.152672\linewidth-4.5802pt\relax}@{}}
\caption{Phaistos order under the face-and-role-preserving null, followed by duplicate-group ablation.}\label{tab:main-3}\\
\toprule
Sequence & Estimator & Observed & Null mean & z & p \\
\midrule
\endfirsthead
\toprule
Sequence & Estimator & Observed & Null mean & z & p \\
\midrule
\endhead
\bottomrule
\endlastfoot
Complete & LZW & 5.72 & 6.71 & 8.88 & \(2.00\times10^{-4}\) \\
Complete & LZ78 phrase rate & 3.81 & 4.00 & 3.58 & \(8.00\times10^{-4}\) \\
Complete & KT1 & 5.03 & 5.48 & 13.69 & \(2.00\times10^{-4}\) \\
Complete & KT2 & 5.29 & 5.49 & 24.30 & \(2.00\times10^{-4}\) \\
Complete & CTM-BDM & 6.19 & 6.39 & 2.53 & \(5.00\times10^{-3}\) \\
Duplicates removed & LZW & 6.04 & 6.77 & 6.85 & \(2.00\times10^{-4}\) \\
Duplicates removed & LZ78 phrase rate & 3.83 & 4.01 & 3.07 & \(2.80\times10^{-3}\) \\
Duplicates removed & KT1 & 5.20 & 5.49 & 8.86 & \(2.00\times10^{-4}\) \\
Duplicates removed & KT2 & 5.45 & 5.49 & 5.23 & \(6.00\times10^{-4}\) \\
Duplicates removed & CTM-BDM & 6.88 & 6.86 & -0.23 & 0.59 \\
\end{longtable}
}

Complete: n = 242, with face and positional-role preservation.
Duplicates removed: n = 208, 53 groups, pooled positional-role
preservation. Coding costs are per sign; LZ78 and CTM-BDM are proxies.
Null draws: 5,000 for the statistical/dictionary measures and 1,000 for
CTM-BDM.

BDM values are complexity proxies normalised by token count, not
certified lossless code lengths. Accordingly, their absolute scale is
not interpreted as an entropy rate or compared numerically with a
Shannon limit. 

The adjacent-sign analysis gives within-compartment
adjacent-sign mutual information of 3.236 bits on the pooled text. The
matched null mean is already 2.522 bits because a short sequence and a
large alphabet produce substantial finite-sample dependence estimates.
The relevant excess is therefore 0.714 bits, not the uncorrected 3.236.
All 4,999 face-and-role-preserving surrogates give smaller values.
Canonically relabelled Re-Pair grammar induction saves 29 symbols
against a null mean of 2.859, again with \(p=0.0002\). Re-Pair's units
are grammar symbols; this result is not a 29-bit code saving.

\FloatBarrier
{
\small
\setlength{\tabcolsep}{3pt}
\renewcommand{\arraystretch}{1.12}
\begin{longtable}{@{}>{\raggedright\arraybackslash}p{\dimexpr 0.166667\linewidth-5.0000pt\relax}>{\raggedright\arraybackslash}p{\dimexpr 0.166667\linewidth-5.0000pt\relax}>{\raggedright\arraybackslash}p{\dimexpr 0.166667\linewidth-5.0000pt\relax}>{\raggedright\arraybackslash}p{\dimexpr 0.166667\linewidth-5.0000pt\relax}>{\raggedright\arraybackslash}p{\dimexpr 0.166667\linewidth-5.0000pt\relax}>{\raggedright\arraybackslash}p{\dimexpr 0.166667\linewidth-5.0000pt\relax}@{}}
\caption{Phaistos local dependence and grammar compression.}\label{tab:main-4}\\
\toprule
Face(s) & Statistic & Observed & Null mean & Excess & p \\
\midrule
\endfirsthead
\toprule
Face(s) & Statistic & Observed & Null mean & Excess & p \\
\midrule
\endhead
\bottomrule
\endlastfoot
A & Adjacent MI (bits) & 3.48 & 2.61 & 0.87 & \(2.00\times10^{-4}\) \\
A & Re-Pair saving (symbols) & 16.00 & 1.63 & 14.37 & \(2.00\times10^{-4}\) \\
B & Adjacent MI (bits) & 3.22 & 2.91 & 0.31 & \(2.00\times10^{-4}\) \\
B & Re-Pair saving (symbols) & 5.00 & 0.33 & 4.67 & \(2.00\times10^{-4}\) \\
AB & Adjacent MI (bits) & 3.24 & 2.52 & 0.71 & \(2.00\times10^{-4}\) \\
AB & Re-Pair saving (symbols) & 29.00 & 2.86 & 26.14 & \(2.00\times10^{-4}\) \\
\end{longtable}
}

The 4,999-draw nulls preserve face membership and initial/medial/final
inventories. Re-Pair uses canonical labels and cannot cross group
boundaries. All six Bonferroni-adjusted probabilities are 0.0012.

Prediction provides a complementary test. With one compartment withheld
at a time, pooled first-order coding improves over zero-order coding by
0.906 bits per internal transition. Exact duplicated compartments can
still appear in training under this procedure. Removing every copy of
the withheld compartment from training reduces the gain to 0.488 bits
per transition, while retaining a positive first-order gain on each
face. The pooled second-order increment changes from 0.133 to -0.0013
bits per transition. Thus, part of the apparent higher-order
predictability is attributable to copied groups, while first-order
constraints extend beyond complete-group duplication. These are
fixed-smoothing predictive contrasts.

\FloatBarrier
{
\small
\setlength{\tabcolsep}{3pt}
\renewcommand{\arraystretch}{1.12}
\begin{longtable}{@{}>{\raggedright\arraybackslash}p{\dimexpr 0.152047\linewidth-6.3860pt\relax}>{\raggedright\arraybackslash}p{\dimexpr 0.146199\linewidth-6.1404pt\relax}>{\raggedright\arraybackslash}p{\dimexpr 0.116959\linewidth-4.9123pt\relax}>{\raggedright\arraybackslash}p{\dimexpr 0.116959\linewidth-4.9123pt\relax}>{\raggedright\arraybackslash}p{\dimexpr 0.116959\linewidth-4.9123pt\relax}>{\raggedright\arraybackslash}p{\dimexpr 0.116959\linewidth-4.9123pt\relax}>{\raggedright\arraybackslash}p{\dimexpr 0.116959\linewidth-4.9123pt\relax}>{\raggedright\arraybackslash}p{\dimexpr 0.116959\linewidth-4.9123pt\relax}@{}}
\caption{Predictive code lengths with complete-group duplication controlled.}\label{tab:main-5}\\
\toprule
Face(s) & Withheld unit & Pairs & Order 0 & Order 1 & Order 2 & Gain 1 & Gain 2 \\
\midrule
\endfirsthead
\toprule
Face(s) & Withheld unit & Pairs & Order 0 & Order 1 & Order 2 & Gain 1 & Gain 2 \\
\midrule
\endhead
\bottomrule
\endlastfoot
A & One group & 92 & 5.08 & 3.96 & 3.78 & 1.12 & 0.18 \\
A & Duplicate family & 92 & 5.39 & 4.94 & 5.00 & 0.46 & -0.06 \\
B & One group & 89 & 5.18 & 4.84 & 4.76 & 0.34 & 0.08 \\
B & Duplicate family & 89 & 5.23 & 4.99 & 4.92 & 0.24 & 0.07 \\
AB & One group & 181 & 5.22 & 4.32 & 4.18 & 0.91 & 0.13 \\
AB & Duplicate family & 181 & 5.36 & 4.88 & 4.88 & 0.49 & \(-1.30\times10^{-3}\) \\
\end{longtable}
}

Costs and gains are bits per internal transition. Gain 1 = order 0 minus
order 1; gain 2 = order 1 minus order 2. Fixed smoothing strength is 10.
No null p value is inferred from these contrasts.

\begin{figure}[tbp]
\centering
\includegraphics[width=0.97\linewidth,height=0.58\textheight,keepaspectratio]{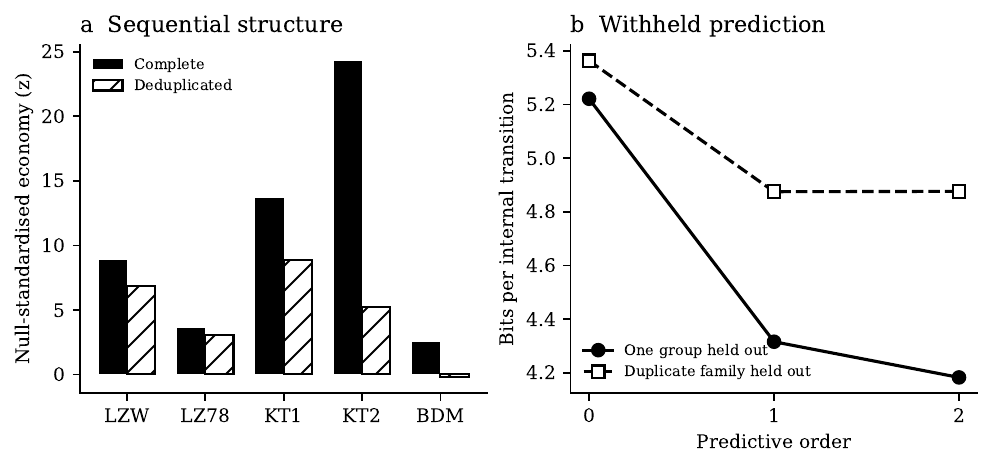}
\caption{Phaistos order survives some, but not all, repetition controls. (a) Null-standardised economy for the complete and deduplicated text. The full-text null preserves face and role, whereas the ablation null preserves pooled role. (b) Pooled predictive costs when one group or its entire duplicate family is withheld. The stricter split retains first-order gain while removing the second-order increment.}
\label{fig:2}
\end{figure}

\subsection{The Phaistos divider arrangement occupies a broad
perturbation
basin}\label{the-phaistos-divider-arrangement-occupies-a-broad-perturbation-basin}

Pairing each sign with a boundary-after indicator makes segmentation an
explicit part of the encoded state. Randomly permuting the multiset of
authentic group lengths retains the lengths and sign stream while moving
the dividers. On both faces, the authentic arrangement has lower LZW and
KT1 costs than all 3,000 such surrogates; CTM-BDM gives nominal
lower-tail probabilities of 0.0290 on A and 0.0350 on B (Table~\ref{tab:main-6}). The
shared perturbation gives concordant directions across the three
information families, with different effect sizes and finite-block
sensitivities reported explicitly.

\FloatBarrier
{
\small
\setlength{\tabcolsep}{3pt}
\renewcommand{\arraystretch}{1.12}
\begin{longtable}{@{}>{\raggedright\arraybackslash}p{\dimexpr 0.166667\linewidth-5.0000pt\relax}>{\raggedright\arraybackslash}p{\dimexpr 0.166667\linewidth-5.0000pt\relax}>{\raggedright\arraybackslash}p{\dimexpr 0.166667\linewidth-5.0000pt\relax}>{\raggedright\arraybackslash}p{\dimexpr 0.166667\linewidth-5.0000pt\relax}>{\raggedright\arraybackslash}p{\dimexpr 0.166667\linewidth-5.0000pt\relax}>{\raggedright\arraybackslash}p{\dimexpr 0.166667\linewidth-5.0000pt\relax}@{}}
\caption{Authentic Phaistos segmentation against permutations of the observed length multiset.}\label{tab:main-6}\\
\toprule
Face & Estimator & Observed & Null mean & z & p \\
\midrule
\endfirsthead
\toprule
Face & Estimator & Observed & Null mean & z & p \\
\midrule
\endhead
\bottomrule
\endlastfoot
A & LZW & 5.99 & 6.54 & 4.68 & \(3.33\times10^{-4}\) \\
A & LZ78 & 3.92 & 4.11 & 1.97 & \(4.80\times10^{-2}\) \\
A & KT0 & 6.03 & 6.17 & 3.85 & \(1.33\times10^{-3}\) \\
A & KT1 & 5.87 & 6.17 & 6.76 & \(3.33\times10^{-4}\) \\
A & KT2 & 6.20 & 6.42 & 5.78 & \(3.33\times10^{-4}\) \\
A & CTM-BDM & 14.33 & 14.48 & 1.97 & \(2.90\times10^{-2}\) \\
B & LZW & 6.66 & 7.22 & 4.80 & \(3.33\times10^{-4}\) \\
B & LZ78 & 4.19 & 4.22 & 0.44 & 0.47 \\
B & KT0 & 6.18 & 6.26 & 2.01 & \(3.20\times10^{-2}\) \\
B & KT1 & 6.26 & 6.42 & 5.11 & \(3.33\times10^{-4}\) \\
B & KT2 & 6.41 & 6.47 & 3.68 & \(4.00\times10^{-3}\) \\
B & CTM-BDM & 14.73 & 14.87 & 1.83 & \(3.50\times10^{-2}\) \\
\end{longtable}
}

Scores are normalised per sign in the 90-state sign-boundary
representation. There are 3,000 nulls for the statistical/dictionary
measures and 1,000 for CTM-BDM. Listed p values are within-test values,
before cross-estimator multiplicity adjustment.

Moving an increasing number of internal dividers produces a rising mean
cost in all three primary estimator families. On A, KT1 increases by
0.0257, 0.1151, 0.1960, 0.2859 and 0.3128 bits per sign after 1, 5, 10,
20 and 30 replacements. On B, the corresponding increases at 1, 5, 10,
20 and 25 replacements are 0.0153, 0.0681, 0.1124, 0.1542 and 0.1588.
About 14\% of one-divider KT1 alternatives are simpler than the
authentic arrangement; at five replacements this fraction is below 0.3\%
on both faces. The result is a broad basin in the sampled perturbation
landscape, not a proof that every authentic boundary is individually
optimal.

\begin{figure}[tbp]
\centering
\includegraphics[width=0.97\linewidth,height=0.58\textheight,keepaspectratio]{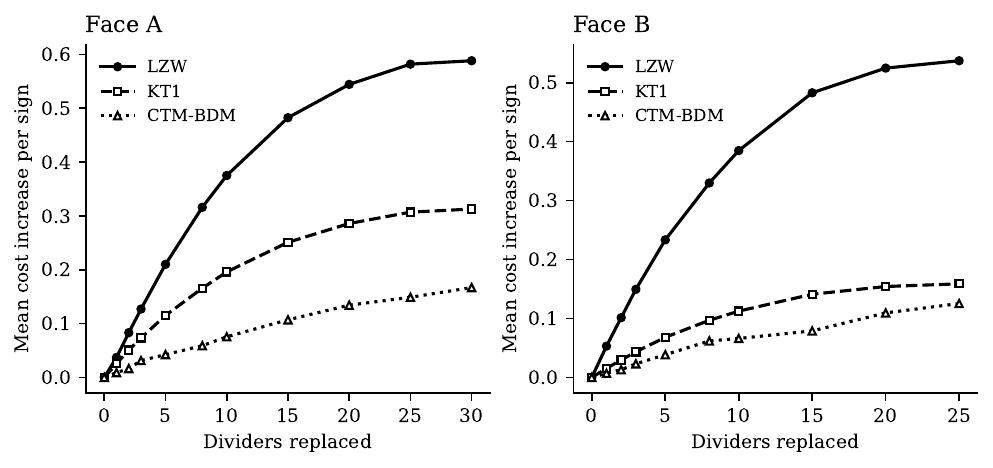}
\caption{Phaistos divider perturbation landscapes. Replacing more authentic dividers increases mean measured cost on both faces. Curves show means, not uncertainty bands: 3,000 draws per distance for LZW/KT1 and 300 for CTM-BDM. CTM values are per-sign complexity proxies. Some one-divider alternatives remain simpler than the authentic arrangement.}
\label{fig:3}
\end{figure}

A fully concealed local test gives different outcomes on the two faces.
Ranking individual boundary insertions by a consensus of CTM-BDM, LZW
and KT1 recovers 15 of 30 A dividers among the top 30 gaps, compared
with an expected random overlap of \(30^2/122=7.38\). B yields 7 of 29
against \(29^2/118=7.13\), and is not significant. A globally economical
segmentation need not be recoverable by an independent one-gap-at-a-time
score.

\subsection{Cross-face prediction and positional geometry identify
distributed boundary
information}\label{cross-face-prediction-and-positional-geometry-identify-distributed-boundary-information}

Learning a boundary score on A and evaluating it on B gives AUC 0.763,
average precision 0.584 and 16 of 29 dividers recovered. Reversing
training and evaluation gives AUC 0.864, average precision 0.772 and 20
of 30 recovered. The corresponding length-permutation tests
independently randomise the training or target dividers. Their Monte
Carlo floor is approximately 0.0002, so these values are reported as
equal to the floor, not smaller than it.

\FloatBarrier
{
\small
\setlength{\tabcolsep}{3pt}
\renewcommand{\arraystretch}{1.12}
\begin{longtable}{@{}>{\raggedright\arraybackslash}p{\dimexpr 0.178082\linewidth-6.4110pt\relax}>{\raggedright\arraybackslash}p{\dimexpr 0.136986\linewidth-4.9315pt\relax}>{\raggedright\arraybackslash}p{\dimexpr 0.136986\linewidth-4.9315pt\relax}>{\raggedright\arraybackslash}p{\dimexpr 0.136986\linewidth-4.9315pt\relax}>{\raggedright\arraybackslash}p{\dimexpr 0.136986\linewidth-4.9315pt\relax}>{\raggedright\arraybackslash}p{\dimexpr 0.136986\linewidth-4.9315pt\relax}>{\raggedright\arraybackslash}p{\dimexpr 0.136986\linewidth-4.9315pt\relax}@{}}
\caption{Phaistos boundary reconstruction, including exclusion of frequent boundary-associated signs.}\label{tab:main-7}\\
\toprule
Test & AUC & AP & Positive / gaps & Top K & Train-null p & Target-null p \\
\midrule
\endfirsthead
\toprule
Test & AUC & AP & Positive / gaps & Top K & Train-null p & Target-null p \\
\midrule
\endhead
\bottomrule
\endlastfoot
A: blind consensus & 0.72 & 0.43 & 30/122 & 15 & NA & \(2.00\times10^{-4}\) \\
B: blind consensus & 0.56 & 0.28 & 29/118 & 7 & NA & 0.16 \\
A to B & 0.76 & 0.58 & 29/118 & 16.00 & \(2.00\times10^{-4}\) & \(2.00\times10^{-4}\) \\
B to A & 0.86 & 0.77 & 30/122 & 20.00 & \(2.00\times10^{-4}\) & \(2.00\times10^{-4}\) \\
A to B: exclude 02 & 0.70 & 0.40 & 25/109 & 12.60 & \(8.40\times10^{-3}\) & \(8.00\times10^{-4}\) \\
B to A: exclude 02 & 0.68 & 0.32 & 17/95 & 5.00 & \(1.40\times10^{-3}\) & \(3.20\times10^{-3}\) \\
A to B: exclude 02,12 & 0.70 & 0.37 & 24/106 & 9.64 & \(1.54\times10^{-2}\) & \(1.40\times10^{-3}\) \\
B to A: exclude 02,12 & 0.72 & 0.33 & 15/78 & 4.64 & \(1.20\times10^{-3}\) & \(1.60\times10^{-3}\) \\
\end{longtable}
}

K equals the eligible positive count. Fractional overlap averages ties
at the cutoff. Blind-consensus p values permute labels and are shown in
the final column. Transfer p values independently randomise training or
target length multisets. AUC significance in the 02/12 exclusion tests does not imply
significant top-\(K\) recovery under every null.

Excluding gaps adjacent to sign 02, or to either 02 or 12, retains
predictive discrimination. The information is therefore distributed
across other local contexts. These exclusions change the eligible
positives and negatives; top-\(K\) counts must be read using the reduced
denominators in the table. Cross-face prediction also allows exact
shared passages in training and testing, so it establishes transfer
between faces rather than independence from every common formula. The
duplicate-family predictive experiment above answers a distinct,
stricter question about internal transitions.

\begin{figure}[htbp]
\centering
\includegraphics[
width=\linewidth,
keepaspectratio
]{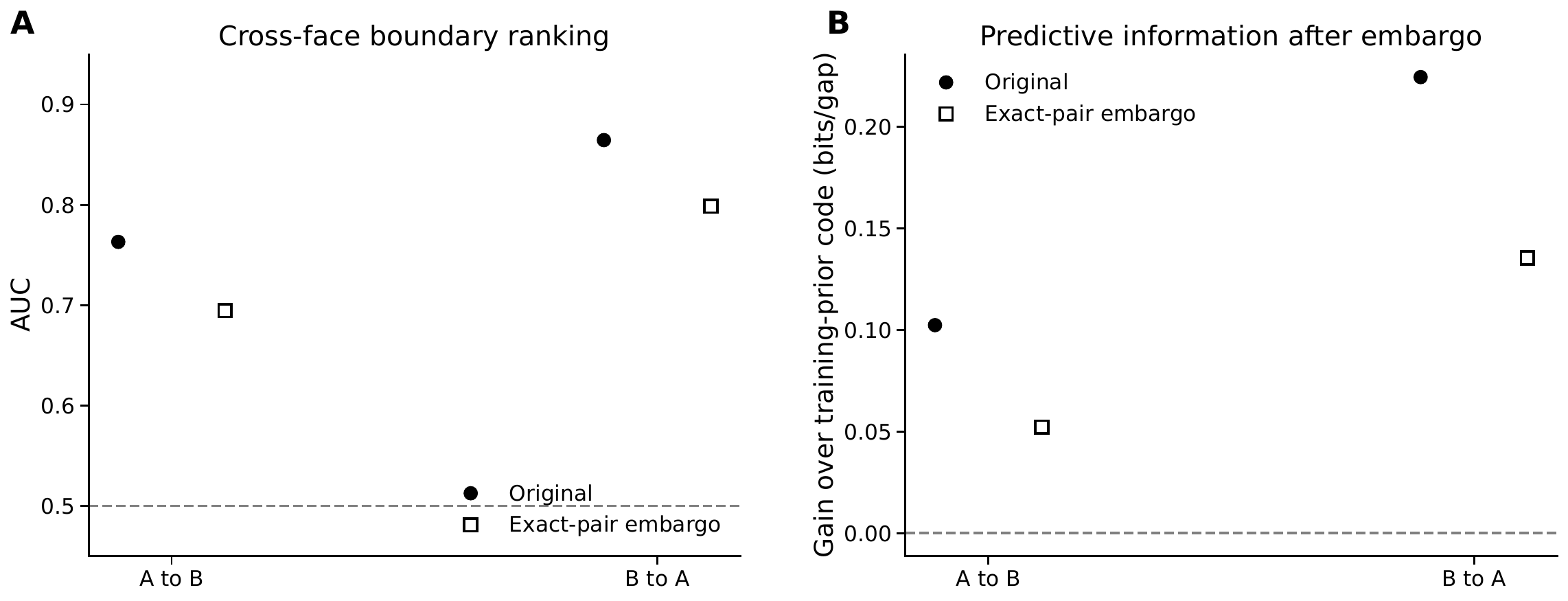}

\caption{\textbf{Exact adjacent-pair embargo of Phaistos cross-face boundary transfer.}
(A) Boundary-ranking AUC for the original cross-face predictor and after excluding every training gap whose ordered sign pair occurs on the target face. The dashed line indicates chance AUC of 0.5.
(B) Predictive code gain relative to the corresponding training-prevalence baseline. Although exact-pair exclusion reduces discrimination and coding efficiency, both directional transfers retain positive code gain. Matched training- and target-divider randomisations confirm significant AUC discrimination after Holm correction. The embargo uses target sign identities but not target boundary labels.}

\label{fig:phaistos-pair-embargo}
\end{figure}

To determine whether cross-face boundary transfer depends on exact local reuse, we introduced a stricter adjacent-pair embargo that removes every training gap whose ordered sign pair occurs anywhere on the target face. For A-to-B transfer, 31 of 122 training gaps were excluded, leaving 91 training and 118 test gaps. AUC declined from 0.763 to 0.695, average precision from 0.584 to 0.517, and predictive code gain from 0.102 to 0.052 bits per gap. For B-to-A transfer, 21 of 118 training gaps were excluded, leaving 97 training and 122 test gaps. AUC declined from 0.864 to 0.799, average precision from 0.772 to 0.611, and code gain from 0.224 to 0.136 bits per gap. Both directions retained significant discrimination under independently randomised training and target divider systems, with Holm-adjusted AUC probabilities of 0.0012 and 0.0052 for A-to-B, and 0.0052 and 0.0008 for B-to-A, respectively (Fig.~\ref{fig:phaistos-pair-embargo}; Supplementary Table~\ref{tab:s24}). These results demonstrate that cross-face predictive information persists beyond exact adjacent-pair overlap. The attenuation may reflect both removal of shared local material and reduced training support.

For every sufficiently frequent sign, we also construct a two-coordinate
role vector from its smoothed initial and final odds. The centred
cross-face cosine is 0.430 for 36 sign types, with \(p=0.0002\) under
independent randomisation of the two divider systems. Removing sign 02
gives 0.297 (\(p=0.0064\)); removing both 02 and 12 gives 0.300
(\(p=0.0070\)). This is a geometry of positional roles in a feature
space, not the geometric position of impressions on clay.

\begin{figure}[tbp]
\centering
\includegraphics[width=0.97\linewidth,height=0.58\textheight,keepaspectratio]{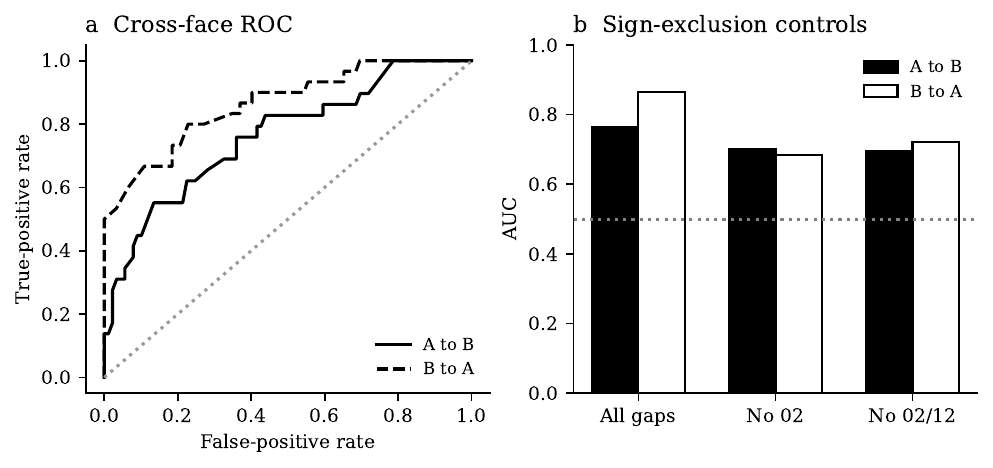}
\caption{Phaistos boundary information transfers between faces. (a) ROC curves from the local categorical classifier. (b) AUC after excluding gaps adjacent to sign 02 or either 02 or 12. The horizontal reference is 0.5; the matched-null means and probabilities are specific to each test. Exact shared passages are not excluded from these transfer experiments.}
\label{fig:4}
\end{figure}

\subsection{Face-specific sign mixtures coexist with economical
transition
reuse}\label{face-specific-sign-mixtures-coexist-with-economical-transition-reuse}

A unigram change-point scan evaluates splits after positions 30 through
212 of the combined sequence. Its maximum is after sign 125, with a
23.331-bit gain, two positions from the archaeological face break at
123. The maximum statistic has \(p=0.000400\) against shuffled-token
scans. The authentic break ranks 12th among 183 candidate token
positions and third among 46 eligible group boundaries. A fixed split
after 31 groups remains unusual under intact-group reordering
(\(z=5.06\), \(p=0.000200\)); the maximum over the group-boundary scan
has the weaker, search-corrected \(p=0.00120\) under that null.

At the authentic break, separate zero-order models save 18.395 bits
relative to a pooled zero-order model. The corresponding first-order
comparison favours the pooled sequential code by 16.387 bits. The
smoothed Jensen-Shannon divergence of the face-level distributions is
0.184 bits. These results support different sign mixtures together with
useful reuse of transition information. They do not identify a unique
grammar: sparse transition counts, shared formulas and prior
regularisation also contribute to the pooled model's advantage. The
complementary boundary-transfer result strengthens the narrower conclusion
that the faces share local structural constraints.

\FloatBarrier
{
\small
\setlength{\tabcolsep}{3pt}
\renewcommand{\arraystretch}{1.12}
\begin{longtable}{@{}>{\raggedright\arraybackslash}p{\dimexpr 0.302326\linewidth-5.4419pt\relax}>{\raggedright\arraybackslash}p{\dimexpr 0.232558\linewidth-4.1860pt\relax}>{\raggedright\arraybackslash}p{\dimexpr 0.232558\linewidth-4.1860pt\relax}>{\raggedright\arraybackslash}p{\dimexpr 0.232558\linewidth-4.1860pt\relax}@{}}
\caption{Face comparison and change-point evidence.}\label{tab:main-8}\\
\toprule
Comparison & Observed gain / score & Null or reference & p \\
\midrule
\endfirsthead
\toprule
Comparison & Observed gain / score & Null or reference & p \\
\midrule
\endhead
\bottomrule
\endlastfoot
KT0: pooled minus separate & 18.40 bits & Token reassignment & \(4.00\times10^{-4}\) \\
KT1: pooled minus separate & -16.39 bits & Token reassignment & \(1.60\times10^{-3}\) \\
LZW: pooled minus separate & 24 bits & Token reassignment & 0.62 \\
CTM: pooled minus separate & -729.11 score units & Token reassignment & \(8.16\times10^{-2}\) \\
Best token split: 125 & 23.33 bits & Maximum over 183 splits & \(4.00\times10^{-4}\) \\
True split: after 31 groups & 18.40 bits & Fixed split; group-order null & \(2.00\times10^{-4}\) \\
Best group-boundary split: 192 & 19.39 bits & Maximum; group-order null & \(1.20\times10^{-3}\) \\
Face distribution JSD & 0.18 bits & Jeffreys smoothing; descriptive & NA \\
\end{longtable}
}

Positive coding contrast favours a reset at the face break. Negative
favours pooled reuse. CTM is a proxy with substantial
length/decomposition overhead. Fixed and maximum-over-splits
probabilities answer different questions.

\subsection{Repeated material explains the principal recurrence without
identifying a cipher
key}\label{repeated-material-explains-the-principal-recurrence-without-identifying-a-cipher-key}

The categorical Fourier spectrum is invariant to the arbitrary sign
labels. On A its highest power is 2.563 at a period of 4.1 signs, with a
maximum-over-frequencies \(p=0.0188\) against intact-group reorderings.
B is ordinary under the same test (\(p=0.943\)). Deleting the second
copy of a 15-sign repeated passage makes the A spectral maximum
nonsignificant against intact-group reorderings (\(p=0.3262\)); removing
the broader formulaic region gives \(p=0.4034\). These ablations alter
the available sequence length and frequency grid, so each altered text
receives its own null distribution.

Phase mutual information asks a different question: whether sign
identity is associated with position modulo a candidate integer period.
A selects period 25 with search-corrected \(p=0.0010\). There are 19
exact matches among 98 sign pairs separated by 25 positions; 15 arise
from the exact passage at zero-based positions 50--64 and 75--89.
Removing the second copy while preserving the original indices gives a
maximum-statistic \(p=0.1015\). The four remaining lag-25 matches occur
among 83 remaining pairs, a rate of 0.0482.

B's phase scan produces a marginal maximum at 37 (\(p=0.0480\)), which
does not survive even a three-analysis Bonferroni adjustment for the
phase comparisons reported here. The Fourier period of 4.1 and the phase
period of 25 are not contradictory estimates of a single frequency: the
former captures aggregate frequency-domain structure, while the latter
captures category preferences at repeated offsets. Together with the
ablations, the data support recurrence linked to identifiable passages.
They do not exclude substitution ciphers or all encryption schemes,
which can preserve such patterns.

\FloatBarrier
{
\small
\setlength{\tabcolsep}{3pt}
\renewcommand{\arraystretch}{1.12}
\begin{longtable}{@{}>{\raggedright\arraybackslash}p{\dimexpr 0.234234\linewidth-5.6216pt\relax}>{\raggedright\arraybackslash}p{\dimexpr 0.225225\linewidth-5.4054pt\relax}>{\raggedright\arraybackslash}p{\dimexpr 0.180180\linewidth-4.3243pt\relax}>{\raggedright\arraybackslash}p{\dimexpr 0.180180\linewidth-4.3243pt\relax}>{\raggedright\arraybackslash}p{\dimexpr 0.180180\linewidth-4.3243pt\relax}@{}}
\caption{Relabelling-invariant spectral and phase tests, with repeated-passage ablation.}\label{tab:main-9}\\
\toprule
Scan & Sequence & Peak period & Peak statistic & Scan p \\
\midrule
\endfirsthead
\toprule
Scan & Sequence & Peak period & Peak statistic & Scan p \\
\midrule
\endhead
\bottomrule
\endlastfoot
Fourier & A complete & 4.10 & 2.56 & \(1.88\times10^{-2}\) \\
Fourier & B complete & 2.71 & 1.38 & 0.94 \\
Fourier & A: remove copy & 2.00 & 1.85 & 0.33 \\
Fourier & A: remove region & 2.00 & 1.72 & 0.40 \\
Phase MI & A complete & 25 & 5.20 & \(1.00\times10^{-3}\) \\
Phase MI & B complete & 37 & 3.34 & \(4.80\times10^{-2}\) \\
Phase MI & A: masked copy & 37 & 3.10 & 0.10 \\
\end{longtable}
}

Fourier statistic is raw summed power; its null permutes intact groups
(4,999 draws). Phase statistic is independently calibrated maximum z,
with 1,999 evaluation draws. Additional four-test Fourier and three-test
phase family corrections are given in Methods. Removing the copy deletes
zero-based {[}75,90); the wider interval is {[}50,87).

\begin{figure}[h]
\centering
\includegraphics[width=0.97\linewidth,height=0.58\textheight,keepaspectratio]{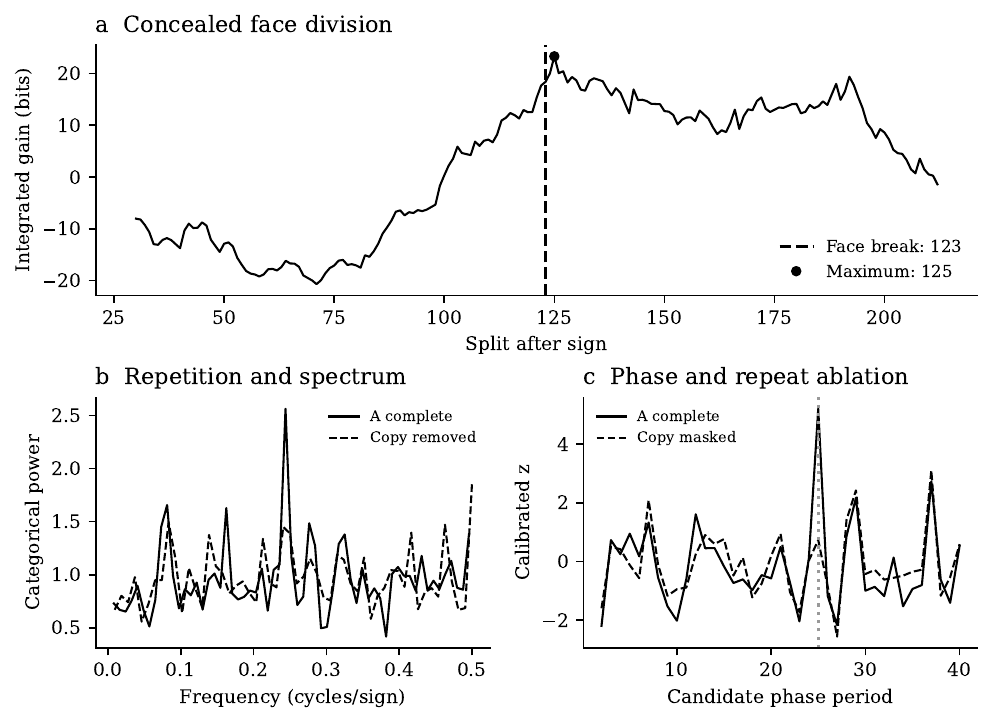}
\caption{Change-point and recurrence structure on Phaistos. (a) The zero-order split scan has its maximum two signs from the known face break. (b) The Fourier spectrum changes after deletion of one exact 15-sign copy, with a separate null at the altered length. (c) The phase-25 signal weakens when that copy is masked while retaining the original position indices. The three panels measure different structural properties.}
\label{fig:5}
\end{figure}

\subsection{Phaistos does not yield additional cross-winding information
in the tested
embeddings}\label{phaistos-does-not-yield-additional-cross-winding-information-in-the-tested-embeddings}

Idealised equal-arc Archimedean spirals with 2.5--6.5 turns define
candidate nonlocal neighbour relations. We evaluate identical-sign
proximity, mutual information between edge endpoints and positional-role
alignment. The conservative null reorders complete compartments while
preserving their internal contents. No complete-face spiral statistic
survives the search over turn counts (Table~\ref{tab:main-10}); at the nominal
four-turn model all six standard scores are below their null means.

\FloatBarrier
{
\small
\setlength{\tabcolsep}{3pt}
\renewcommand{\arraystretch}{1.12}
\begin{longtable}{@{}>{\raggedright\arraybackslash}p{\dimexpr 0.166667\linewidth-5.0000pt\relax}>{\raggedright\arraybackslash}p{\dimexpr 0.166667\linewidth-5.0000pt\relax}>{\raggedright\arraybackslash}p{\dimexpr 0.166667\linewidth-5.0000pt\relax}>{\raggedright\arraybackslash}p{\dimexpr 0.166667\linewidth-5.0000pt\relax}>{\raggedright\arraybackslash}p{\dimexpr 0.166667\linewidth-5.0000pt\relax}>{\raggedright\arraybackslash}p{\dimexpr 0.166667\linewidth-5.0000pt\relax}@{}}
\caption{Phaistos spiral scans under intact-group reordering.}\label{tab:main-10}\\
\toprule
Face & Statistic & Best turns & Best z & Scan p & z at 4 turns \\
\midrule
\endfirsthead
\toprule
Face & Statistic & Best turns & Best z & Scan p & z at 4 turns \\
\midrule
\endhead
\bottomrule
\endlastfoot
A & Same-sign edges & 5.80 & 2.59 & 0.30 & -0.50 \\
A & Endpoint MI & 2.50 & 2.45 & 0.15 & -2.25 \\
A & Role alignment & 3.60 & 2.57 & 0.18 & -0.75 \\
B & Same-sign edges & 5.20 & 2.92 & 0.16 & -1.30 \\
B & Endpoint MI & 5.40 & 2.60 & \(9.19\times10^{-2}\) & -0.34 \\
B & Role alignment & 2.60 & 1.69 & 0.78 & -1.77 \\
\end{longtable}
}

Each p value accounts for the turn-count scan under intact-group
reordering. Four turns is an idealised reference geometry, not a
measured ground-truth coordinate fit.

An apparent rectangular width-25 preference is similarly
repetition-sensitive. After removal of the second 15-sign copy, the
group-order-corrected rectangular scan gives \(p=0.653\); after removal
of the broader formulaic interval it gives \(p=0.887\). CTM-BDM scans of
alternative column and radial traversals are also nonsignificant on the
complete faces: rectangular and spiral probabilities are 0.781 and 0.957
on A, and 0.349 and 0.990 on B. Synthetic planted controls recover a
width-20 rectangle and a four-turn spiral with \(p=0.001996\) against
their token-permutation controls. This demonstrates sensitivity to those
planted alternatives, not guaranteed power against every possible
ancient geometric code.

The negative result is specific to idealised coordinates and tested
statistics. Measured impression coordinates, crowding, sign orientation
and production sequence could support different geometric hypotheses. At
present, segmentation and local sign dependencies are more strongly
supported than additional cross-winding organisation.

\begin{figure}[tbp]
\centering
\includegraphics[width=0.97\linewidth,height=0.58\textheight,keepaspectratio]{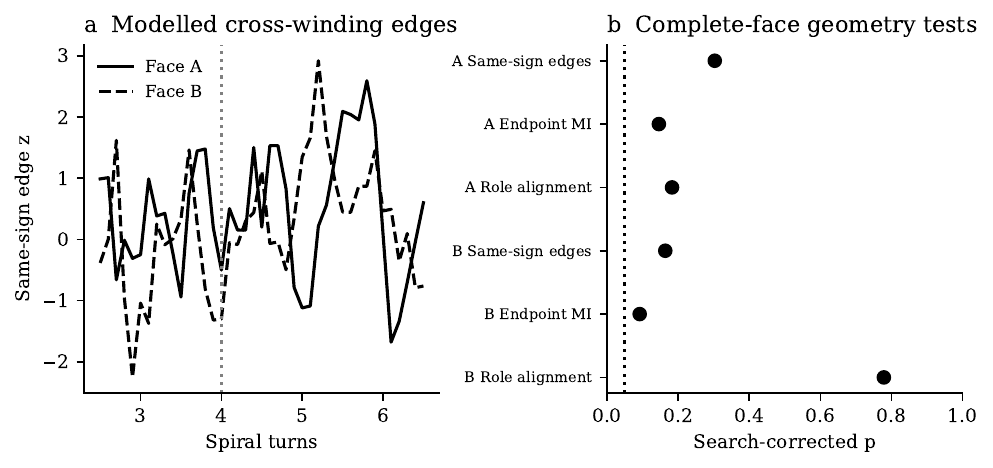}
\caption{Idealised spiral geometry adds no significant complete-face signal under the tested group-order controls. (a) Same-sign edge standard scores across candidate turn counts. Four turns is a reference, not a coordinate fit. (b) Maximum-scan probabilities for the six complete-face tests. The vertical line indicates 0.05 before any additional family correction.}
\label{fig:6}
\end{figure}

\subsection{Khipu, Indus and manuscript layouts extend the
reconstruction
task}\label{khipu-indus-and-manuscript-layouts-extend-the-reconstruction-task}

The comparative analysis withholds whole records or record collections
from training. Every classifier predicts a concealed spatial relation
from categorical forms, without receiving the target coordinates as
predictors. Table~\ref{tab:main-11} reports AUC, average precision, prevalence and
matched-null probabilities; Table~\ref{tab:main-12} expresses the same fixed predictors
as probabilistic coding models. These tasks differ in their units,
inventories and baseline difficulty, so their AUCs are not rankings of
decipherment progress.

\FloatBarrier
{
\small
\setlength{\tabcolsep}{3pt}
\renewcommand{\arraystretch}{1.12}
\begin{longtable}{@{}>{\raggedright\arraybackslash}p{\dimexpr 0.142857\linewidth-5.1429pt\relax}>{\raggedright\arraybackslash}p{\dimexpr 0.142857\linewidth-5.1429pt\relax}>{\raggedright\arraybackslash}p{\dimexpr 0.142857\linewidth-5.1429pt\relax}>{\raggedright\arraybackslash}p{\dimexpr 0.142857\linewidth-5.1429pt\relax}>{\raggedright\arraybackslash}p{\dimexpr 0.142857\linewidth-5.1429pt\relax}>{\raggedright\arraybackslash}p{\dimexpr 0.142857\linewidth-5.1429pt\relax}>{\raggedright\arraybackslash}p{\dimexpr 0.142857\linewidth-5.1429pt\relax}@{}}
\caption{Prediction on withheld cultural records, with object-bootstrap uncertainty.}\label{tab:main-11}\\
\toprule
Record & AUC & 95\% interval & AP & Prevalence & Null AUC & p \\
\midrule
\endfirsthead
\toprule
Record & AUC & 95\% interval & AP & Prevalence & Null AUC & p \\
\midrule
\endhead
\bottomrule
\endlastfoot
Rongorongo & 0.51 & 0.42--0.59 & 0.02 & 0.02 & 0.50 & 0.38 \\
Indus & 0.81 & 0.74--0.87 & 0.41 & 0.17 & 0.54 & \(1.00\times10^{-3}\) \\
Voynich & 0.84 & 0.82--0.86 & 0.47 & 0.12 & 0.51 & \(1.00\times10^{-3}\) \\
Khipu & 0.86 & 0.78--0.92 & 0.73 & 0.39 & 0.50 & \(1.00\times10^{-3}\) \\
Egyptian & 0.95 & 0.88--0.98 & 0.94 & 0.49 & 0.64 & \(1.00\times10^{-3}\) \\
\end{longtable}
}

AUC intervals use 1,999 resamples of the withheld objects with the
trained model fixed. Nulls permute labels within
inscriptions/pages/files/tablets, or within cords for khipu. Only three
tablets enter the Rongorongo interval. All five tests use 999 null
draws.

Khipu provide the clearest change in physical organisation from a linear
inscription. The source contains 54,403 cords and 110,677 knots. The
eligible test retains 532 khipu and 15,558 cords with at least two
recorded knot clusters at distinct positive positions. We predict which
eligible cluster is farthest along its cord from raw knot-type counts
and cluster size. Whole khipu are withheld; 7,707 test clusters include
3,028 distal positives. AUC is 0.858, with a bootstrap interval of
0.779--0.918 obtained by resampling the 107 held-out khipu. The code
gain is 0.298 bits per cluster relative to the training-prevalence
baseline. The decoded field \texttt{knot\_value\_type} is excluded
because its interpretation can depend on spatial rank, which would make
this test circular. Recorded positions supply labels only. The result
reconstructs a coarse spatial role from knot form; it does not newly
infer the numerical base or the complete cord tree.

Indus requires stricter controls for short and repeated inscriptions.
Among 5,679 source records, 75 unique complete multiline sequences have
an unambiguous recorded layout under our filters. Mohenjo-daro is
withheld as the test site, and sequences shared across sites are
excluded from this comparison. The resulting test includes 292 gaps in
47 inscriptions, of which 49 are row breaks. AUC is 0.812, with a
bootstrap interval of 0.742--0.871. The predictor's code gain is only
0.0092 bits per gap, despite its strong ranking performance. This
distinction matters: the local signs identify likely breaks better than
the matched null, but the raw naive-Bayes probabilities are only
slightly more economical than a constant training-based probability. The
input's speculative translation columns are never used, and the result
concerns recorded slash delimiters rather than calibrated object
coordinates.

Voynich supplies a larger layout test. The selected transcription
contains 207 body pages, 4,130 lines and 34,963 word-like units. We
represent a word by a categorical tuple containing its first two
characters, last two characters and length capped at 12. These are
features of the EVA transcription, not phonetic assumptions. Every fifth
sorted page is designated for evaluation; held-out pages with exact line
duplicates in training are excluded. The final test comprises 6,190 gaps
on 42 pages, including 719 line breaks. AUC is 0.843, its bootstrap
interval is 0.818--0.860, and code gain is 0.114 bits per gap. The
result is compatible with linguistic organisation, scribal generation or
layout-dependent variation; it does not discriminate these mechanisms by
itself.

Egyptian joins form a known-script spatial control. We extract explicit
adjacent Gardiner-style sign-code pairs joined by a colon, star or dash
from Manuel de Codage editions. The colon labels a vertical join. Exact
duplicated extracted pair lists are removed; 186 eligible files contain
40,986 pairs. The held-out test has 9,144 pairs in 38 files, with 4,450
vertical joins. AUC is 0.945, with a bootstrap interval of 0.881--0.980,
and code gain is 0.660 bits per pair. The within-file permutation mean
is 0.640, not 0.5, because the null retains each file's class balance
and associated score distribution. Different editions of the same
ancient work can still share passages. This control establishes selected
digital-layout recoverability, rather than independent reconstruction of
every ancient quadrat.

\FloatBarrier
{
\small
\setlength{\tabcolsep}{3pt}
\renewcommand{\arraystretch}{1.12}
\begin{longtable}{@{}>{\raggedright\arraybackslash}p{\dimexpr 0.142857\linewidth-5.1429pt\relax}>{\raggedright\arraybackslash}p{\dimexpr 0.142857\linewidth-5.1429pt\relax}>{\raggedright\arraybackslash}p{\dimexpr 0.142857\linewidth-5.1429pt\relax}>{\raggedright\arraybackslash}p{\dimexpr 0.142857\linewidth-5.1429pt\relax}>{\raggedright\arraybackslash}p{\dimexpr 0.142857\linewidth-5.1429pt\relax}>{\raggedright\arraybackslash}p{\dimexpr 0.142857\linewidth-5.1429pt\relax}>{\raggedright\arraybackslash}p{\dimexpr 0.142857\linewidth-5.1429pt\relax}@{}}
\caption{Information gain and probability quality for the fixed comparative predictors.}\label{tab:main-12}\\
\toprule
Record & Test candidates & Bootstrap units & Baseline cost & Model cost & Code gain & Brier \\
\midrule
\endfirsthead
\toprule
Record & Test candidates & Bootstrap units & Baseline cost & Model cost & Code gain & Brier \\
\midrule
\endhead
\bottomrule
\endlastfoot
Rongorongo & 4059 & 3 & 0.11 & 0.16 & -0.05 & 0.02 \\
Indus & 292 & 47 & 0.66 & 0.65 & \(9.20\times10^{-3}\) & 0.15 \\
Voynich & 6190 & 42 & 0.52 & 0.40 & 0.11 & 0.08 \\
Khipu & 7707 & 107 & 0.97 & 0.67 & 0.30 & 0.14 \\
Egyptian & 9144 & 38 & 1.14 & 0.48 & 0.66 & 0.10 \\
\end{longtable}
}

Costs and gains are bits per candidate. The constant baseline uses the
smoothed training prevalence. Predictive probabilities are naive-Bayes
outputs without test-set recalibration. Brier score is mean squared
probability error. Tasks differ, so absolute gains are not a common
semantic scale.

\begin{figure}[h]
\centering
\includegraphics[width=0.97\linewidth,height=0.58\textheight,keepaspectratio]{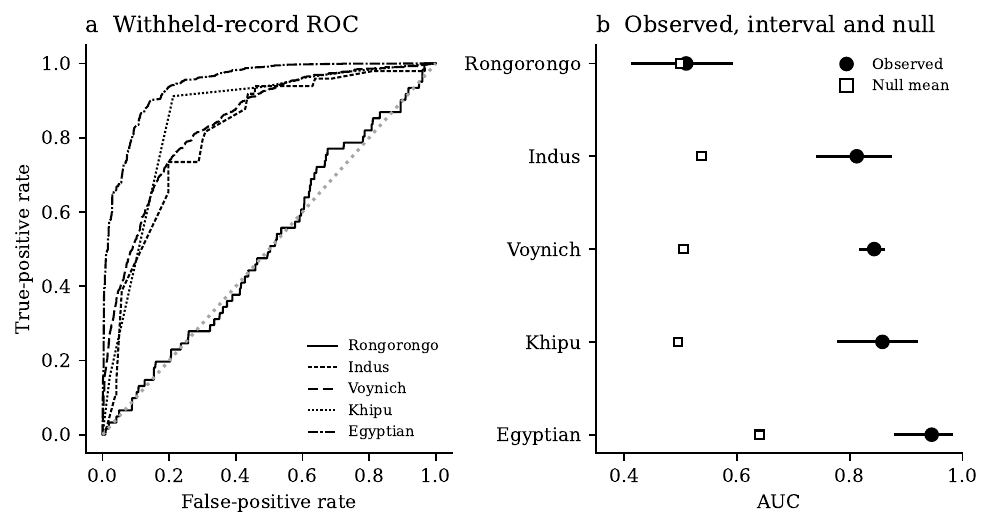}
\caption{Spatial prediction on withheld cultural records. (a) ROC curves for five distinct tasks. (b) Observed AUC, object-bootstrap 95\% interval and matched-null mean. Bootstrap resampling keeps the trained model fixed. The three-tablet Rongorongo interval is particularly limited. The tasks and prevalences differ, so the plot is not a ranking of decipherment progress.}
\label{fig:7}
\end{figure}

\subsection{Rongorongo distinguishes traversal regularity from
transferable line
boundaries}\label{rongorongo-distinguishes-traversal-regularity-from-transferable-line-boundaries}

Thirteen Rongorongo tablets contribute 229 nonempty transcribed lines
and 11,666 tokens. Consistently oriented lines have a first-order coding
cost of 8.117 bits per token, compared with 8.221 under independent
random reversal of lines. LZW similarly gives 8.900 against 9.156. Both
tests have \(p=0.001\) with 999 randomisations. Mean KT1 cost rises as a
small fraction of lines is reversed, reaches its greatest sampled value
near half reversal, and falls again when almost all lines are reversed.
The fully reversed corpus costs 8.115 bits per token, slightly below the
original orientation. The landscape supports consistency of traversal,
not a unique absolute reading direction.

For a stricter boundary-transfer test, tablets H, P and Q, which contain
parallel material, are withheld together. The other tablets supply
training data. On 4,059 test gaps with 61 line breaks, AUC is 0.510 and
the matched-null \(p=0.382\). The predictor also loses 0.0508 bits per
gap relative to its constant-probability baseline. Thus, a traversal can
preserve strong sequential regularity without furnishing transferable
line endings. Only three tablets form this evaluation family, limiting
its uncertainty estimate and generality.

\begin{figure}[tbp]
\centering
\includegraphics[width=0.97\linewidth,height=0.58\textheight,keepaspectratio]{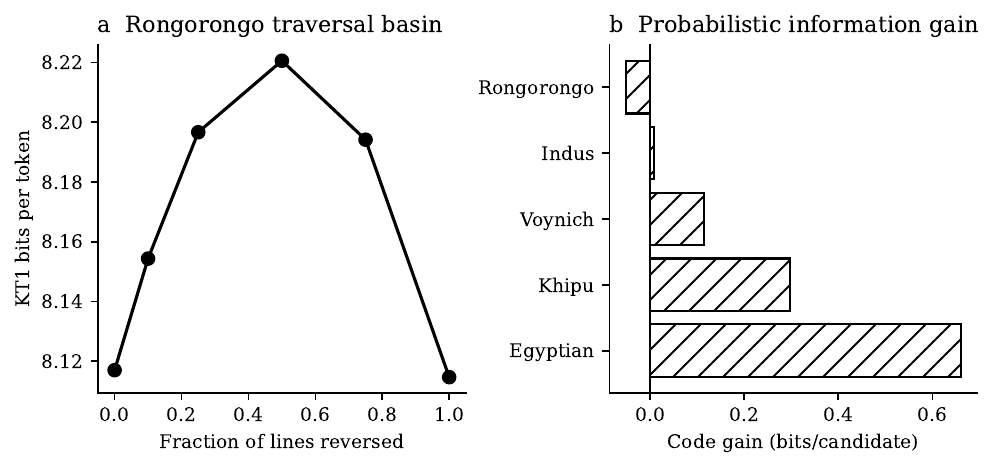}
\caption{Traversal regularity and probabilistic prediction measure different information. (a) Rongorongo cost is lowest near either consistent orientation and highest near mixed orientation. Interior means use 99 reversal subsets; the endpoints are deterministic. (b) Fixed-model code gain relative to a training-prevalence baseline. Indus has strong ranking but small code gain; Rongorongo loses information under this probability model.}
\label{fig:8}
\end{figure}

\subsection{Finite-block and transcription sensitivity delimit the
Phaistos
conclusions}\label{finite-block-and-transcription-sensitivity-delimit-the-phaistos-conclusions}

The CTM-BDM sequential result varies materially with block length. For pooled A+B, role-preserving tests give \(p=0.251\), 0.000800
and 0.0122 for lengths 4, 5 and 6. Face-wise six-bit tests are
nonsignificant. After removing duplicate complete groups, LZW and KT1
remain significant but CTM-BDM gives \(p=0.592\). These outcomes are
consistent with different estimators exploiting different parts of the
record, and identify which finite representations carry the measured
programme-frequency signal.

Alternative assignments 03, 08, 20 and 44 at the damaged position leave
LZW unchanged at 5.719 bits per sign and KT1 within 5.0263--5.0273. The
face-split gain remains positive, ranging from 16.073 to 18.395 bits.
These are sensitivity checks over four specified readings; they are not
an exhaustive analysis of every possible damage model.

No individual sign survives the false-discovery correction as
an exceptional occurrence-pattern or deletion candidate. For sign 02,
the CTM occurrence test changes from raw \(p=0.0410\) to \(q=0.433\),
and KT1 from \(p=0.0160\) to \(q=0.720\). Deleting all its occurrences
gives \(p=0.442\), 0.326 and 0.100 under CTM, LZW and KT1. Oblique
strokes yield nominal CTM \(p=0.0477\) and final-sign mutual-information
\(p=0.0503\), with other tested associations nonsignificant. These tests
identify no statistically secure individual structural operator, and say
nothing direct about semantic importance.

Reversal provides a further check. The implemented phase-averaged,
symmetry-completed CTM-BDM is exactly reversal invariant. KT1
reverse-minus-forward differences are -0.00435 bits per sign on A,
+0.00388 on B and -0.000149 on A+B, with \(p=0.148\), 0.196 and 0.917.
The small KT2 asymmetries change sign between faces and remain below
0.005 bits per sign. Direction consequently remains an archaeological
and comparative question rather than a consequence of the global
complexity estimates.

\FloatBarrier
{
\small
\setlength{\tabcolsep}{3pt}
\renewcommand{\arraystretch}{1.12}
\begin{longtable}{@{}>{\raggedright\arraybackslash}p{\dimexpr 0.166667\linewidth-5.0000pt\relax}>{\raggedright\arraybackslash}p{\dimexpr 0.166667\linewidth-5.0000pt\relax}>{\raggedright\arraybackslash}p{\dimexpr 0.166667\linewidth-5.0000pt\relax}>{\raggedright\arraybackslash}p{\dimexpr 0.166667\linewidth-5.0000pt\relax}>{\raggedright\arraybackslash}p{\dimexpr 0.166667\linewidth-5.0000pt\relax}>{\raggedright\arraybackslash}p{\dimexpr 0.166667\linewidth-5.0000pt\relax}@{}}
\caption{Phaistos estimator and transcription sensitivity.}\label{tab:main-13}\\
\toprule
Block length & A+B order p & A order p & B order p & A divider p & B divider p \\
\midrule
\endfirsthead
\toprule
Block length & A+B order p & A order p & B order p & A divider p & B divider p \\
\midrule
\endhead
\bottomrule
\endlastfoot
4 & 0.25 & 0.45 & 0.83 & 0.21 & 0.45 \\
5 & \(8.00\times10^{-4}\) & \(2.62\times10^{-2}\) & 0.36 & \(2.00\times10^{-2}\) & 0.44 \\
6 & \(1.22\times10^{-2}\) & 0.47 & 0.39 & \(3.47\times10^{-2}\) & \(3.57\times10^{-2}\) \\
\end{longtable}
}

Separate block-length sensitivity analyses. Order nulls preserve pooled
positional roles; divider nulls permute lengths. These p values are not
the face-and-role primary null in Table~\ref{tab:main-3}.

\subsection{A common AID experiment tests content--structure coupling
across
records}\label{a-common-aid-experiment-tests-contentstructure-coupling-across-records}

\begin{figure}[tbp]
\centering
\includegraphics[width=0.97\linewidth,height=0.58\textheight,keepaspectratio]{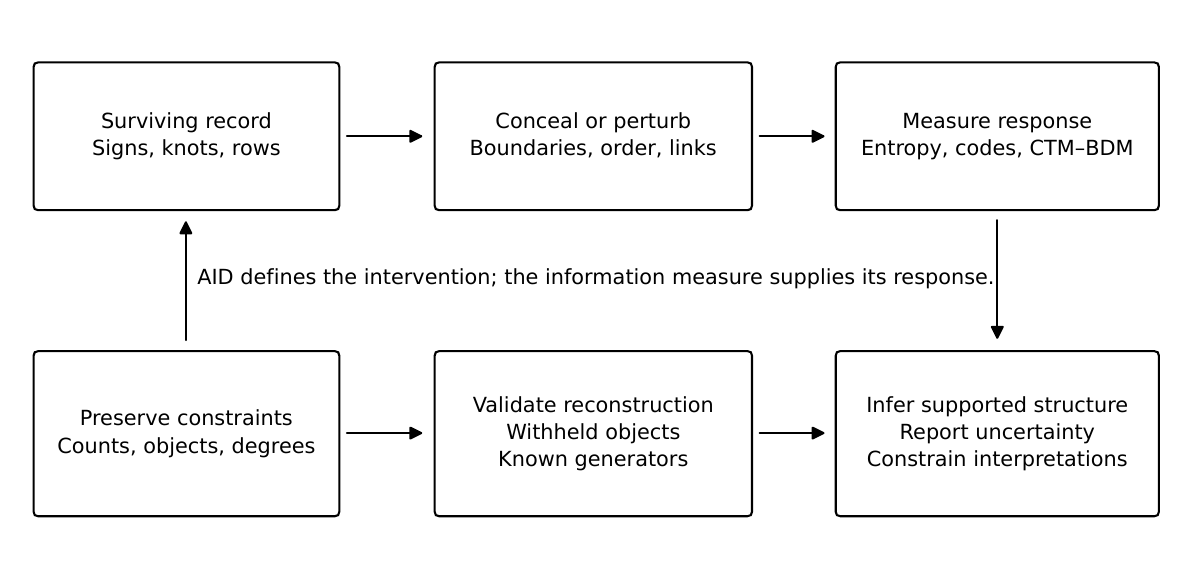}
\caption{AID links representations, computable interventions, information responses and validation. The schematic specifies an experimental workflow. It does not depict a recovered historical causal graph. CTM-BDM connects the scoring stage to finite programme-output distributions, while the same workflow accommodates classical and predictive information.}
\label{fig:9}
\end{figure}

The Phaistos divider basin motivates an intervention that can be applied
without imposing a single physical shape on every corpus. We keep local
features fixed and exchange binary structural roles within each held-out
object, or within each cord for khipu. Conditional entropy, an
integrated KT code and the original frozen predictive code measure the
same perturbed arrangements. These are three views of one intervention,
rather than three independent discoveries. Table~\ref{tab:main-14} specifies the
additional experiments and their preserved constraints; Figure~\ref{fig:9} makes
the common design explicit.

\FloatBarrier
{
\small
\setlength{\tabcolsep}{3pt}
\renewcommand{\arraystretch}{1.12}
\begin{longtable}{@{}>{\raggedright\arraybackslash}p{\dimexpr 0.285714\linewidth-5.1429pt\relax}>{\raggedright\arraybackslash}p{\dimexpr 0.274725\linewidth-4.9451pt\relax}>{\raggedright\arraybackslash}p{\dimexpr 0.219780\linewidth-3.9560pt\relax}>{\raggedright\arraybackslash}p{\dimexpr 0.219780\linewidth-3.9560pt\relax}@{}}
\caption{Phaistos-guided comparative experiments.}\label{tab:main-14}\\
\toprule
Target & Records & Intervention & Validation \\
\midrule
\endfirsthead
\toprule
Target & Records & Intervention & Validation \\
\midrule
\endhead
\bottomrule
\endlastfoot
Structural role & All five comparisons & Within-object label swaps & 999 constrained nulls \\
Local reuse & All five comparisons & Embargo shared feature pairs & Original held-out objects \\
Frequent features & All five comparisons & Exclude training-selected values & Changed test subset \\
Traversal & Rongorongo & Independent line reversal & H/P/Q family ablation \\
Row alignment & Rongorongo, Indus, Voynich & Coherent interior row shifts & Endpoint/order controls \\
Cord topology & Khipu & Reassign parent identities & Degree, fibre, twist controls \\
Attachment recovery & Khipu & Conceal child--parent allocation & Constrained inverse assignment \\
Generator mechanism & Synthetic rows and trees & Set copying probability to zero & Known mechanism and null controls \\
\end{longtable}
}

Spectral and cipher-period tests remain targeted to Phaistos. The
additional cases are not pooled into an artificial continuous
inscription. AID is the intervention framework; the score and null are
chosen for the relation being tested.

\FloatBarrier
{
\small
\setlength{\tabcolsep}{3pt}
\renewcommand{\arraystretch}{1.12}
\begin{longtable}{@{}>{\raggedright\arraybackslash}p{\dimexpr 0.166667\linewidth-5.0000pt\relax}>{\raggedright\arraybackslash}p{\dimexpr 0.166667\linewidth-5.0000pt\relax}>{\raggedright\arraybackslash}p{\dimexpr 0.166667\linewidth-5.0000pt\relax}>{\raggedright\arraybackslash}p{\dimexpr 0.166667\linewidth-5.0000pt\relax}>{\raggedright\arraybackslash}p{\dimexpr 0.166667\linewidth-5.0000pt\relax}>{\raggedright\arraybackslash}p{\dimexpr 0.166667\linewidth-5.0000pt\relax}@{}}
\caption{Authentic structural roles under three information costs.}\label{tab:main-15}\\
\toprule
Record & Cost & Observed / candidate & Null excess & p & Holm p \\
\midrule
\endfirsthead
\toprule
Record & Cost & Observed / candidate & Null excess & p & Holm p \\
\midrule
\endhead
\bottomrule
\endlastfoot
Rongorongo & Entropy & 0.02 & \(9.99\times10^{-3}\) & \(1.00\times10^{-3}\) & \(1.50\times10^{-2}\) \\
Rongorongo & KT & 0.72 & \(8.57\times10^{-3}\) & \(1.00\times10^{-3}\) & \(1.50\times10^{-2}\) \\
Rongorongo & Frozen & 0.16 & \(1.22\times10^{-3}\) & 0.40 & 0.40 \\
Indus & Entropy & \(6.85\times10^{-3}\) & 0.06 & \(1.00\times10^{-3}\) & \(1.50\times10^{-2}\) \\
Indus & KT & 0.92 & 0.05 & \(1.00\times10^{-3}\) & \(1.50\times10^{-2}\) \\
Indus & Frozen & 0.65 & 0.22 & \(1.00\times10^{-3}\) & \(1.50\times10^{-2}\) \\
Voynich & Entropy & 0.01 & 0.03 & \(1.00\times10^{-3}\) & \(1.50\times10^{-2}\) \\
Voynich & KT & 0.95 & 0.02 & \(1.00\times10^{-3}\) & \(1.50\times10^{-2}\) \\
Voynich & Frozen & 0.40 & 0.41 & \(1.00\times10^{-3}\) & \(1.50\times10^{-2}\) \\
Khipu & Entropy & 0.58 & 0.38 & \(1.00\times10^{-3}\) & \(1.50\times10^{-2}\) \\
Khipu & KT & 0.59 & 0.38 & \(1.00\times10^{-3}\) & \(1.50\times10^{-2}\) \\
Khipu & Frozen & 0.67 & 1.30 & \(1.00\times10^{-3}\) & \(1.50\times10^{-2}\) \\
Egyptian & Entropy & 0.11 & 0.52 & \(1.00\times10^{-3}\) & \(1.50\times10^{-2}\) \\
Egyptian & KT & 0.45 & 0.50 & \(1.00\times10^{-3}\) & \(1.50\times10^{-2}\) \\
Egyptian & Frozen & 0.48 & 1.43 & \(1.00\times10^{-3}\) & \(1.50\times10^{-2}\) \\
\end{longtable}
}

Null excess is mean null cost minus authentic cost; positive values
favour the authentic feature--role association. Conditional entropy and
conditional KT are fitted to the evaluated labels and are calibrated by
randomisation. Only the frozen code uses probabilities learned
exclusively from the training set. Holm adjustment covers the 15 tests
in this table.

Conditional entropy and KT distinguish the authentic assignment in all
five corpora, with adjusted \(p=0.015\). Frozen predictive coding does
so for Indus, Voynich, khipu and Egyptian, but not Rongorongo
(\(p=0.396\)). The distinction is substantive: local organisation can be
present in held-out tablets without the model trained on the other
tablets predicting its labels. Within-test conditional organisation and
cross-record transfer answer different questions.

\begin{figure}[tbp]
\centering
\includegraphics[width=0.97\linewidth,height=0.58\textheight,keepaspectratio]{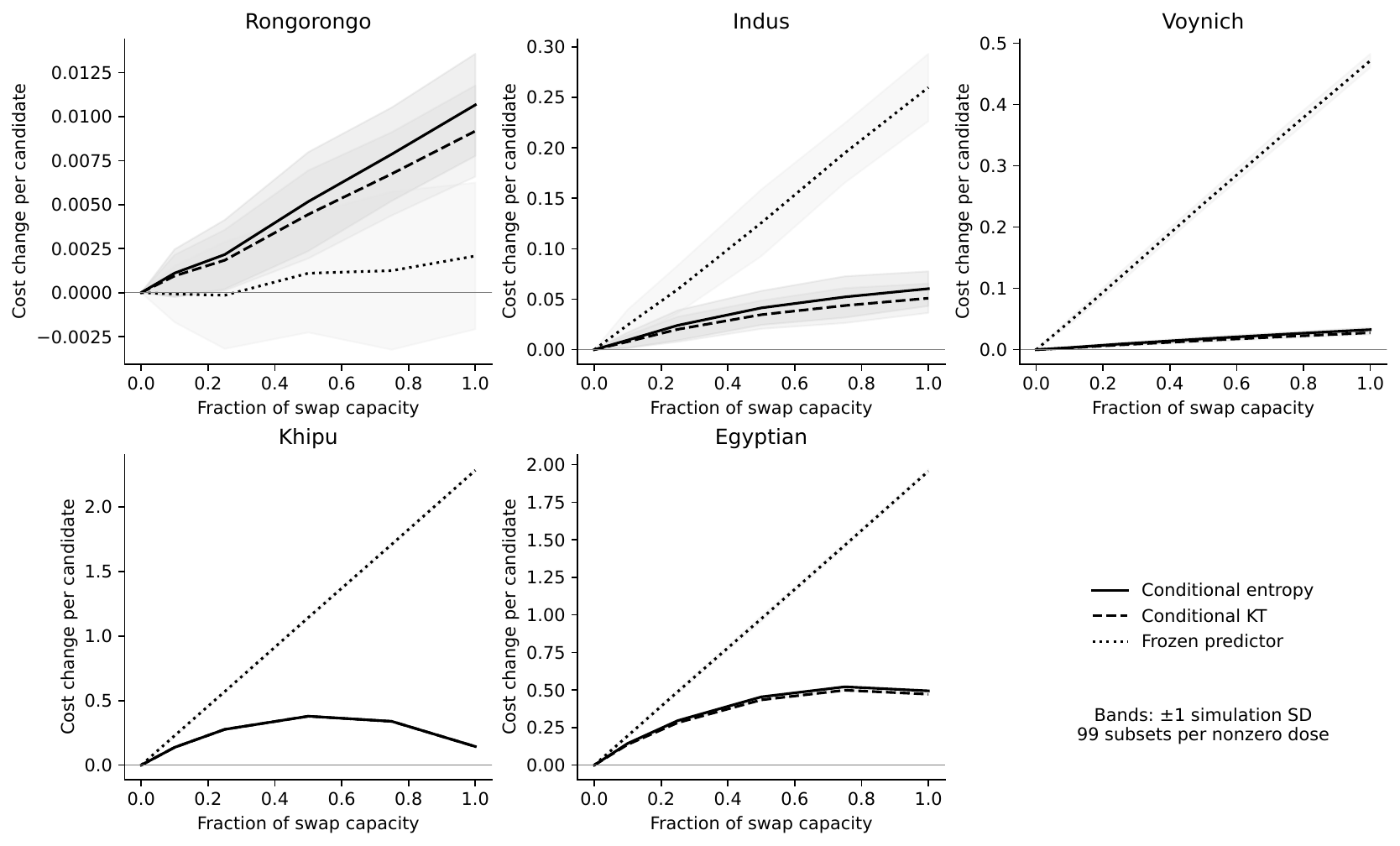}
\caption{The same structural-role intervention under three costs. Each panel preserves its original features and within-stratum label counts. Curves show mean cost change per candidate; bands show one simulation standard deviation over 99 subsets at each nonzero dose. Dose is the fraction of available positive--negative swaps, not the fraction of all candidates. The full-swap endpoint differs from a uniform random-label null. Frozen log-loss responses are partly determined by score separation, as derived in Methods.}
\label{fig:10}
\end{figure}

Graded exchanges produce interpretable landscapes (Figure~\ref{fig:10}). The
independent predictive response generally rises as authentic positive
and negative roles are exchanged. Conditional costs need not rise
monotonically to the full-swap endpoint: exchanging nearly every
possible role can expose an alternative regular assignment. For binary
labels, wholesale complementation would leave empirical conditional
entropy unchanged. Accordingly, we do not identify every nonmonotone
curve as failure or demand that all measures reproduce the exact
Phaistos basin. The intervention determines which alternatives are being
compared.

\subsection{Transfer persists beyond shared local pairs, with
corpus-specific
limits}\label{transfer-persists-beyond-shared-local-pairs-with-corpus-specific-limits}

Phaistos showed that duplicate-compartment control and exclusion of
frequent boundary signs can separate reusable constraints from obvious
copying. The comparative analogue removes every training feature pair
that appears anywhere in the test set, irrespective of its label, then
refits the same classifier. A separate test removes test candidates
containing either feature's most frequent training value. These
exclusions assess exact local combinations and dominant features; they
do not remove all near-parallel text families.

\FloatBarrier
{
\small
\setlength{\tabcolsep}{3pt}
\renewcommand{\arraystretch}{1.12}
\begin{longtable}{@{}>{\raggedright\arraybackslash}p{\dimexpr 0.172185\linewidth-6.1987pt\relax}>{\raggedright\arraybackslash}p{\dimexpr 0.165563\linewidth-5.9603pt\relax}>{\raggedright\arraybackslash}p{\dimexpr 0.132450\linewidth-4.7682pt\relax}>{\raggedright\arraybackslash}p{\dimexpr 0.132450\linewidth-4.7682pt\relax}>{\raggedright\arraybackslash}p{\dimexpr 0.132450\linewidth-4.7682pt\relax}>{\raggedright\arraybackslash}p{\dimexpr 0.132450\linewidth-4.7682pt\relax}>{\raggedright\arraybackslash}p{\dimexpr 0.132450\linewidth-4.7682pt\relax}@{}}
\caption{Transfer after shared-pair embargo and common-feature exclusion.}\label{tab:main-16}\\
\toprule
Record & Test & Train n & Test n & AUC & Gain (bits) & Holm p \\
\midrule
\endfirsthead
\toprule
Record & Test & Train n & Test n & AUC & Gain (bits) & Holm p \\
\midrule
\endhead
\bottomrule
\endlastfoot
Rongorongo & Pair embargo & 5226 & 4059 & 0.51 & -0.04 & 0.73 \\
Rongorongo & Common excluded & 7594 & 3678 & 0.51 & -0.05 & 0.73 \\
Indus & Pair embargo & 110 & 292 & 0.76 & -0.08 & \(1.00\times10^{-2}\) \\
Indus & Common excluded & 145 & 226 & 0.75 & -0.11 & \(1.00\times10^{-2}\) \\
Voynich & Pair embargo & 24610 & 6190 & 0.84 & 0.12 & \(1.00\times10^{-2}\) \\
Voynich & Common excluded & 28566 & 5867 & 0.85 & 0.12 & \(1.00\times10^{-2}\) \\
Khipu & Pair embargo & 22 & 7707 & 0.65 & 0.08 & \(1.00\times10^{-2}\) \\
Khipu & Common excluded & 30155 & 2661 & 0.59 & 0.36 & \(1.00\times10^{-2}\) \\
Egyptian & Pair embargo & 9252 & 9144 & 0.89 & 0.58 & \(1.00\times10^{-2}\) \\
Egyptian & Common excluded & 31842 & 6914 & 0.92 & 0.44 & \(1.00\times10^{-2}\) \\
\end{longtable}
}

Gain is the average saving relative to the appropriate
training-prevalence code. Negative values mean the fitted probabilities
are less useful than that constant baseline, even when AUC is above
chance. AUC tests use within-stratum permutations and a ten-test Holm
family. Interpretation should use the unrounded estimates; the displayed
values are rounded for presentation.

Voynich and Egyptian retain both discrimination and positive predictive
information after pair embargo: AUCs are approximately 0.843 and 0.888,
with gains of 0.117 and 0.581 bits per candidate. Indus retains ranking
(AUC 0.761) but loses 0.075 bits per gap under the refitted
probabilities. This identifies calibration and training support as
concrete limitations of the present model rather than erasing the
boundary association.

Khipu retains AUC 0.654 and a gain of 0.075 bits after pair embargo, but
only 22 training candidates remain from the original 30,155. Its coarse
feature alphabet makes this a severe support-depletion experiment. It
should therefore not be described as robust learning of a rich unseen
knot grammar. Excluding common features retains smaller but significant
discrimination (AUC 0.595) on a different subset. Rongorongo remains
indistinguishable from its matched null under both exclusions. These
results locate the next modelling tasks more precisely than a single
pooled performance ranking.

\begin{figure}[tbp]
\centering
\includegraphics[width=0.97\linewidth,height=0.58\textheight,keepaspectratio]{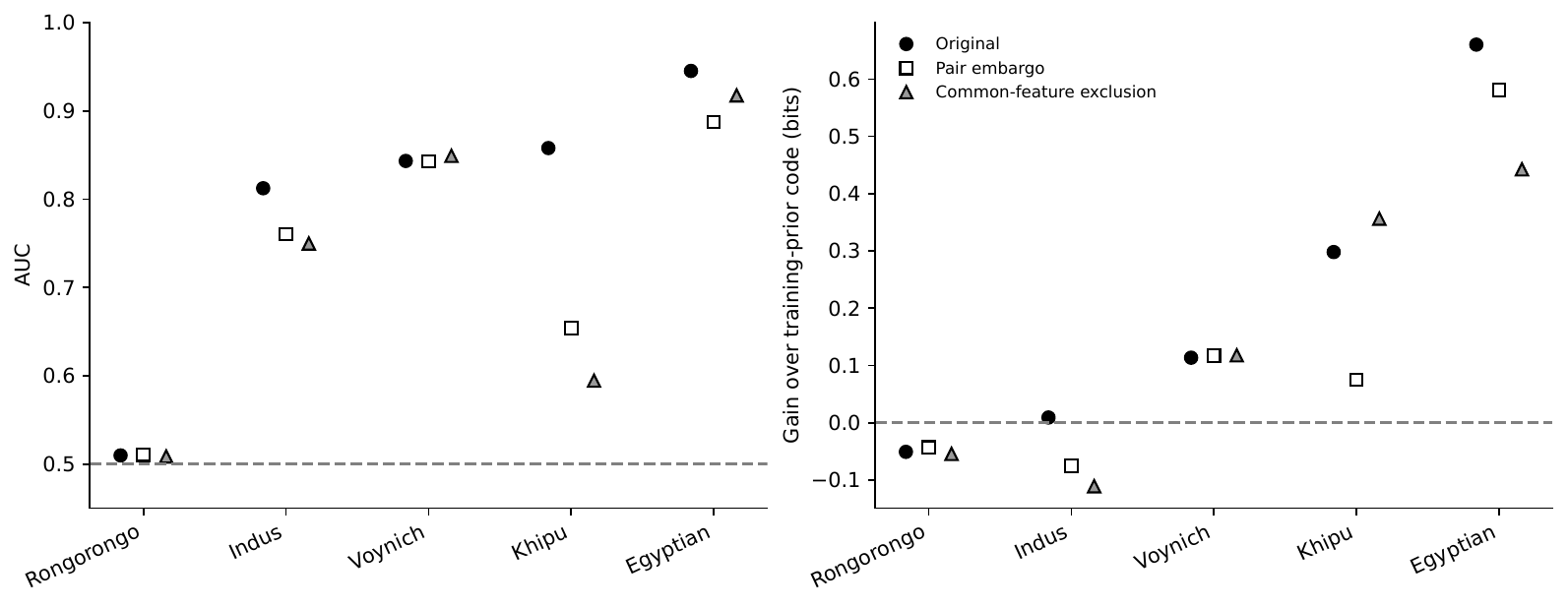}
\caption{Exclusion tests retain different parts of predictive information. Left, AUC measures ranking. Right, probability-code gain measures savings over the training-prior baseline. Circles are original results, open squares pair embargo, and triangles common-feature exclusion. Exclusion changes training support or test prevalence; differences are not controlled estimates of a single causal effect. The khipu embargo leaves only 22 training rows.}
\label{fig:11}
\end{figure}

\paragraph{Grouped-split stability of comparative structural prediction.}

\begin{figure}[htbp]
\centering
\includegraphics[
width=\linewidth,
height=0.78\textheight,
keepaspectratio
]{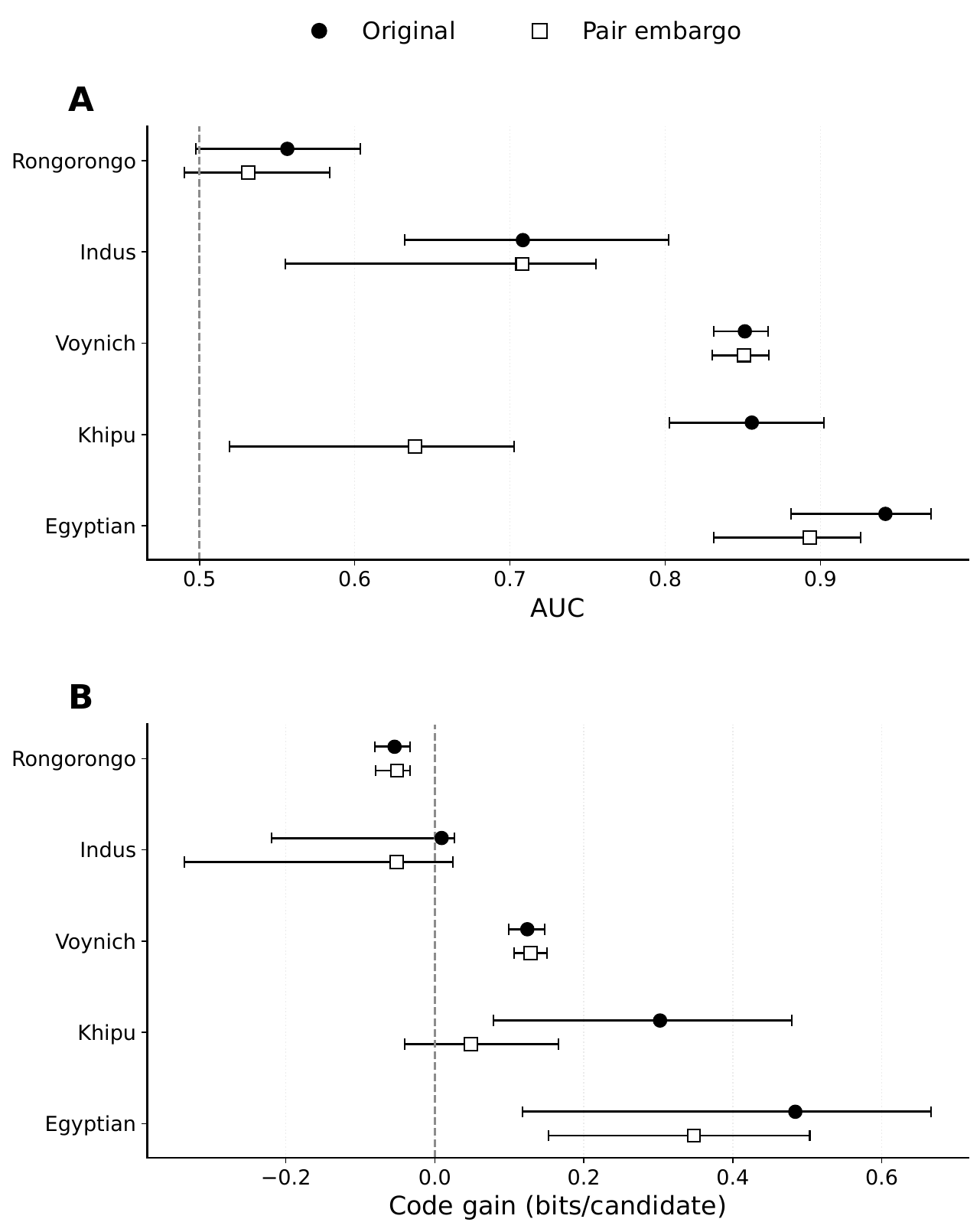}

\caption{\textbf{Grouped-split stability of comparative structural prediction.}
(A) Median area under the receiver operating characteristic curve (AUC) across grouped hold-out evaluations, comparing the original model (filled circles) with exact-pair embargo (open squares).
(B) Corresponding median predictive code gain relative to the training-prevalence baseline, expressed in bits per candidate. Horizontal error bars represent empirical 2.5th--97.5th percentiles across evaluated splits, not confidence intervals. Dashed vertical lines indicate AUC = 0.5 and zero code gain, respectively; the AUC reference is not a corpus-specific constrained-null expectation. Ten repetitions of five-fold grouped evaluation produced 50 splits for Rongorongo, Voynich, khipu and Egyptian; Indus used five eligible leave-one-site-out evaluations. Rongorongo's H/P/Q family was grouped, and Voynich evaluation excluded pages sharing exact lines with training data. The khipu embargo retained only 0.081\% of training candidates at the median. Repeated folds are dependent, and these results describe partition sensitivity rather than independent-replicate uncertainty.}

\label{fig:r2-grouped-stability}
\end{figure}

To evaluate sensitivity to the selection of withheld records, we repeated the comparative prediction experiments using grouped hold-out splits (Fig.~\ref{fig:r2-grouped-stability}; Supplementary Table~\ref{tab:r2-grouped-splits}). Ten repetitions of five-fold grouped evaluation yielded 50 splits for Rongorongo, Voynich, khipu and Egyptian, while five eligible leave-one-site-out evaluations were performed for Indus. Reimplementation of the original splits gave AUC discrepancies below 0.001, confirming numerical agreement with the original implementation.

Voynich retained nearly identical median AUCs before and after exact-pair embargo (0.851 and 0.851), with corresponding predictive code gains of 0.124 and 0.128 bits per candidate. Egyptian AUC decreased from 0.942 to 0.893, while code gain remained positive (0.484 and 0.348 bits per candidate). Khipu exhibited a larger reduction in median AUC (0.856 to 0.639) and code gain (0.302 to 0.049 bits per candidate). However, the khipu embargo retained a median of only 25 training candidates, approximately 0.081\% of the original training set, making this an extreme support-depletion experiment rather than an isolated measure of sensitivity to shared pairs.

Indus retained a similar median AUC (0.708 in both conditions), but median code gain declined from 0.009 to -0.051 bits per candidate; only two of five embargoed site evaluations had positive code gain. Rongorongo showed median AUCs of 0.557 and 0.532, with negative median code gains in both conditions (-0.054 and -0.051 bits per candidate). These results distinguish stable ranking from useful probabilistic prediction and show corpus-dependent sensitivity to the selection of withheld records and exclusion of overlapping local features.

\subsection{Rongorongo traversal survives removal of the parallel-tablet
family}\label{rongorongo-traversal-survives-removal-of-the-parallel-tablet-family}

We next reuse Phaistos's multi-estimator strategy on a structurally
appropriate question: consistent traversal of Rongorongo lines. The same
999 independent line-reversal masks are evaluated by KT1, LZW and
one-hot CTM-BDM at block lengths 4, 5 and 6. Removing the H/P/Q parallel
family leaves 165 lines and 7,604 tokens, providing a substantive
shared-family ablation rather than deleting exact duplicate whole lines,
of which there are none in this transcription.

\FloatBarrier
{
\small
\setlength{\tabcolsep}{3pt}
\renewcommand{\arraystretch}{1.12}
\begin{longtable}{@{}>{\raggedright\arraybackslash}p{\dimexpr 0.198473\linewidth-5.9542pt\relax}>{\raggedright\arraybackslash}p{\dimexpr 0.190840\linewidth-5.7252pt\relax}>{\raggedright\arraybackslash}p{\dimexpr 0.152672\linewidth-4.5802pt\relax}>{\raggedright\arraybackslash}p{\dimexpr 0.152672\linewidth-4.5802pt\relax}>{\raggedright\arraybackslash}p{\dimexpr 0.152672\linewidth-4.5802pt\relax}>{\raggedright\arraybackslash}p{\dimexpr 0.152672\linewidth-4.5802pt\relax}@{}}
\caption{Rongorongo orientation under identical reversal interventions.}\label{tab:main-17}\\
\toprule
Selection & Measure & Observed / token & Null excess & p & Holm p \\
\midrule
\endfirsthead
\toprule
Selection & Measure & Observed / token & Null excess & p & Holm p \\
\midrule
\endhead
\bottomrule
\endlastfoot
All lines & KT1 & 8.12 & 0.10 & \(1.00\times10^{-3}\) & \(1.00\times10^{-2}\) \\
All lines & LZW & 8.90 & 0.26 & \(1.00\times10^{-3}\) & \(1.00\times10^{-2}\) \\
All lines & BDM 4 & 2.52 & \(1.95\times10^{-4}\) & 0.31 & 1.00 \\
All lines & BDM 5 & 3.41 & \(1.28\times10^{-3}\) & 0.18 & 1.00 \\
All lines & BDM 6 & 4.76 & \(1.38\times10^{-4}\) & 0.48 & 1.00 \\
H/P/Q removed & KT1 & 8.07 & 0.08 & \(1.00\times10^{-3}\) & \(1.00\times10^{-2}\) \\
H/P/Q removed & LZW & 8.93 & 0.18 & \(1.00\times10^{-3}\) & \(1.00\times10^{-2}\) \\
H/P/Q removed & BDM 4 & 3.41 & \(1.57\times10^{-5}\) & 0.46 & 1.00 \\
H/P/Q removed & BDM 5 & 4.52 & \(-1.05\times10^{-3}\) & 0.78 & 1.00 \\
H/P/Q removed & BDM 6 & 5.95 & \(-4.39\times10^{-3}\) & 0.94 & 1.00 \\
\end{longtable}
}

All lines: 229 lines and 11,666 tokens. Family removed: 165 lines and
7,604 tokens. BDM uses overlapping within-line binary windows pooled by
sign channel. It is a complexity proxy, not an emitted code rate. The
ten comparisons constitute one Holm family.

\begin{figure}[tbp]
\centering
\includegraphics[width=0.97\linewidth,height=0.58\textheight,keepaspectratio]{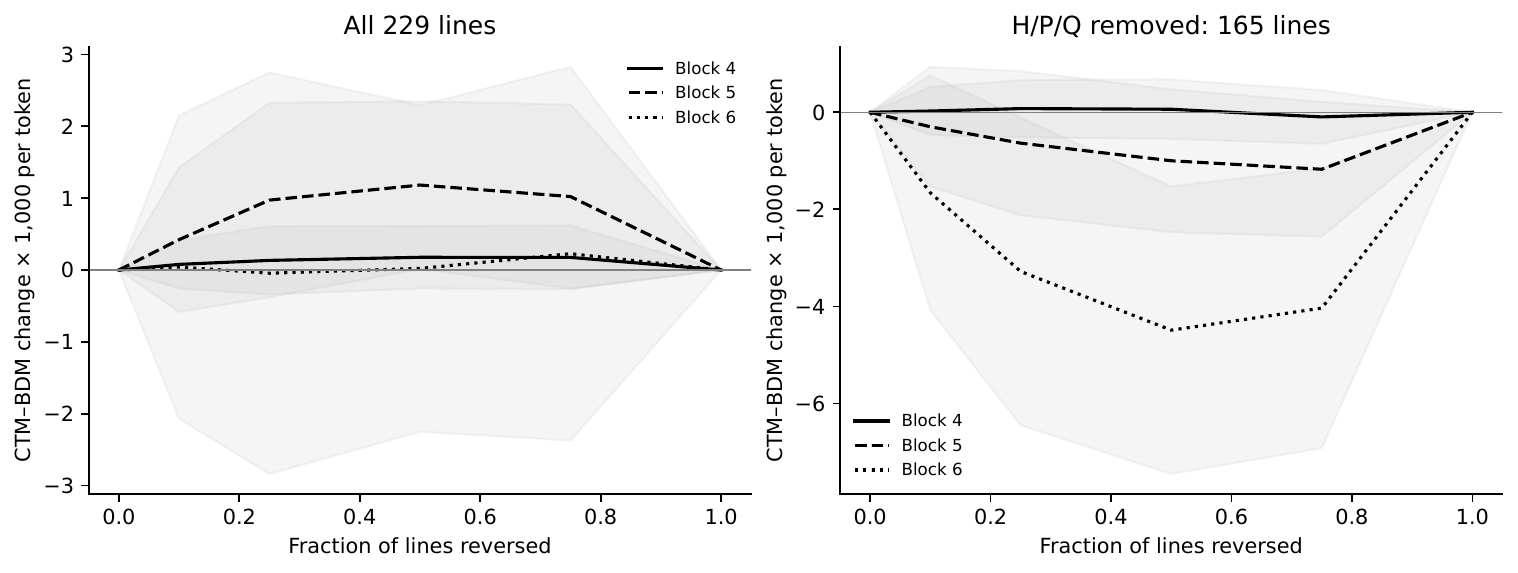}
\caption{Finite CTM-BDM traversal responses at three block lengths. Curves are mean changes relative to the conventional transcription; bands show one simulation standard deviation. A complete reversal returns exactly to zero by construction. Left, all lines; right, H/P/Q removed. The table reports matched comparisons with KT1 and LZW under identical random masks. These curves do not identify an absolute historical reading direction.}
\label{fig:12}
\end{figure}

KT1 and LZW prefer consistent traversal both before and after family
removal (all adjusted \(p=0.010\)). The three finite CTM-BDM scales do
not distinguish the authentic orientation from mixed-line nulls in this
experiment (all adjusted \(p=1\)). They nevertheless provide an exact
mathematical control: reversing every line leaves the implemented score
unchanged. This follows from the symmetry of the CTM table and the
bijection between every window and its reversal (Figure~\ref{fig:12}). The result
characterises which descriptions respond to the tested traversal change;
it is not a general ranking of CTM-BDM against other methods. In
particular, the successful calibrated BDM reconstruction of Arecibo
addresses a different spatial dependency.

\subsection{Row layout alone does not supply a universal geometric
signal}\label{row-layout-alone-does-not-supply-a-universal-geometric-signal}

A direct extension of the Phaistos cross-winding test asks whether
aligned interiors of neighbouring rows preferentially contain the same
categories. We use relative token centres, conceal alignment through
coherent circular shifts of each row interior, and retain endpoints,
inventories and cyclic interior order. This targets cross-row
information that is additional to simple row-end preferences.

\FloatBarrier
{
\small
\setlength{\tabcolsep}{3pt}
\renewcommand{\arraystretch}{1.12}
\begin{longtable}{@{}>{\raggedright\arraybackslash}p{\dimexpr 0.142857\linewidth-5.1429pt\relax}>{\raggedright\arraybackslash}p{\dimexpr 0.142857\linewidth-5.1429pt\relax}>{\raggedright\arraybackslash}p{\dimexpr 0.142857\linewidth-5.1429pt\relax}>{\raggedright\arraybackslash}p{\dimexpr 0.142857\linewidth-5.1429pt\relax}>{\raggedright\arraybackslash}p{\dimexpr 0.142857\linewidth-5.1429pt\relax}>{\raggedright\arraybackslash}p{\dimexpr 0.142857\linewidth-5.1429pt\relax}>{\raggedright\arraybackslash}p{\dimexpr 0.142857\linewidth-5.1429pt\relax}@{}}
\caption{Relative cross-row alignment under interior-shift nulls.}\label{tab:main-18}\\
\toprule
Record & Selection & Row pairs & Edges & Equality & Null mean & Holm p \\
\midrule
\endfirsthead
\toprule
Record & Selection & Row pairs & Edges & Equality & Null mean & Holm p \\
\midrule
\endhead
\bottomrule
\endlastfoot
Rongorongo & All & 213 & 10670 & 0.02 & 0.02 & 0.49 \\
Rongorongo & Deduplicated & 213 & 10670 & 0.02 & 0.02 & 0.49 \\
Indus & All & 9 & 23 & 0 & 0 & 1.00 \\
Indus & Deduplicated & 9 & 23 & 0 & 0 & 1.00 \\
Voynich & All & 3592 & 25241 & 0.01 & 0.01 & 1.00 \\
Voynich & Deduplicated & 3592 & 25241 & 0.01 & 0.01 & 1.00 \\
\end{longtable}
}

Voynich equality refers to the existing word-form tuple, not full-word
identity. Exact ordered row-pair deduplication removes no additional
eligible pairs in these selected samples; independent null draws account
for minor differences in their probabilities. Indus has zero observed
and null equality, making this particular statistic uninformative.
Spatial Holm adjustment jointly covers these six comparisons, two
attachment tests and one recovery test.

Rongorongo's authentic equality rate is 0.02493 against 0.02277 under
the null, but this is not significant after the defined correction.
Voynich's word-form equality is similarly unexceptional. Only nine Indus
row pairs meet the interior-length requirement, yielding 23 edges; all
observed and randomised equality scores are zero. For Indus, this is a
failure of the statistic to obtain an informative comparison from the
available selection, rather than evidence that its writing lacked
geometric organisation. More flexible sign correspondences or measured
coordinates would constitute new hypotheses requiring new validation.

\begin{figure}[tbp]
\centering
\includegraphics[width=0.97\linewidth,height=0.58\textheight,keepaspectratio]{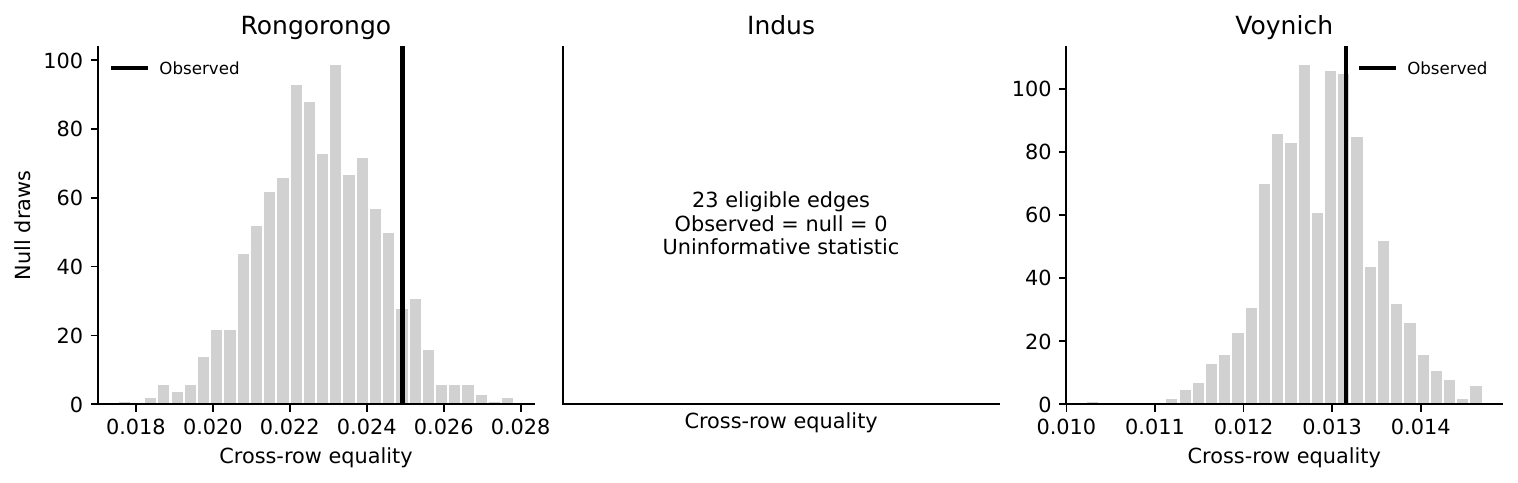}
\caption{Cross-row equality against coherent interior-shift nulls. Histograms contain 999 null realisations; vertical lines mark authentic alignment. The Indus selection yields no equality under any sampled arrangement, so a histogram would conceal its zero-variance limitation. These tests concern the stated relative-coordinate model, not every possible spatial relation.}
\label{fig:13}
\end{figure}

\subsection{Khipu knot forms carry information about actual cord
attachments}\label{khipu-knot-forms-carry-information-about-actual-cord-attachments}

The direct topological extension concerns branching relations.
Unlike row-break labels, the khipu database records explicit cord-parent
identities. We represent each knotted cord by its raw counts of S, L and
E knot forms, and compare the cosine similarity of connected parent and
child profiles with degree-preserving reassignments. The operation keeps
every child at its recorded depth, on its own khipu, and with one
parent; every parent's number of child attachments is retained. The
stricter model additionally matches the child's fibre and twist.

\begin{figure}[tbp]
\centering
\includegraphics[width=0.97\linewidth,height=0.58\textheight,keepaspectratio]{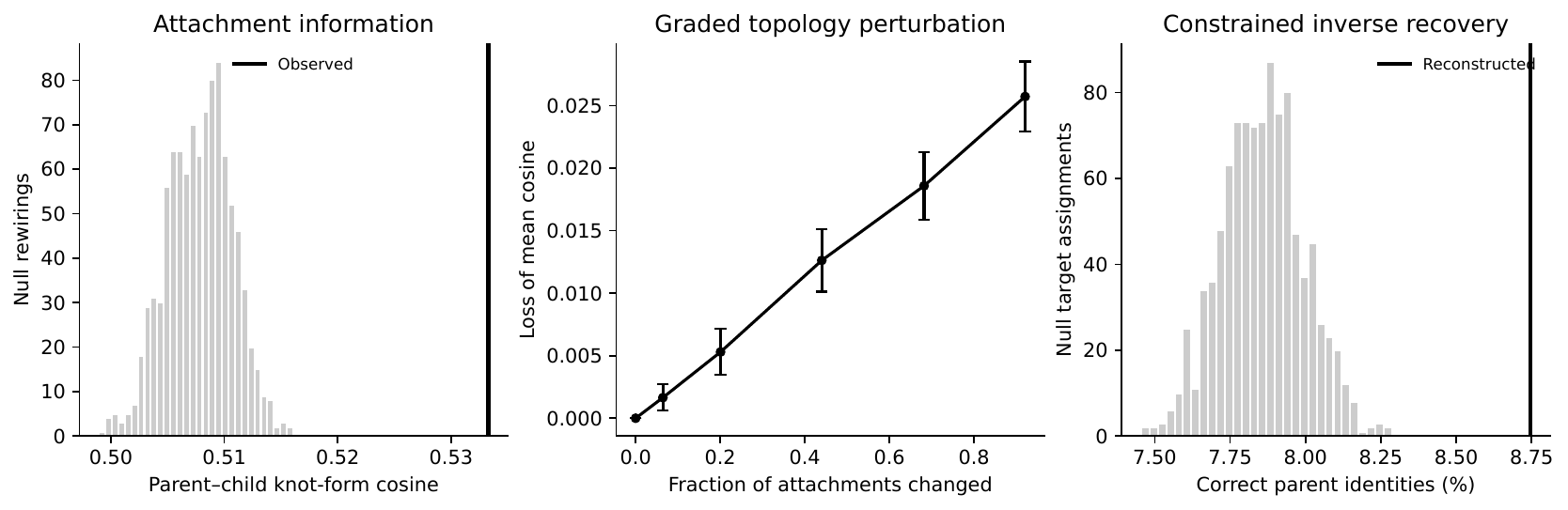}
\caption{Khipu attachment information and its inverse test under fibre/twist controls. Left, authentic parent--child cosine compared with 999 constrained rewirings. Centre, mean affinity loss against the actual fraction of changed attachments; error bars show one simulation standard deviation. Right, parent recovery compared with 999 constrained target assignments. The statistically detectable inverse gain is small in absolute terms and conditions on known parent capacities.}
\label{fig:14}
\end{figure}

Of 8,959 eligible knotted parent--child links before rewiring
restrictions, 8,812 admit the basic comparison and 8,780 admit the
stricter comparison. Under the latter, observed affinity is 0.53329
against a null mean of 0.50784 (\(z=8.89\); adjusted \(p=0.009\)).
Reassigning progressively more attachments reduces the mean affinity
(Figure~\ref{fig:14}). Thus knot-form content is associated with the authentic
branching relations beyond the degrees, depth and material categories
retained by this null. Manufacturing, accounting and copying mechanisms
remain possible explanations; cosine affinity alone does not identify
their historical direction of causation.

\FloatBarrier
{
\small
\setlength{\tabcolsep}{3pt}
\renewcommand{\arraystretch}{1.12}
\begin{longtable}{@{}>{\raggedright\arraybackslash}p{\dimexpr 0.142857\linewidth-5.1429pt\relax}>{\raggedright\arraybackslash}p{\dimexpr 0.142857\linewidth-5.1429pt\relax}>{\raggedright\arraybackslash}p{\dimexpr 0.142857\linewidth-5.1429pt\relax}>{\raggedright\arraybackslash}p{\dimexpr 0.142857\linewidth-5.1429pt\relax}>{\raggedright\arraybackslash}p{\dimexpr 0.142857\linewidth-5.1429pt\relax}>{\raggedright\arraybackslash}p{\dimexpr 0.142857\linewidth-5.1429pt\relax}>{\raggedright\arraybackslash}p{\dimexpr 0.142857\linewidth-5.1429pt\relax}@{}}
\caption{Authentic cord-parent affinity under constrained topological perturbation.}\label{tab:main-19}\\
\toprule
Constraints & Khipu & Links & Cosine & Null mean & z & Holm p \\
\midrule
\endfirsthead
\toprule
Constraints & Khipu & Links & Cosine & Null mean & z & Holm p \\
\midrule
\endhead
\bottomrule
\endlastfoot
Khipu + depth & 311 & 8812 & 0.53 & 0.51 & 9.07 & \(9.00\times10^{-3}\) \\
Also fibre + twist & 310 & 8780 & 0.53 & 0.51 & 8.89 & \(9.00\times10^{-3}\) \\
\end{longtable}
}

All comparisons preserve parent outdegrees, child indegrees, khipu
identity and depth. Fibre/twist strata are defined by the child. Only
strata with at least two distinct eligible parents enter. Affinity uses
raw knot counts, without interpreted numerical values.

A significant attachment contrast does not guarantee accurate inverse
reconstruction. We therefore conceal the child--parent assignment and
solve the maximum-affinity allocation subject to the retained parent
degrees and strict strata. Averaged over 99 independent tie-breaking
solutions, the reconstruction identifies 8.747\% of authentic parents,
compared with 7.866\% expected from the retained degrees alone. Against
999 constrained target permutations, the observed ensemble recall has
\(z=6.52\) and adjusted \(p=0.009\) (Table~\ref{tab:main-20}).

The gain is approximately 0.88 percentage points: genuine partial
recovery under the stated randomisation model, but far from complete
reconstruction. This is a conditional inverse task with parent
candidates and capacities supplied, not discovery of an unknown complete
tree. Its value is that the same raw content yields both a perturbation
signal and an above-null reconstruction of the hidden relation. 

\FloatBarrier
{
\small
\setlength{\tabcolsep}{3pt}
\renewcommand{\arraystretch}{1.12}
\begin{longtable}{@{}>{\raggedright\arraybackslash}p{\dimexpr 0.142857\linewidth-5.1429pt\relax}>{\raggedright\arraybackslash}p{\dimexpr 0.142857\linewidth-5.1429pt\relax}>{\raggedright\arraybackslash}p{\dimexpr 0.142857\linewidth-5.1429pt\relax}>{\raggedright\arraybackslash}p{\dimexpr 0.142857\linewidth-5.1429pt\relax}>{\raggedright\arraybackslash}p{\dimexpr 0.142857\linewidth-5.1429pt\relax}>{\raggedright\arraybackslash}p{\dimexpr 0.142857\linewidth-5.1429pt\relax}>{\raggedright\arraybackslash}p{\dimexpr 0.142857\linewidth-5.1429pt\relax}@{}}
\caption{Constrained recovery of concealed parent identities.}\label{tab:main-20}\\
\toprule
Links & Tie solutions & Recall & Degree chance & Null mean & z & Holm p \\
\midrule
\endfirsthead
\toprule
Links & Tie solutions & Recall & Degree chance & Null mean & z & Holm p \\
\midrule
\endhead
\bottomrule
\endlastfoot
8780 & 99 & 0.09 & 0.08 & 0.08 & 6.52 & \(9.00\times10^{-3}\) \\
\end{longtable}
}

Recall averages 99 assignments obtained by tiny independent tie-breaking
jitters. Single-solution recalls range from 0.08166 to 0.09237.
Candidate parent identities, degrees and strict strata are retained.
Chance is degree-weighted rather than the reciprocal of a global parent
count.

\clearpage

\subsection{Known executable mechanisms separate geometric dependence
from its
carrier}\label{known-executable-mechanisms-separate-geometric-dependence-from-its-carrier}

Finally, we intervene on a known generator rather than only on a
surviving representation. Synthetic row records and ternary trees either
copy the aligned predecessor or parent with probability \(c=0.8\), or
generate independent uniform symbols at \(c=0\). Changing \(c\) edits
the executable mechanism while keeping topology, dimensions, alphabet
and expected one-token frequencies fixed. This supplies a concrete
finite example of a computable hypothesis intervention in software
space.

\FloatBarrier
{
\small
\setlength{\tabcolsep}{3pt}
\renewcommand{\arraystretch}{1.12}
\begin{longtable}{@{}>{\raggedright\arraybackslash}p{\dimexpr 0.200000\linewidth-4.8000pt\relax}>{\raggedright\arraybackslash}p{\dimexpr 0.200000\linewidth-4.8000pt\relax}>{\raggedright\arraybackslash}p{\dimexpr 0.200000\linewidth-4.8000pt\relax}>{\raggedright\arraybackslash}p{\dimexpr 0.200000\linewidth-4.8000pt\relax}>{\raggedright\arraybackslash}p{\dimexpr 0.200000\linewidth-4.8000pt\relax}@{}}
\caption{Calibration under interventions on known copying mechanisms.}\label{tab:main-21}\\
\toprule
Geometry & Copy probability & Records & \(p\leq 0.05\) & Mean excess \\
\midrule
\endfirsthead
\toprule
Geometry & Copy probability & Records & \(p\leq 0.05\) & Mean excess \\
\midrule
\endhead
\bottomrule
\endlastfoot
Rows & 0.00 & 99 & 1 & \(3.38\times10^{-4}\) \\
Rows & 0.80 & 99 & 99 & 0.70 \\
Tree & 0.00 & 99 & 6 & \(3.68\times10^{-3}\) \\
Tree & 0.80 & 99 & 99 & 0.31 \\
\end{longtable}
}

Each of 99 independent records has 199 nulls. Row controls: six rows of
width 20, alphabet 12. Tree controls: ternary depth four, alphabet
three. The positive condition changes dependency on geometry; the
geometry itself remains present in both conditions.

All 99 coupled records are detected for each geometry. Under independent
symbols, the corresponding counts are one of 99 row records and six of
99 tree records. These finite controls show sensitivity to a specified
geometry-dependent mechanism and approximately controlled false-positive
behaviour for its independent counterpart. They do not establish
universal recovery of all geometric codes. Together with the historical
tests, they sharpen the central claim: recoverable spatial information
depends on coupling between content and organisation, and cannot be
inferred merely from the visible shape of a carrier.

\begin{figure}[tbp]
\centering
\includegraphics[width=0.97\linewidth,height=0.58\textheight,keepaspectratio]{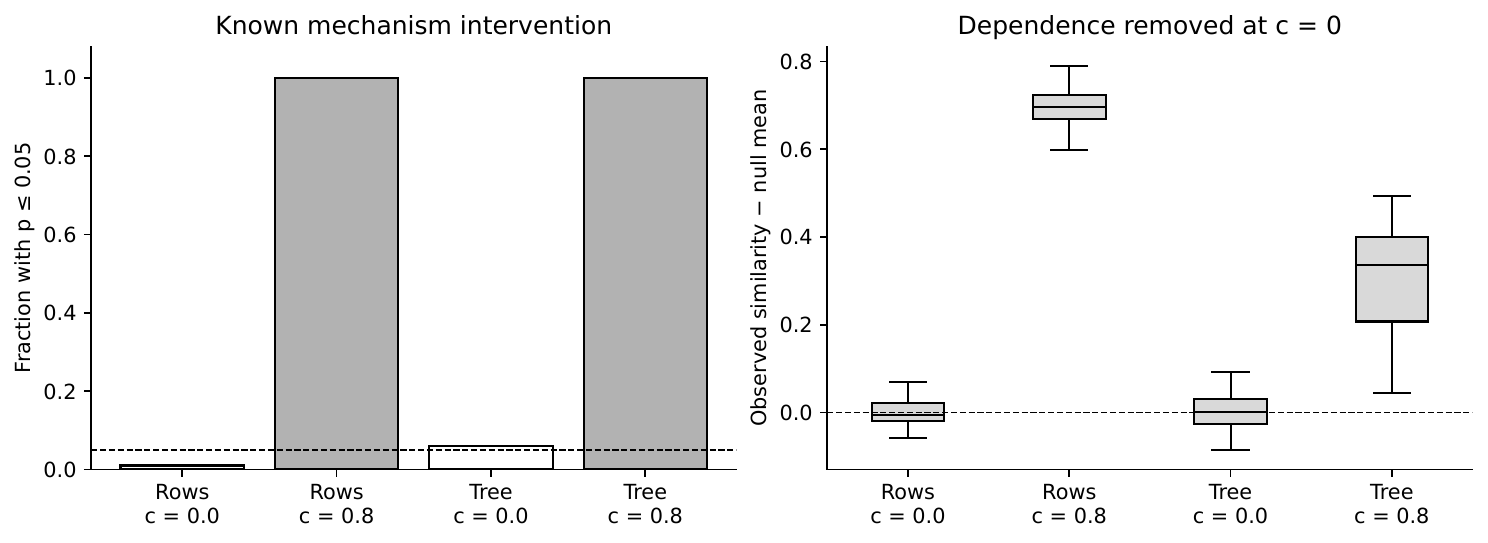}
\caption{An executable intervention removes dependence while retaining geometry. Left, empirical rejection fractions for 99 records per condition, with a reference at the nominal 0.05 threshold. Right, distributions of excess similarity relative to each record's constrained null. The copying probability is a known mechanism parameter; this causal interpretation is available for the synthetic controls because the generator is specified.}
\label{fig:15}
\end{figure}

\clearpage

\section{Discussion}\label{discussion}

\subsection{What extends from Arecibo to historical one-way
records}\label{what-extends-from-arecibo-to-historical-one-way-records}

The common experimental object is a relation that has been concealed. In
Arecibo that relation is a rectangular arrangement of bits; in Phaistos
it includes compartment membership; in khipu it is a distal position or
a cord-parent attachment; in manuscripts it is a line or vertical join.
This makes the extension more precise than applying an information
measure to several unrelated corpora. In each case, the question is
whether symbolic data constrain a structural variable that the algorithm
is not given directly.

The strongest comparative lesson is selective recoverability. Arecibo's
intended width is recoverable under the calibrated finite estimator.
Phaistos supplies strong compartment and local-dependency information,
but its idealised cross-winding tests remain negative. Khipu knot forms
predict a coarse spatial rank. Voynich forms predict line breaks, while
Rongorongo's consistent traversal does not produce reliable boundary
transfer to the parallel-text family. These outcomes show that visible
geometry, recoverable structure and readable meaning are distinct
empirical properties.

Several complementary Phaistos analyses converge without every
estimator agreeing. Positional controls
remove simple endpoint-frequency explanations; duplicate-family
prediction removes complete-compartment copying from the training set;
boundary transfer tests cross-face organisation; and spectral ablation
identifies a concrete source of long-range recurrence. At the same time,
the six-bit CTM response is not significant on either face separately
and the pooled second-order predictive increment vanishes under the
stricter duplicate-family split. This pattern supports constrained local
organisation with substantial formulaic reuse. It does not determine
whether the compartments encode speech, names, lists, ritual formulas or
another symbolic practice.

\subsection{Geometry can be a carrier, constraint, address or
operator}\label{geometry-can-be-a-carrier-constraint-address-or-operator}

Four generative roles provide useful hypotheses. Geometry acts as a
carrier when it holds marks whose relevant dependencies follow a path.
It acts as a constraint when available space changes which expression is
chosen or how it is abbreviated. It acts as an address when a location
selects the significance of a mark, as spatial rank can do in numerical
records. It acts as an operator when a relation such as above, beside,
enclosed by or attached to changes how components combine. These
categories concern mechanisms; they are not labels established merely by
a significant AUC.

Mutual information alone cannot choose among the causal possibilities,
because \(I(X;G)=I(G;X)\). Content determining layout and layout
determining content can yield the same association. A promising next
step is therefore a paired natural experiment: compare repeated or
closely parallel content on differently shaped carriers, and compare
similar carriers carrying different content. If symbol substitutions
track available line capacity after content-family control, that
supports a geometric constraint. If a spatial relation predicts a
numerical function across independent records, that supports an address
or operator. AID perturbations can specify the contrasts, while
epigraphic or physical evidence identifies which contrasts are
historically meaningful.

\subsection{Structural evidence can restrict decipherment without
supplying phonetic
values}\label{structural-evidence-can-restrict-decipherment-without-supplying-phonetic-values}

For Phaistos, a proposed reading should explain the transferred boundary
preferences and repeated modules on both faces, including contexts
without signs 02 and 12. It should also avoid treating the lag-25
repetition as independent evidence for a periodic key. For Indus, a
row-final hypothesis should predict withheld row breaks after
controlling inscription family, site and available space. The small
probabilistic code gain suggests that better-calibrated, externally
validated models are needed before turning ranking performance into
strong quantitative evidence for a specific sign function.

For khipu, the new attachment perturbation and constrained recovery
identify branching relations that are partly encoded in raw knot forms,
without using interpreted numerical values. The next decisive task is
reconstruction of arithmetic aggregation across independent cord
families after withholding all fields derived from interpreted values.
The present 0.88-percentage-point attachment gain supplies a measurable
starting constraint, rather than an arithmetic decipherment. For
Rongorongo, parallel passages permit tests of substitutions and
omissions at aligned structural positions, provided entire passage
families remain separated between training and evaluation. A previously identified calendrical passage, selected in advance,
could provide a labelled control. For Voynich, the useful
question is whether line-conditioned form changes persist within scribal
hand, illustration type and near-parallel lexical contexts.

A candidate sign mapping \(\phi\) can ultimately be assessed by a
penalised predictive comparison,

\begin{equation}
\label{eq:8}
\Delta L_\phi=L_0(X_{\mathrm{test}})-\left[L(\phi)+L(X_{\mathrm{test}}\mid Y_{\mathrm{train}},\phi)\right].
\end{equation}

\(Y\) is an external corpus and \(L(\phi)\) charges for specifying the
mapping. A positive gain would mean that an explicit mapping explains
withheld material economically after its flexibility is accounted for.
This is a proposed next-stage criterion, not a phonetic inference
computed in the present study. The same separation between training and
evaluation must include any search over languages, mappings and model
settings.

\subsection{Further spatial configurations and evidential
limits}\label{further-spatial-configurations-and-evidential-limits}

Maya inscriptions offer ordered blocks with internal arrangements;
cuneiform tablets offer ruled columns, tabular accounts and changes in
orientation; Ogham offers edge-oriented strokes; and cylindrical or
circular carriers require distinctions between a linearised reading, a
cyclic order and a periodic process. These cases could challenge the
reconstruction framework in ways that another flat sign stream cannot.
They are proposed extensions and are not counted among the analysed
datasets.

The current evidence comes from heterogeneous digital transcriptions.
Egyptian operators are editorial encodings of layout; Indus row
delimiters are not measured coordinates; Rongorongo order is already
normalised; and the khipu labels use recorded cluster positions. These
are legitimate structural targets, but recovering them is not equivalent
to recovering an untouched ancient object from raw photographs.
Physical-coordinate studies and independent epigraphic checks would
strengthen the archaeological interpretation.

The numerical programme is also exploratory. The comparative AUC
intervals resample held-out objects while holding the trained predictor
fixed; they do not incorporate retraining variability or all
transcription uncertainty. Search corrections cover the scans specified
in each experiment, and the sign-wise correction covers its defined
family. They do not retroactively correct every modelling choice made
across the research programme. The most persuasive conclusions are
therefore the specific reproducible contrasts supported by distinct
methods and transparent negative results.

The grouped-split analysis further clarifies the limits of structural generalisation. Voynich retains similar discrimination and positive predictive code gain across alternative page partitions after exact-pair exclusion, indicating that its observed line-boundary association is not restricted to the original fixed split or exact local feature-pair reuse. Egyptian retains positive predictive information but exhibits reduced discrimination after exclusion, consistent with a contribution from recurring sign-pair combinations. Indus illustrates that ranking and probabilistic information must be distinguished: its median AUC remains above 0.5 after embargo, whereas median code gain becomes negative. Rongorongo similarly lacks positive predictive code gain across the evaluated partitions.

The khipu results require particular qualification. Although positive median discrimination and code gain persist after pair embargo, the intervention eliminates more than 99.9\% of training candidates. Consequently, the observed reduction cannot be attributed solely to the removal of shared local information; it also reflects extreme loss of training support. More informative tests will require richer raw knot-form representations and overlap controls that preserve adequate sample sizes. Overall, repeated grouped evaluation characterises sensitivity to the choice of withheld records but does not establish semantic decipherment, eliminate transcription uncertainty or provide independent experimental replication.

\section{Methods}\label{methods}

\subsection{Data provenance and units of
independence}\label{data-provenance-and-units-of-independence}

All Phaistos numerical tables are generated from the Phaistos
machine-readable outputs or from the new classical calculations supplied
here. Its provisional damaged slot is assigned sign 03 in the main
analysis. Catalogue item 46 in the stored transcription denotes an
oblique stroke, which is removed from the 45-type sign stream and
retained as a group annotation. Pooled sequential codes concatenate the
two face streams; within-group MI and predictive codes do not create
transitions across dividers. The analysis scripts record these different
conventions explicitly.

The comparative sources are the Open Khipu Repository database, Rongopy
tablet transcriptions, a derivative Indus inscription CSV, the
Zandbergen-Landini EVA transcription and Manuel de Codage text editions
\cite{openKhipu,rongopy,indusCorpus,reddy2011voynich,evaVoynich,rosmorducMdcDoc,rosmorducMdcTexts}. Exact reproduction requires the source revisions and SHA-256
fingerprints of the analysed input files. Khipu predictors use only counts of raw S, L and E knot types and
cluster size capped at 12; these denote single, long and figure-eight
knot forms. Distances and identifiers define targets, grouping and
splits only. The farthest eligible cluster is labelled positive after
excluding cords with fewer than two eligible clusters or duplicated
cluster positions.

Indus processing requires complete records, more than one nonempty
slash-delimited row and no unknown \texttt{000} sign. Identical
flattened sequences are collapsed, retaining their site sets and
observed layouts. Only unambiguous layouts enter the site holdout.
Voynich processing retains body-text lines, resolves the recorded first
alternative in alternative transcriptions, removes marked uncertain
words and uses the stated form tuples. Egyptian extraction uses only
explicit nonoverlapping adjacent sign-code pairs separated by colon,
star or dash; aliases, inferred operators and full nested parses are
excluded. These inclusion rules define the analysed corpora and should
be distinguished from total surviving historical corpora.

\subsection{Entropy, dependence and finite predictive
codes}\label{entropy-dependence-and-finite-predictive-codes}

For empirical category proportions \(\hat p_s=n_s/n\), Shannon entropy
is

\begin{equation}
\label{eq:9}
\widehat H_0(X)=-\sum_s\hat p_s\log_2\hat p_s.
\end{equation}

A bit is a base-two unit of information. Entropy measures uncertainty in
a distribution; it does not assess the meaning of a sign. The entropy
rate of a stationary process is a limiting conditional uncertainty,

\begin{equation}
\label{eq:10}
h=\lim_{m\to\infty}H(X_t\mid X_{t-1},\ldots,X_{t-m}).
\end{equation}

No such limit is identified from the Disc. The quantities labelled KT0,
KT1 and KT2 are finite Bayesian coding costs, where order 0 uses sign
counts, order 1 uses the preceding sign and order 2 uses two preceding
signs \cite{shannon1948,krichevsky1981}.

For a context \(c\) and alphabet size \(q\), the predictive probability
with symmetric Dirichlet smoothing is

\begin{equation}
\label{eq:11}
\begin{aligned}\widehat P(x_t=s\mid c)&=\frac{n_{c,s}+\alpha}{n_c+q\alpha},\quad\alpha=\tfrac12,\\ L_{\mathrm{KT}}&=-\sum_t\log_2\widehat P(x_t\mid c_t).\end{aligned}
\end{equation}

Counts are updated after each prediction. The prior gives unseen
categories nonzero probability. Integrating the transition probabilities
gives the equivalent context-wise factor

\begin{equation}
\label{eq:12}
P(X\mid c)=\frac{\Gamma(q\alpha)}{\Gamma(n_c+q\alpha)}\prod_s\frac{\Gamma(n_{c,s}+\alpha)}{\Gamma(\alpha)}.
\end{equation}

\(\Gamma\) is the extension of factorial to noninteger arguments. The
Rongorongo comparison starts every line with a shared start context and
estimates transitions without crossing line ends. Phaistos sequential
coding uses the contexts of its explicit flattened stream. Thus the
absolute costs across the two corpora are not directly comparable.

The separate group-withholding predictor uses hierarchical smoothing
with fixed strength 10. Its unigram distribution is Jeffreys-smoothed; a
first-order row is shrunk towards that distribution, and a second-order
row towards the first-order row:

\begin{equation}
\label{eq:13}
p_1(b\mid a)=\frac{n_{ab}+10p_0(b)}{n_a+10},\qquad p_2(c\mid a,b)=\frac{n_{abc}+10p_1(c\mid b)}{n_{ab}+10}.
\end{equation}

Only successors inside a held-out group are scored, excluding its first
sign. In duplicate-family withholding, every training group identical to
the target group is excluded. Parameters are not tuned on the reported
test predictions.

For two categorical variables \(U,V\), empirical mutual information is

\begin{equation}
\label{eq:14}
\widehat I(U;V)=\sum_{u,v}\hat p_{uv}\log_2\frac{\hat p_{uv}}{\hat p_u\hat p_v}.
\end{equation}

The within-compartment calculation pools adjacent pairs. A finite sample
can yield positive MI even under an independent generator, particularly
when most pairs are rare. We therefore report both raw MI and excess
above the matched-null mean. The null, not a zero-MI reference, supplies
the significance comparison.

\subsection{Dictionary and grammar
compression}\label{dictionary-and-grammar-compression}

LZW builds a dictionary as it scans a sequence, replacing previously
encountered strings by dictionary references \cite{welch1984}. We reuse the
token-level implementation. For successive dictionary sizes
\(d_j\) at emitted references, its dynamic-width cost has the form

\begin{equation}
\label{eq:15}
L_{\mathrm{LZW}}=\sum_{j=1}^{m}\operatorname{ceil}\bigl(\log_2\max(d_j,2)\bigr).
\end{equation}

The ceiling function rounds up to the nearest integer. The stored
implementation fixes the alphabet before coding. Reported Phaistos
scores use alphabet sizes 45 for signs and 90 for sign-boundary pairs.
Rongorongo line coding includes a separator category and divides by the
number of original sign tokens, so separator costs remain in its
numerator. Categorical identities are canonicalised where required;
lengths are compared within the same representation and null family.

For LZ78 \cite{ziv1978}, the reported phrase statistic is \(c(n)\log_2c(n)/n\), where
\(c(n)\) is the number of parsed phrases. This is a finite-sample
entropy-rate proxy, not the same quantity as the implementation's
emitted-bit count. Re-Pair \cite{larsson1999} repeatedly replaces the most frequent
adjacent pair by a new grammar symbol, respecting compartment
boundaries. If \(r\) rules are created and the final start strings
contain \(m\) symbols, the reported grammar size and saving are

\begin{equation}
\label{eq:16}
G=m+2r,\qquad S_{\mathrm{RP}}=n-G.
\end{equation}

The two-symbol rule cost is deliberately a grammar-size convention. It
omits a full binary encoding of rule identities and therefore must not
be called a minimum-description-length value in bits. In the new
calculation, first-occurrence canonical relabelling precedes Re-Pair,
making deterministic tie handling invariant to renaming of the original
sign catalogue.

\subsection{CTM, BDM and their categorical
implementation}\label{ctm-bdm-and-their-categorical-implementation}

The algorithmic coding theorem relates complexity to universal
algorithmic probability,

\begin{equation}
\label{eq:17}
K(x)=-\log_2m_U(x)+O(1),
\end{equation}

where \(m_U(x)\) sums the probabilities of self-delimiting programmes
that output \(x\) \cite{liVitanyi2008,solerToscano2014}. CTM estimates short-block complexity from
finite programme-output frequencies. BDM extends such estimates to
larger objects by decomposing them into blocks and adding a multiplicity
term \cite{zenil2018bdm}.

The Phaistos CTM calculation enumerated \(16^6=16,777,216\) deterministic
three-state, two-symbol machines in the specified machine convention,
with a 21-step bound. Outputs are the visited tape intervals at halt.
For each length, counts are averaged over reversal, complement and
reverse-complement orbits, and then Jeffreys-smoothed. If
\(\bar N_{\ell,b}\) is the symmetrised count of binary block \(b\) of
length \(\ell\),

\begin{equation}
\label{eq:18}
\widehat P_\ell(b)=\frac{\bar N_{\ell,b}+1/2}{\sum_u\bar N_{\ell,u}+2^{\ell-1}},\qquad \widehat K_\ell(b)=-\log_2\widehat P_\ell(b).
\end{equation}

This is a length-conditioned finite CTM estimator. Its numerical values
are not those of the larger databases used by PyBDM. The main scalar scores are computed from the machine-output counts.
This finite enumeration does not solve general Kolmogorov complexity.

\FloatBarrier
{
\small
\setlength{\tabcolsep}{3pt}
\renewcommand{\arraystretch}{1.12}
\begin{longtable}{@{}>{\raggedright\arraybackslash}p{\dimexpr 0.333333\linewidth-4.0000pt\relax}>{\raggedright\arraybackslash}p{\dimexpr 0.333333\linewidth-4.0000pt\relax}>{\raggedright\arraybackslash}p{\dimexpr 0.333333\linewidth-4.0000pt\relax}@{}}
\caption{Phaistos short-machine output database.}\label{tab:main-22}\\
\toprule
Binary block length & Halting outputs & Distinct strings represented \\
\midrule
\endfirsthead
\toprule
Binary block length & Halting outputs & Distinct strings represented \\
\midrule
\endhead
\bottomrule
\endlastfoot
2 & 5,604,128 & 4/4 \\
3 & 1,710,992 & 8/8 \\
4 & 237,368 & 16/16 \\
5 & 17,848 & 32/32 \\
6 & 1,456 & 63/64 \\
\end{longtable}
}

Raw machine-output counts before symmetry completion and
length-conditioned smoothing. The one-bit table is set to one bit for
each binary value. Reproduction requires the transition convention, halting convention
and corresponding machine-output counts.

For a specified block partition, BDM is

\begin{equation}
\label{eq:19}
\operatorname{BDM}(x)=\sum_{b\in\mathcal B(x)}\left[\widehat K(b)+\log_2n_b\right],
\end{equation}

where \(\mathcal B(x)\) contains the distinct blocks and \(n_b\) is each
multiplicity. The categorical Phaistos estimator constructs one binary
indicator channel per sign,

\begin{equation}
\label{eq:20}
Z_s(t)=\mathbf1[x_t=s],\qquad C_b(X)=\frac1b\sum_{o=0}^{b-1}\sum_{s=1}^{q}\operatorname{BDM}_{b,o}(Z_s).
\end{equation}

Here \(b\) is block length and \(o\) the block offset; residual prefixes
and suffixes are retained using the appropriate shorter-block tables.
For sequences shorter than \(b\), a single offset is used. Summation
over all channels ensures exact invariance under arbitrary sign-label
permutations. Symmetry completion and offset averaging give reversal
invariance in this implementation. These are explicit properties of the
estimator, not empirical evidence that an inscription was read in both
directions.

The Arecibo control uses PyBDM's two-dimensional binary estimator with
its default \(4\times4\) PartitionIgnore decomposition \cite{pybdmDocs}.
Incomplete edge blocks are discarded. Every tested width therefore has
its own retained subset and padding pattern. Shape normalisation uses
the implementation's minimum and maximum bounds,

\begin{equation}
\label{eq:21}
C^{\mathrm{norm}}(A)=\frac{C(A)-C_{\min}(\operatorname{shape}A)}{C_{\max}(\operatorname{shape}A)-C_{\min}(\operatorname{shape}A)}.
\end{equation}

Normalisation does not recover discarded edge information. Independent
shape-matched randomisation controls part of this bias, while the
retained raw scan exposes the method's sensitivity.

\subsection{Boundary perturbation, prediction and positional-role
geometry}\label{boundary-perturbation-prediction-and-positional-role-geometry}

Let \(b_t=1\) when a compartment ends after sign \(t\). Pairing
\((x_t,b_t)\) produces an alphabet of at most 90 states. For a
perturbation replacing \(d\) authentic internal dividers by \(d\)
previously unoccupied gaps, define

\begin{equation}
\label{eq:22}
\mathcal L_C(d)=\mathbb E\left[\frac{C(X,B^{(d)})-C(X,B)}{n}\right].
\end{equation}

The replacement distance is half the Hamming distance between internal
boundary vectors, not necessarily a one-position physical shift. Means
are estimated over the sampled random perturbations. A separate null
permutes the authentic group-length multiset. Neither experiment assumes
that a compartment is a word.

Blind insertion scores each candidate gap against an otherwise
unsegmented representation. The consensus averages ranks across CTM-BDM,
LZW and KT1; it does not estimate a phonetic model. Cross-face
prediction instead trains a categorical naive-Bayes boundary score. With
left and right features \(u,v\), the score is

\begin{equation}
\label{eq:23}
s(u,v)=\log\frac{\pi_1}{\pi_0}+\log\frac{P(u\mid b=1)}{P(u\mid b=0)}+\log\frac{P(v\mid b=1)}{P(v\mid b=0)}.
\end{equation}

\(\pi\) is the class prior and the feature distributions receive
additive smoothing. Phaistos uses smoothing 1; the comparative
implementation uses 1/2. Dropping the constant prior does not alter AUC.
In the comparative probability analysis it is restored using the
training prevalence, giving \(q=1/(1+e^{-s})\). Khipu's two inputs are
knot-type profile and capped cluster size; Voynich's inputs are the
tuples for the words on either side. The conditional-independence
assumption is an approximation, especially where features overlap.

\paragraph{Exact-pair embargo and constrained randomisation.}

For each directional Phaistos transfer experiment, we identified all ordered adjacent-sign pairs occurring among target-face candidate gaps. Every training candidate containing one of these pairs was excluded, irrespective of its boundary label. We then refitted the existing categorical naive-Bayes boundary model with the original additive smoothing parameter of one. The test set remained unchanged, and its boundary labels were not used to construct the embargo.

We evaluated AUC, average precision, Brier score and predictive code gain relative to each model's own smoothed training-prevalence baseline. For each direction, 4,999 matched surrogate segmentations independently permuted the authentic group-length multiset of either the training or target face while preserving its sign stream. The embargo was maintained under these randomisations. Upper-tail Monte Carlo probabilities were calculated as $(1+\#\{T_{\mathrm{null}}\geq T_{\mathrm{observed}}\})/5000$.

The four primary AUC tests, comprising two transfer directions and independent training- and target-divider nulls, formed a separate Holm family. Code-gain null probabilities were exploratory and were not included in this family. The embargo controls exact ordered-pair overlap but does not eliminate longer shared passages, overlapping sign marginals or uncertainty arising from the reduced training sample.

Phaistos positional-role vectors use counts \(n_{s,I},n_{s,M},n_{s,F}\)
and \(n_s\):

\begin{equation}
\label{eq:24}
r_s=\left(\log\frac{n_{s,I}+1}{n_s-n_{s,I}+2},\ \log\frac{n_{s,F}+1}{n_s-n_{s,F}+2}\right).
\end{equation}

Only types with at least two total occurrences across faces are
retained. The two coordinates for retained signs are flattened into a
vector on each face and centred by its overall mean. Concordance is
their cosine,

\begin{equation}
\label{eq:25}
\rho_{\mathrm{role}}=\frac{\widetilde r_A\cdot\widetilde r_B}{\|\widetilde r_A\|\,\|\widetilde r_B\|}.
\end{equation}

Independent permutations of the two length multisets supply the
role-concordance null. This procedure retains each face's sequence while
altering its compartment roles.

\subsection{Predictive discrimination, information gain and
uncertainty}\label{predictive-discrimination-information-gain-and-uncertainty}

For positive scores \(s_i^+\) and negative scores \(s_j^-\), AUC is

\begin{equation}
\label{eq:26}
\operatorname{AUC}=\frac{1}{n_+n_-}\sum_{i,j}\left[\mathbf1(s_i^+>s_j^-)+\tfrac12\mathbf1(s_i^+=s_j^-)\right].
\end{equation}

It measures ranking, not calibration or the fraction of authentic
boundaries recovered. Average precision sums precision at score
thresholds, weighted by increases in recall; its reference level depends
on positive prevalence. Top-\(K\) recovery counts positives among the
\(K\) highest scores. The Phaistos transfer implementation averages
selection within a tied cutoff, so some reported overlaps are fractional
expectations.

To express predictive information in bits, we use the binary log loss
for label \(y_i\) and predicted probability \(q_i\),

\begin{equation}
\label{eq:27}
\begin{aligned}\ell(y_i,q_i)&=-y_i\log_2q_i-(1-y_i)\log_2(1-q_i),\\ \Delta\ell&=\frac1N\sum_i[\ell(y_i,\pi_{\mathrm{train}})-\ell(y_i,q_i)].\end{aligned}
\end{equation}

A positive \(\Delta\ell\) means the fixed model encodes the withheld
labels more economically than the constant training-based predictor. It
is not an unbiased estimator of mutual information. Brier score,
\(N^{-1}\sum_i(q_i-y_i)^2\), checks squared probability error. These
metrics explain why ranking and information gain can differ.

The uncertainty analysis resamples held-out physical or editorial
units 1,999 times and reports the 2.5th and 97.5th AUC percentiles.
Khipu are resampled as whole khipu, not as individual cords; Indus as
inscriptions; Voynich as pages; Egyptian as files; and Rongorongo as
tablets. The trained predictor remains fixed. This captures variation
across the evaluation units represented in the sample, but not full
training, catalogue or archaeological uncertainty.

\subsection{Change points and face-model
evidence}\label{change-points-and-face-model-evidence}

For counts \(\mathbf n\), let \(D(\mathbf n)\) be the log integrated
probability from equation \eqref{eq:12}. The zero-order split score
after position \(c\) is

\begin{equation}
\label{eq:28}
G(c)=\frac{D(\mathbf n_{1:c})+D(\mathbf n_{c+1:n})-D(\mathbf n_{1:n})}{\log 2}.
\end{equation}

Positive values favour separate integrated sign-frequency models.
Scanning many positions requires comparing the maximum with maxima from
surrogate scans. The fixed authentic split and the best scanned split
therefore receive different null tests. The comparison
\(L(A+B)-L(A)-L(B)\) likewise has a positive sign when resetting at the
face break shortens a code. At first order it includes finite prior and
sequential-reset effects; it is not an unrestricted causal comparison of
two grammars.

With face-frequency vectors \(p_A,p_B\) smoothed by 1/2 per type and
\(m=(p_A+p_B)/2\), their Jensen-Shannon divergence is

\begin{equation}
\label{eq:29}
\begin{aligned}\operatorname{JSD}(A,B)&=\tfrac12D_{\mathrm{KL}}(p_A\|m)+\tfrac12D_{\mathrm{KL}}(p_B\|m),\\ D_{\mathrm{KL}}(p\|q)&=\sum_s p_s\log_2(p_s/q_s).\end{aligned}
\end{equation}

The smoothing convention is essential for reproducing the reported
0.184-bit value.

\subsection{Categorical Fourier, phase and geometric
scans}\label{categorical-fourier-phase-and-geometric-scans}

The centred indicator spectrum of the sign sequence is

\begin{equation}
\label{eq:30}
P(k)=\frac1n\sum_{s=1}^{q}\left|\sum_{t=0}^{n-1}\big(\mathbf1[x_t=s]-\hat p_s\big)e^{-2\pi i kt/n}\right|^2.
\end{equation}

\(i\) is the imaginary unit, \(k\) a frequency bin and \(n/k\) its
period in sign positions. Renaming signs merely permutes the summed
channels and leaves \(P(k)\) unchanged. The maximum over nonzero
real-FFT bins is compared with corresponding maxima from role-preserving
or intact-group randomisations. The transform is computed on the
compacted sequence after deletion ablations, with an independently
constructed null for that length.

Phase analysis uses

\begin{equation}
\label{eq:31}
I_p=I(X_t;\ t\bmod p),\qquad A(\ell)=\frac1{n-\ell}\sum_{t=0}^{n-\ell-1}\mathbf1[x_t=x_{t+\ell}].
\end{equation}

\(I_p\) measures categorical association with cycle phase, while
\(A(\ell)\) counts the proportion of identical signs at lag \(\ell\). Phase periods range
from 2 to 40, subject to the half-length limit. In the masked A phase
experiment, the original position indices are retained when positions
75--89 are excluded. This prevents deletion from artificially rephasing
the surviving signs. Each phase scan uses 999 independent calibration
shuffles and 1,999 evaluation shuffles. The lag-25 decomposition is an
exact count, not an additional significance test.

For an ideal Archimedean spiral, \(r(\theta)=a+b\theta\) and coordinates
are \((r\cos\theta,r\sin\theta)\). The implementation resamples
equal arc length over the chosen turn count. Candidate edges join nearby
occurrences while excluding short sequential separations. For an edge
set \(E\), identical-sign proximity is

\begin{equation}
\label{eq:32}
S_E=\frac1{|E|}\sum_{(u,v)\in E}\mathbf1[x_u=x_v],
\end{equation}

with endpoint MI calculated from the paired edge labels. The edge set
can change with turn count, so every candidate receives its own
surrogate distribution before the maximum is assessed. Rectangular tests
compare positions at a candidate row-width separation; alternative
traversal tests reorder the same sign identities by column or radius
before coding. These constructions are models of geometry, not
coordinates measured from the Disc.

\subsection{Randomisation, multiplicity and perturbation
interpretation}\label{randomisation-multiplicity-and-perturbation-interpretation}

For a statistic whose larger values indicate structure, the Monte Carlo
tail probability is

\begin{equation}
\label{eq:33}
p=\frac{1+\sum_{r=1}^{B}\mathbf1[T_r\geq T_{\mathrm{obs}}]}{B+1}.
\end{equation}

For compression costs, the inequality is reversed. A lower-tail standard
score is \((\mu_0-C_{\mathrm{obs}})/\sigma_0\); its sign is chosen so
that positive means more economical than the null. Standard scores are
descriptive, and their tail probabilities are computed from permutations
rather than a presumed Gaussian distribution.

For independently calibrated scan values \(z_j\) at candidate
configurations \(j\), the search statistic is \(T=\max_jz_j\). Each
evaluation surrogate undergoes the same full scan. The Arecibo test
follows this procedure with 99 calibration and 199 evaluation shuffles;
the Phaistos phase tests use the larger budgets stated above. The
Phaistos geometry tests use standardised maximum procedures with
test-specific simulation budgets. None of these scan corrections covers unrelated
hypotheses outside its defined family.

The five comparative AUC tests and two Rongorongo coding tests form a
seven-test reported family: Bonferroni correction turns the six raw
\(p=0.001\) values into 0.007. The added Phaistos MI and Re-Pair family
has six tests, each giving \(p=0.0002\), so its Bonferroni values are
0.0012. The three role-concordance probabilities become 0.0006, 0.0192
and 0.0210. The four intact-group Fourier tests give a minimum
Bonferroni probability of 0.0752; the complete A Fourier maximum is thus
evidence within its scan, not a separately family-wide discovery. The
three phase tests give corrected probabilities 0.0030, 0.144 and 0.3045.
These explicit family adjustments coexist with the broader exploratory
qualification.

For the sign-wise analyses, Benjamini-Hochberg adjusted values are
obtained from sorted probabilities as
\(q_{(i)}=\min_{j\geq i}\{mp_{(j)}/j,1\}\) \cite{benjamini1995}. A sign deletion
changes length and context as well as identity, so its matched
fixed-count null is essential. Perturbation signatures describe
sensitivity of the observed representation. They are not interventions
on the ancient maker and do not alone identify a historical cause
\cite{zenil2019causal}.

\subsection{Comparative role interventions and information
landscapes}\label{comparative-role-interventions-and-information-landscapes}

The additional comparative experiments retain exactly the original
training and test candidates. Each candidate has a fixed categorical
feature pair \(Z_i=(u_i,v_i)\), a binary structural role \(Y_i\), and an
object or cord identifier defining its randomisation stratum. We
evaluate three costs on the test set. The empirical conditional-entropy
cost and the integrated conditional KT cost are

\begin{equation}
\label{eq:40}
\begin{aligned}C_H(Y\mid Z)&=-\sum_z\sum_{y=0}^1 n_{zy}\log_2\frac{n_{zy}}{n_z},\\C_{\rm KT}(Y\mid Z)&=-\sum_z\log_2\frac{\mathrm B(n_{z1}+1/2,n_{z0}+1/2)}{\mathrm B(1/2,1/2)}.\end{aligned}
\end{equation}

Here \(n_{zy}\) counts candidates having feature pair \(z\) and label
\(y\), \(n_z=n_{z0}+n_{z1}\), and terms with zero count are zero. The
beta function is \(\mathrm B(a,b)=\Gamma(a)\Gamma(b)/\Gamma(a+b)\). The
empirical entropy cost summarises the observed conditional distribution.
KT integrates a Bernoulli probability separately for each categorical
context under a Jeffreys prior. Both costs are recomputed after each
perturbation; neither is a held-out prediction when estimated from the
test labels themselves. Sparse contexts can have very small empirical
entropy even in random data, so the matched null is essential.

The third cost uses probabilities \(q_i\) from the original training
data only,

\begin{equation}
\label{eq:41}
C_{\rm pred}(Y\mid Z)=-\sum_i\left[Y_i\log_2q_i+(1-Y_i)\log_2(1-q_i)\right].
\end{equation}

These probabilities are kept fixed under every perturbation. All three
costs are divided by the same number of candidates. We generate 999
independent null arrangements by permuting labels within each test
inscription, tablet, page or Egyptian file, and within each cord for
khipu. This preserves each stratum's positive-label count, all features,
the train/test split and the amount of data. The null tests whether the
authentic pairing of local content and structural role has greater
measured organisation than these alternatives.

For a graded perturbation, stratum \(g\) has capacity
\(m_g=\min(n_{g1},n_{g0})\). We randomly pair that many distinct
positive and negative positions within each stratum, pool the eligible
pairs and select a global budget without replacement. At dose
\(f\in\{0,0.1,0.25,0.5,0.75,1\}\), the number selected is the nearest
integer to \(f\sum_gm_g\). Every selected pair exchanges its labels, so
every perturbed role vector has the same within-stratum prevalence. The
achieved distance is

\begin{equation}
\label{eq:42}
d(Y,Y^{(f)})=\tfrac12\sum_i|Y_i-Y_i^{(f)}|=\operatorname{round}\!\left(f\sum_gm_g\right).
\end{equation}

At each nonzero dose, 99 independent subsets give a mean response and
standard deviation. The full-swap endpoint is maximally distant within
this construction; it is not the uniform random-label null. A basin need
not be monotone for every cost, because a sufficiently extensive
exchange can create a different regular role convention.

The frozen predictive response has a useful exact interpretation.
Exchanging a positive position \(i\) with a negative position \(j\)
changes its cost by

\begin{equation}
\label{eq:43}
\delta C_{\rm pred}(i,j)=\log_2\frac{q_i}{1-q_i}-\log_2\frac{q_j}{1-q_j}.
\end{equation}

This follows by subtracting the two original log losses from the two
exchanged log losses. A smooth mean predictive response is therefore
partly an algebraic consequence of score separation. We use it to
interpret the perturbation landscape, and assess independent
generalisation through the original withheld prediction and the new
exclusion tests.

\subsection{Exclusion of shared pairs and common
features}\label{exclusion-of-shared-pairs-and-common-features}

To test transfer beyond exact local combinations, let
\(\mathcal Z_{\rm test}\) be the set of feature pairs anywhere in the
original test set. The pair-embargo training set is

\begin{equation}
\label{eq:44}
\mathcal D_{\rm train}^{\rm embargo}=\{(z,y,g)\in\mathcal D_{\rm train}:z\notin\mathcal Z_{\rm test}\}.
\end{equation}

The exclusion uses test features but no test labels. It is a
deliberately stringent transfer experiment, not a model tuned using the
target outcomes. We refit the same marginal-feature naive-Bayes model on
the remaining training candidates and score the unchanged test set.
Every exact repeated feature pair across training and test is removed.
This does not remove every possible longer passage or near-duplicate
family. The khipu features are particularly coarse, so this exclusion
removes nearly the entire training set; its remaining support is
reported explicitly.

An independent common-feature test identifies the most frequent value of
each of the two features using training data alone, then excludes test
candidates containing either of those two values. The original trained
predictor is retained. The denominator and label prevalence change, so
we report sample sizes and code gain against the corresponding model's
training-prevalence baseline as well as AUC and average precision. Every
estimable exclusion test uses 999 within-stratum label permutations. Its
descriptive AUC interval uses 499 resamples of whole held-out objects,
with khipu resampled as whole khipu rather than individual cords. As in
the original analysis, these intervals condition on the trained model.

\subsection{Grouped hold-out stability and exact-pair embargo}

We evaluated whether comparative structural prediction depended on the original selection of withheld records by repeating the existing marginal-feature naive-Bayes analysis under alternative grouped partitions. The same corpus filters, feature definitions, smoothing and predictive coding procedures were retained. Ten repetitions of five-fold grouped evaluation were performed for Rongorongo, Voynich, khipu and Egyptian, yielding 50 evaluation splits per corpus. A fixed random seed (20260914) was used.

Grouping respected the relevant physical or editorial units. Rongorongo tablets H, P and Q were assigned to a single group to prevent parallel-family leakage. Voynich pages were withheld as complete units, and evaluation pages containing any exact line present in training were excluded separately within each split. Khipu were partitioned as whole objects, and Egyptian files were partitioned after deduplication of identical extracted pair lists. Indus was evaluated using leave-one-site-out validation, excluding from training any inscriptions also associated with the withheld site and retaining only test inscriptions unique to that site. Sites with fewer than ten eligible test gaps were excluded, leaving five evaluable sites.

For each split, the original predictor was fitted on the corresponding training data. A second predictor was fitted after removing all training candidates whose exact feature pair occurred anywhere in the test set. This embargo used test features but not test labels. Both predictors evaluated the same test candidates, and probability-code gains were calculated against their respective smoothed training-prevalence baselines.

Discrimination and probability quality were assessed using AUC, average precision, Brier score and predictive code gain. We additionally recorded training-set retention, the proportion of splits with AUC above 0.5 and the proportion with positive code gain. Results were summarised using medians and empirical 2.5th--97.5th percentiles across evaluated splits. These percentiles describe sensitivity to partition selection; because repeated folds share observations and training data, they are not independent-replicate confidence intervals. No inferential probabilities were assigned to this exploratory stability experiment. The original splits were reimplemented as a numerical consistency check.

\subsection{Rongorongo: the same traversal intervention under three
information
families}\label{rongorongo-the-same-traversal-intervention-under-three-information-families}

The traversal comparison uses the original 229 nonempty lines,
preserving sign identities and line membership. For each of 999 null
replicates, every line is independently reversed with probability one
half. The same reversal masks are supplied to KT1, LZW and the three
CTM-BDM block lengths. A second analysis excludes the entire H/P/Q
parallel-tablet family. Removing exact duplicate whole lines would
remove no lines, so it would not constitute a substantive repetition
control for this corpus.

For CTM-BDM, we use the three-state output counts, the same
reverse/complement symmetrisation and the same shorter-block CTM values
as for Phaistos. The comparative decomposition uses overlapping windows
of length \(b\in\{4,5,6\}\), with stride one, within each line. A line
shorter than \(b\) contributes a single whole-line window. No window
crosses a line boundary. For sign channel \(s\) and binary block \(w\),
let \(N_{s,w}^{(b)}\) be its multiplicity pooled across all line
windows. Then

\begin{equation}
\label{eq:45}
C_b^{\rm pool}(X)=\frac1n\sum_s\sum_{w:N_{s,w}^{(b)}>0}\left[\widehat K(w)+\log_2N_{s,w}^{(b)}\right].
\end{equation}

All sign channels, including all-zero blocks, are included. The one-hot
construction is exactly invariant under relabelling the sign catalogue.
Overlapping windows also enforce a useful global symmetry. If every line
is reversed, \(N_{s,w}\) becomes \(N_{s,w^{\rm rev}}\); because the CTM
table has equal values for these two blocks, the entire sum is
unchanged. This identity is checked numerically. Mixed orientations can
change pooled block reuse even though a complete reversal cannot. The
denominator is the number of original tokens; overlapping BDM remains a
complexity proxy, rather than an emitted lossless code rate.

This explicit decomposition differs from the single-stream offset
average used in the Phaistos experiment. An initial development
calculation pooled offset-specific partitions across unequal line
lengths; its complete-reversal contrast exposed partition-phase
dependence. We therefore used the overlapping construction for the
reported traversal test. The overlapping construction removes this partition-phase dependence. This change enforces the intended mathematical symmetry and
was not selected by its significance level. For each BDM block length,
99 reversal subsets at fractions 0.1, 0.25, 0.5 and 0.75 describe the
landscape; the original and fully reversed endpoints are deterministic.

\subsection{Cross-row alignment and coherent geometric
perturbations}\label{cross-row-alignment-and-coherent-geometric-perturbations}

For an adjacent row pair of lengths \(a\) and \(b\), assign relative
centres \(u_i=(i+1/2)/a\) and \(v_j=(j+1/2)/b\), using zero-based
indices. Only interior tokens are eligible; both first and last tokens
remain outside the test. For each interior token in the first row,
choose the nearest interior centre in the next row,

\begin{equation}
\label{eq:46}
j^*(i)=\operatorname*{arg\,min}_{1\leq j\leq b-2}|u_i-v_j|,\qquad S_{\rm row}=\frac1{|E|}\sum_{(i,j)\in E}\mathbf1[x_i=x_j].
\end{equation}

Ties in nearest distance use the first index, consistently in observed
and null calculations. Both rows must contain at least four tokens. The
statistic pools equality over the selected edges. Rongorongo and Indus
use their transcribed sign categories. Voynich uses the existing
word-form tuple comprising the first two characters, last two characters
and length capped at 12, so equality there concerns word form rather
than literal full-word identity.

Each of 999 geometric nulls circularly shifts every eligible row
interior independently, while keeping its endpoints fixed. The same
sampled shift is reused wherever that row participates in adjacent-row
comparisons. This makes the perturbation coherent across the whole
record. It retains row membership, inventory, endpoint positions and
cyclic interior order, while altering which interior categories face one
another. The equality statistic is recalculated without changing the
nearest-centre rule. We also remove exact duplicate ordered whole-row
pairs and repeat the test. Relative centres provide an explicit layout
model; the transcriptions do not supply measured impression coordinates,
spacing or row curvature.

\subsection{Khipu attachment perturbation and constrained
reconstruction}\label{khipu-attachment-perturbation-and-constrained-reconstruction}

The database's \texttt{PENDANT\_FROM} field supplies a cord's parent
identifier. We retain links between cords on the same khipu, with child
depth exactly one greater than parent depth and at least one recorded S,
L or E knot on each endpoint. Let \(\mathbf c_i\) be the three raw
counts for cord \(i\). Their normalised cosine affinity and its mean
over eligible attachments are

\begin{equation}
\label{eq:47}
s(i,j)=\frac{\mathbf c_i\cdot\mathbf c_j}{\|\mathbf c_i\|\,\|\mathbf c_j\|},\qquad S_T=\frac1{|E_T|}\sum_{(i,j)\in E_T}s(i,j).
\end{equation}

Cosine is a structural-affinity statistic ranging from zero to one for
these nonnegative vectors; it measures profile alignment rather than
information in bits. In this AID experiment it supplies a directly
interpretable response to a computable topological perturbation and a
score for inverse assignment. No interpreted numerical value enters
these vectors. The first null permutes parent identifiers among child
edges within each khipu and child depth. The stricter null further
stratifies by the child's recorded fibre and twist. Only strata with at
least two distinct eligible parents contribute. The operation preserves
every child's one-parent constraint, every parent's outdegree and the
depth layering, hence it cannot create a cycle. It changes actual
attachment identities while retaining these aspects of topology and
material context. We use 999 null rewirings. At six graded doses, we
select a stated fraction of child edges within each stratum and permute
their parent identifiers; because a permutation can leave entries
unchanged, the actual fraction of changed attachments is measured and
reported separately.

The inverse experiment conceals the parent identities assigned to
individual children while retaining candidate parents, their outdegrees
and the same strict strata. For child-parent allocation variables
\(A_{ij}\in\{0,1\}\) and parent capacities \(d_j\), the reconstruction
solves

\begin{equation}
\label{eq:48}
\max_A\sum_{i,j}A_{ij}s(i,j),\qquad \sum_jA_{ij}=1,\qquad \sum_iA_{ij}=d_j.
\end{equation}

We duplicate parent slots according to their capacities and solve the
resulting linear assignment problem. Repeated knot profiles create tied
optima. Ninety-nine independent uniform cost jitters between zero and
\(10^{-10}\) specify tie-breaking without using the authentic
assignment; reported recovery averages these 99 solutions. The same
prediction ensemble is compared with 999 independently rewired target
assignments. Random assignment under the capacities has expected
parent-identity recall

\begin{equation}
\label{eq:49}
r_0=\frac1n\sum_g\frac{\sum_{j\in g}d_j^2}{n_g}.
\end{equation}

Here \(n_g\) is the number of child slots in stratum \(g\). The squared
degrees account for a parent occupying more than one slot. This is
reconstruction conditional on retained topological information, rather
than unconstrained recovery of an unknown tree. It tests whether
knot-form content helps choose the missing attachment identities beyond
the information in the retained degrees and strata.

\subsection{Executable-generator controls and
multiplicity}\label{executable-generator-controls-and-multiplicity}

We construct row and tree records with known generating mechanisms. Each
record's root or first row is uniform over its alphabet. Every
subsequent token either copies its aligned predecessor or parent, with
probability \(c\), or takes a fresh uniform value. In executable form
the local rule is

\begin{equation}
\label{eq:50}
X_v=\begin{cases}X_{\pi(v)},&Q_v<c,\\ U_v,&Q_v\geq c,\end{cases}\qquad Q_v\sim\operatorname{Unif}[0,1],\quad U_v\sim\operatorname{Unif}\{1,\ldots,q\}.
\end{equation}

The map \(\pi(v)\) selects the previous row's aligned token or the
parent node. The mechanism intervention changes \(c\) from 0.8 to 0,
retaining the dimensions, topology, alphabet and uniform expected
one-token marginals. Consequently, the controls distinguish dependence
on geometry from mere marginal sign frequencies. They are a finite,
explicitly specified software-space intervention, not a search over all
computable programmes.

Row controls have six rows of width 20 and alphabet size 12; the
statistic excludes endpoints, as above, and uses pairwise interior
circular-shift nulls. These fixed equal-width row-pair contrasts have
independently uniform relative offsets. Tree controls are rooted ternary
trees of depth four and alphabet size three; rewiring is within child
depth, retaining degrees. The depth-one edges are excluded because they
have only one candidate parent. For each geometry and coupling value, 99
independent records are generated, each with 199 nulls. We report the
fraction with permutation \(p\leq0.05\) and mean excess similarity. This
measures sensitivity and false-positive behaviour for these specified
generators, not power over every possible encoding.

The additional comparative analyses use seed 20260908 plus stable SHA-256-derived
offsets for separate experiments; the stated seeds define the sampling procedures. Holm's
step-down adjustment \cite{holm1979} is applied separately to 15 role-cost tests, ten
estimable exclusion tests, ten traversal comparisons, and nine
spatial/reconstruction tests. If ordered raw probabilities are
\(p_{(1)}\leq\cdots\leq p_{(m)}\), the adjusted values are

\begin{equation}
\label{eq:51}
p_{(j)}^{\rm Holm}=\min\!\left(1,\max_{1\leq k\leq j}(m-k+1)p_{(k)}\right).
\end{equation}

Holm adjustment controls family-wise error for each specified family
without assuming independent tests. The synthetic rejection rates use
their stated nominal threshold because false-positive calibration is the
quantity being measured. As with the earlier analyses, family
corrections do not make the entire adaptive research programme
preregistered. The constrained attachment-recovery experiment was added
after the attachment-affinity result and is identified as an exploratory
inverse test.

\section{Conclusion}\label{conclusion}

The earlier theory proved sufficient asymptotic conditions under which
non-random information retains recoverable geometric and topological
structure, and Arecibo supplied a finite demonstration using a genuine
one-way message. In the present independently calibrated width scan, the
intended 23-column arrangement is the global maximum with \(z=10.174\)
and a search-corrected permutation probability of \(p=0.005\). It
remains among the three leading normalised-BDM candidates in 84.8\% of
trials after exactly 5\% of the transmitted bits are inverted. This
establishes the positive-control case: a concealed geometry can be
reconstructed from content alone, but the answer depends on explicit
treatment of estimator bias, shape normalisation, noise and the full
search over alternative dimensions. 

Phaistos provides the most complete historical result. Relative to
face-and-positional-role-preserving nulls, the pooled inscription
contains 0.714 bits of excess within-compartment adjacent-sign mutual
information (\(p=0.0002\)), while canonical Re-Pair saves 29 grammar
symbols against a null mean of 2.859 (\(p=0.0002\)). When every exact
duplicate of a withheld compartment is removed from training, a
first-order predictor still saves 0.488 bits per internal transition,
whereas its second-order increment falls to \(-0.0013\) bits. The
archaeological divider system is more economical than all 3,000
length-matched surrogates under LZW and KT1 on both faces
(\(p=0.000333\)); the CTM-BDM contrast is also concordant
(\(p=0.0290\) on A and \(p=0.0350\) on B). The mean coding cost rises
monotonically as more authentic dividers are replaced, defining a broad
perturbational basin rather than an isolated optimum.

The strongest reconstruction result is cross-face transfer. Boundary
models trained on one face recover structure on the other with AUC 0.763
for A-to-B and 0.864 for B-to-A, and the discrimination remains 0.696
and 0.721 after every gap adjacent to signs 02 or 12 is removed. 
An additional exact-pair embargo retained AUCs of 0.695 and 0.799, demonstrating transferable boundary information beyond exact adjacent-pair overlap while revealing sensitivity to the removal of shared material and training support.
Cross-face positional-role concordance is 0.430 (\(p=0.0002\)),
remaining significant when those signs are excluded. A change-point scan
places its best division after sign 125, only two positions from the
archaeological face boundary at 123; separate zero-order face models
save 18.395 bits, while pooled first-order coding saves 16.387 bits.
Together these values support two different sign mixtures with reusable
local constraints, but do not identify a unique grammar or semantic
system.

The negative Phaistos results are equally informative. The apparent
A-side recurrence is explained principally by an exact 15-sign repeated
passage: the period-25 phase scan changes from \(p=0.0010\) to
\(p=0.1015\) when one copy is masked while original positions are
retained, and the group-order Fourier scan changes from \(p=0.0188\) to
\(p=0.3262\) when one copy is removed. None of the six complete-face
idealised spiral tests is significant under intact-group controls, and
all six four-turn standard scores lie below their null means. Direction
is unresolved, no individual sign survives false-discovery correction,
and the small CTM effects vary across block lengths. The defensible
reconstruction is therefore a segmented one-dimensional symbolic
topology with face-dependent sign mixtures and transferable local
organisation, not an additional message encoded across the modelled
spiral windings and not a decipherment.

The comparative objects show that this reconstruction principle extends
beyond a stamped linear sequence. Raw knot-form features predict the
distal eligible cluster on held-out khipu with AUC 0.858 and a gain of
0.298 bits per candidate. Recorded Indus row breaks yield AUC 0.812 but
only 0.0092 bits per gap, distinguishing useful ranking from weak
probabilistic information. Voynich line boundaries yield AUC 0.843 and
0.114 bits per gap. The known-script Egyptian control reaches AUC 0.945
and 0.660 bits per join. By contrast, Rongorongo has a significant basin
favouring consistent line orientation under both KT1 and LZW
(\(p=0.001\)), yet held-tablet boundary transfer is indistinguishable
from its matched null (AUC 0.510, \(p=0.382\)) and loses 0.0508 bits per
gap. Geometry, traversal regularity and transferable boundary
information are therefore empirically separable properties.

The Phaistos-guided extensions add a direct test of branching topology.
Across 8,780 eligible khipu attachments with depth, degree, fibre and
twist controls, authentic raw knot-form cosine is 0.53329 against
0.50784 under rewiring (adjusted \(p=0.009\)). Constrained inverse
assignment recovers 8.747\% of parent identities against a degree-based
expectation of 7.866\% (adjusted \(p=0.009\)). This is partial
topological reconstruction supported by a measured perturbation
response. Cross-row alignment does not yield comparable corrected
evidence in Rongorongo or Voynich, and the short eligible Indus
selection provides an uninformative zero-equality statistic.

Transfer beyond exact shared feature pairs retains positive predictive
information in Voynich and Egyptian, while Indus retains ranking but
loses probability-code efficiency. Rongorongo's KT1/LZW traversal
preference survives removal of H/P/Q (adjusted \(p=0.010\)); its three
tested CTM-BDM scales provide an exact global-reversal control without
significant mixed-orientation separation. In executable controls,
changing only the copying mechanism from coupled to independent removes
the geometric signal: all 99 coupled records are detected for each
geometry, compared with one row and six tree false positives among 99
independent controls each.

Repeated grouped evaluation further shows that predictive information persists across alternative withheld records in Voynich and Egyptian, whereas Indus exhibits unstable coding efficiency and khipu pair exclusion causes extreme training-support depletion, delimiting the strength of claims about cross-record generalisation.

AID unifies these results by specifying the relation to perturb, what
the intervention preserves, which information measure records its
effect, and how reconstruction is validated. Its operational use is
agnostic to that measure. Classical information supplies statistical and
predictive descriptions; algorithmic probability adds a connection to
computable generative hypotheses in software space. The controls
demonstrate interventions on known mechanisms, while the archaeological
experiments establish reproducible consequences of changing candidate
structures. This distinction allows causal questions to guide analysis
without converting association into an unsupported claim about an
ancient producer.

The scientific result is a quantitative set of constraints on possible
interpretations. Recoverable geometry is the geometry of the
data-generating organisation, which need not coincide with every visible
dimension of the physical carrier. Candidate decipherments can now be
tested against held-out structural facts: they should predict
compartment boundaries, positional roles, repeated modules, spatial
ranks or layout operators across independent objects, while paying
explicitly for mapping and model complexity. A reading that merely fits
the surviving marks but fails these concealed-structure tests explains
less than the information already demonstrably present in the records.

\section{Data and code availability}\label{data-and-code-availability}

The historical source corpora are available from the repositories cited in
Methods. The analyses use a specified Phaistos transcription, finite
machine-output counts, filtered comparative corpora and the simulation
procedures described above. Exact reproduction requires the analysis scripts,
the Phaistos transcription and its provenance, the machine-output tables,
the precise source revisions and input-file fingerprints, and the numerical
outputs underlying the tables and figures. These materials also include the
44 additional comparative null tests, the four Phaistos exact-pair AUC tests,
the grouped-split evaluations and the executable-generator controls.
Repository links to the historical corpora identify the input sources;
they do not constitute a deposit of the analysis code. The results remain
exploratory, and their archaeological interpretation requires independent
epigraphic assessment.

\appendix
\section{Supplementary tables}
\setcounter{table}{23}
\renewcommand{\thetable}{S\arabic{table}}
\renewcommand{\theHtable}{S\arabic{table}}

\FloatBarrier
{
\small
\setlength{\tabcolsep}{3pt}
\renewcommand{\arraystretch}{1.12}
\begin{longtable}{@{}>{\raggedright\arraybackslash}p{\dimexpr 0.125000\linewidth-5.2500pt\relax}>{\raggedright\arraybackslash}p{\dimexpr 0.125000\linewidth-5.2500pt\relax}>{\raggedright\arraybackslash}p{\dimexpr 0.125000\linewidth-5.2500pt\relax}>{\raggedright\arraybackslash}p{\dimexpr 0.125000\linewidth-5.2500pt\relax}>{\raggedright\arraybackslash}p{\dimexpr 0.125000\linewidth-5.2500pt\relax}>{\raggedright\arraybackslash}p{\dimexpr 0.125000\linewidth-5.2500pt\relax}>{\raggedright\arraybackslash}p{\dimexpr 0.125000\linewidth-5.2500pt\relax}>{\raggedright\arraybackslash}p{\dimexpr 0.125000\linewidth-5.2500pt\relax}@{}}
\caption{Phaistos cross-face prediction before and after exact adjacent-pair exclusion.}\label{tab:s24}\\
\toprule
Transfer & Model & Training gaps & AUC & AP & Gain (bits/gap) & Train-null Holm p & Target-null Holm p \\
\midrule
\endfirsthead
\toprule
Transfer & Model & Training gaps & AUC & AP & Gain (bits/gap) & Train-null Holm p & Target-null Holm p \\
\midrule
\endhead
\bottomrule
\endlastfoot
A to B & Original & 122 & 0.76 & 0.58 & 0.10 & -- & -- \\
A to B & Pair embargo & 91 & 0.70 & 0.52 & 0.05 & \(1.20\times10^{-3}\) & \(5.20\times10^{-3}\) \\
B to A & Original & 118 & 0.86 & 0.77 & 0.22 & -- & -- \\
B to A & Pair embargo & 97 & 0.80 & 0.61 & 0.14 & \(5.20\times10^{-3}\) & \(8.00\times10^{-4}\) \\
\end{longtable}
}

The A-to-B test set contains 118 gaps, and the B-to-A test set contains 122
gaps; the embargo leaves these test sets unchanged. AP denotes average
precision. Code gains use each model's own smoothed training-prevalence
baseline. The four embargoed AUC tests form one Holm family. Dashes indicate
that the original models are not members of this correction family; their
raw null probabilities are reported in Table~\ref{tab:main-7}.

\FloatBarrier
{
\small
\setlength{\tabcolsep}{3pt}
\renewcommand{\arraystretch}{1.12}
\begin{longtable}{@{}>{\raggedright\arraybackslash}p{\dimexpr 0.166667\linewidth-5.0000pt\relax}>{\raggedright\arraybackslash}p{\dimexpr 0.166667\linewidth-5.0000pt\relax}>{\raggedright\arraybackslash}p{\dimexpr 0.166667\linewidth-5.0000pt\relax}>{\raggedright\arraybackslash}p{\dimexpr 0.166667\linewidth-5.0000pt\relax}>{\raggedright\arraybackslash}p{\dimexpr 0.166667\linewidth-5.0000pt\relax}>{\raggedright\arraybackslash}p{\dimexpr 0.166667\linewidth-5.0000pt\relax}@{}}
\caption{Median comparative performance across grouped evaluation splits.}\label{tab:r2-grouped-splits}\\
\toprule
Record & Splits & Original AUC & Embargo AUC & Original gain (bits) & Embargo gain (bits) \\
\midrule
\endfirsthead
\toprule
Record & Splits & Original AUC & Embargo AUC & Original gain (bits) & Embargo gain (bits) \\
\midrule
\endhead
\bottomrule
\endlastfoot
Rongorongo & 50 & 0.56 & 0.53 & -0.05 & -0.05 \\
Indus & 5 & 0.71 & 0.71 & \(9.00\times10^{-3}\) & -0.05 \\
Voynich & 50 & 0.85 & 0.85 & 0.12 & 0.13 \\
Khipu & 50 & 0.86 & 0.64 & 0.30 & 0.05 \\
Egyptian & 50 & 0.94 & 0.89 & 0.48 & 0.35 \\
\end{longtable}
}

Gains are bits per candidate relative to each model's training-prevalence
baseline. Four corpora use ten repetitions of five-fold grouped evaluation;
Indus uses five eligible leave-one-site-out evaluations. Repeated folds share
observations and are not independent experimental replications. The khipu
embargo retains a median of 25 training candidates, approximately 0.081\% of
the original training set. Only two of the five embargoed Indus site
evaluations have positive code gain. The graphical partition-sensitivity
summaries are shown in Figure~\ref{fig:r2-grouped-stability}.


\begin{thebibliography}{99}
\bibitem{zenil2024geometry} Zenil, H., Abrah\~ao, F. S. \& Ozelim, L. C. S. M. Non-Random Data Encodes Its Geometric and Topological Dimensions. arXiv:2405.07803v3 (2024). \url{https://arxiv.org/html/2405.07803v3}
\bibitem{zenil2024arecibo} Zenil, H., Adams, A. \& Abrah\~ao, F. S. An Optimal, Universal and Agnostic Decoding Method for Message Reconstruction, Bio and Technosignature Detection. arXiv:2303.16045v3 (2024). \url{https://arxiv.org/html/2303.16045v3}
\bibitem{heraklionPhaistos} Heraklion Archaeological Museum. The Phaistos Disc. Museum catalogue and object description. \url{https://heraklionmuseum.gr/en/exhibit/the-phaistos-disc/}
\bibitem{urton2005khipu} Urton, G. \& Brezine, C. J. Khipu accounting in ancient Peru. \emph{Science} \textbf{309}, 1065--1067 (2005). \url{https://pubmed.ncbi.nlm.nih.gov/16099983/}
\bibitem{openKhipu} OKR Team. The Open Khipu Repository, version 2.1.0 (2025). Database and documentation. \url{https://github.com/khipulab/open-khipu-repository}; archive: \url{https://doi.org/10.5281/zenodo.18025748}
\bibitem{rongopy} de Souza, J. G. Rongopy: computational experiments and machine-readable Rongorongo transcriptions. \url{https://github.com/jgregoriods/rongopy}
\bibitem{indusCorpus} Indus Valley Script Corpus. Derivative digital corpus with provenance to the Lipi export and ICIT. \url{https://github.com/field-cady/indus_valley_script_corpus}; upstream: \url{https://github.com/yajnadevam/lipi}
\bibitem{reddy2011voynich} Reddy, S. \& Knight, K. What We Know About the Voynich Manuscript. Proceedings of the 5th ACL-HLT Workshop on Language Technology for Cultural Heritage, Social Sciences, and Humanities, 78--86 (2011). \url{https://aclanthology.org/W11-1511/}
\bibitem{evaVoynich} Zandbergen, R. \& Landini, G. EVA transcription, version 3b, 13 May 2025. Analysed copy: \url{https://github.com/Aspect-Research/voynich-autoexploration/blob/master/data/transcriptions/eva_zl3b.txt}
\bibitem{rosmorducMdcDoc} Rosmorduc, S. Manuel de Codage documentation. \url{https://khety.qenherkhopeshef.org/HieroEncoding/MDC/mdc.html}
\bibitem{rosmorducMdcTexts} Rosmorduc, S. Manuel de Codage texts for JSesh. \url{https://github.com/rosmord/MDC-texts}
\bibitem{jackson1999phaistos} Jackson, M. P. A Statistical Study of the Phaistos Disc. \emph{Kadmos} \textbf{38}(1--2), 19--30 (1999). \url{https://doi.org/10.1515/kadm.1999.38.1-2.19}
\bibitem{davis2026phaistos} Davis, B. Syllabotactic analysis of the Phaistos Disk script against Linear A. In \emph{The Undeciphered Aegean Scripts}, 314--339 (Cambridge University Press, 2026). \url{https://doi.org/10.1017/9781009562331.009}
\bibitem{giorgi2026phaistos} Giorgi, F. M. \& Baldacci, F. A Cryptanalytic Test of the Phaistos Disc as a Protein Sequence. \emph{Cryptography} \textbf{10}, 60 (2026). \url{https://doi.org/10.3390/cryptography10040060}
\bibitem{areciboTranscription} ThomasSalty. Arecibo: binary-message transcription and reconstruction code. \url{https://github.com/ThomasSalty/Arecibo/blob/master/be.txt}
\bibitem{shannon1948} Shannon, C. E. A mathematical theory of communication. \emph{Bell System Technical Journal} \textbf{27}, 379--423 and 623--656 (1948).
\bibitem{krichevsky1981} Krichevsky, R. E. \& Trofimov, V. K. The performance of universal encoding. \emph{IEEE Transactions on Information Theory} \textbf{27}, 199--207 (1981).
\bibitem{welch1984} Welch, T. A. A technique for high-performance data compression. \emph{Computer} \textbf{17}, 8--19 (1984).
\bibitem{liVitanyi2008} Li, M. \& Vit\'anyi, P. \emph{An Introduction to Kolmogorov Complexity and Its Applications}, 3rd edn (Springer, 2008).
\bibitem{solerToscano2014} Soler-Toscano, F., Zenil, H., Delahaye, J.-P. \& Gauvrit, N. Calculating Kolmogorov complexity from the output frequency distributions of small Turing machines. \emph{PLoS ONE} \textbf{9}, e96223 (2014). \url{https://doi.org/10.1371/journal.pone.0096223}
\bibitem{zenil2018bdm} Zenil, H. et al. A decomposition method for global evaluation of Shannon entropy and local estimations of algorithmic complexity. \emph{Entropy} \textbf{20}, 605 (2018). \url{https://doi.org/10.3390/e20080605}
\bibitem{pybdmDocs} PyBDM. Block decomposition method implementation and CTM datasets. \url{https://pybdm-docs.readthedocs.io/en/latest/}
\bibitem{benjamini1995} Benjamini, Y. \& Hochberg, Y. Controlling the false discovery rate: a practical and powerful approach to multiple testing. \emph{Journal of the Royal Statistical Society B} \textbf{57}, 289--300 (1995).
\bibitem{zenil2019causal} Zenil, H., Kiani, N. A., Zea, A. A. \& Tegn\'er, J. Causal deconvolution by algorithmic generative models. \emph{Nature Machine Intelligence} \textbf{1}, 58--66 (2019). \url{https://doi.org/10.1038/s42256-018-0005-0}
\bibitem{aidScholarpedia} Algorithmic Information Dynamics. Scholarpedia. \url{https://scholarpedia.org/article/Algorithmic_Information_Dynamics}
\bibitem{pearl2009} Pearl, J. Causal inference in statistics: An overview. \emph{Statistics Surveys} \textbf{3}, 96--146 (2009). \url{https://ftp.cs.ucla.edu/pub/stat_ser/r350.pdf}
\bibitem{ziv1978} Ziv, J. \& Lempel, A. Compression of individual sequences via variable-rate coding. \emph{IEEE Transactions on Information Theory} \textbf{24}(5), 530--536 (1978). \url{https://doi.org/10.1109/TIT.1978.1055934}
\bibitem{larsson1999} Larsson, N. J. \& Moffat, A. Offline dictionary-based compression. In \emph{Proceedings of the Data Compression Conference}, 296--305 (IEEE, 1999). \url{https://doi.org/10.1109/DCC.1999.755679}
\bibitem{holm1979} Holm, S. A simple sequentially rejective multiple test procedure. \emph{Scandinavian Journal of Statistics} \textbf{6}(2), 65--70 (1979). \url{https://www.jstor.org/stable/4615733}
\end{thebibliography}
\end{document}